%% file: main.tex
\documentclass[prx, twocolumn, floatfix, superscriptaddress, nofootinbib, nolongbibliography]{revtex4-2}
\usepackage{graphicx}
\usepackage{amsmath,amssymb,mathtools,bm,dsfont,tikz}
\usepackage{tikz-cd}
\usepackage{amsthm}
\usepackage[notrig]{physics}
\usepackage[colorlinks=true, allcolors=purple]{hyperref}
\usepackage[normalem]{ulem}

\newcommand{\gs}{\text{g.s.}}

\newcommand{\LGT}{\text{LGT}}

\newcommand{\ITE}{\text{ITE}}
\newcommand{\Ising}{\text{I}}

\newcommand{\vh}{\boldsymbol{h}}
\newcommand{\ve}{\boldsymbol{e}}

\newcommand{\vs}{\boldsymbol{s}}

\newcommand{\bh}{\boldsymbol{\eta}}
\newcommand{\bs}{\boldsymbol{\sigma}}

\DeclareMathOperator{\poly}{poly}

\theoremstyle{definition}

\makeatletter 
    
\renewcommand\onecolumngrid{
\do@columngrid{one}{\@ne}%
\def\set@footnotewidth{\onecolumngrid}
\def\footnoterule{\kern-6pt\hrule width 1.5in\kern6pt}%
}

\renewcommand\twocolumngrid{
        \def\footnoterule{
        \dimen@\skip\footins\divide\dimen@\thr@@
        \kern-\dimen@\hrule width.5in\kern\dimen@}
        \do@columngrid{mlt}{\tw@}
}%

\makeatother

\begin{document}

\title{Quantum Error Correction in Nearly Critical Toric Codes}


\author{Zack Weinstein}
\email{zackmw@caltech.edu}
\affiliation{Department of Physics and Institute for Quantum Information and Matter, California Institute of Technology, Pasadena, California 91125, USA}

\author{Samuel J. Garratt}
\affiliation{Department of Physics, Princeton University, Princeton, New Jersey 08544, USA}


\begin{abstract}
We investigate quantum error correction protocols for topological codes deformed away from their stabilizer limits, particularly as they are pushed towards quantum phase transitions into trivial phases. Our focus is primarily on quantum information stored within the ground-state manifold of the transverse-field toric code (TFTC), subjected to independent bit-flip and phase-flip decoherence. Although the transverse field induces strong quantum fluctuations of $m$ anyons in the ground states, we show that projective measurements of the ``bare'' toric code stabilizers can nevertheless be used to construct an efficient decoding algorithm which exhibits a finite threshold to both types of decoherence, even when the transverse field is tuned infinitesimally close to the quantum critical point. Our construction of the optimal postmeasurement decoders---that is, the optimal decoders \emph{assuming} that the stabilizers are first measured---crucially relies on the \emph{stoquastic} nature of the TFTC Hamiltonian, which affords a description of its ground states in terms of a classical lattice gauge theory. Using stoquasticity, we first prove that stabilizer measurements act unitarily on the undecohered TFTC ground states to excellent approximation, and that the optimal postmeasurement decoder can be efficiently implemented via Monte Carlo simulation of a constrained lattice gauge theory. We then construct the optimal postmeasurement decoder for the TFTC states under bit-flip and phase-flip decoherence, where we numerically demonstrate a finite threshold close to the quantum critical point. We provide theoretical evidence (without a rigorous proof) that stabilizer measurements constitute an optimal decoding strategy in this setting. For phase-flip decoherence, we further prove that the optimal postmeasurement decoding threshold is independent of the transverse field strength throughout the topological phase. Finally, we briefly investigate quantum information stored in the ground-state manifold of the ferromagnetic transverse-field Ising model, viewed as a Hamiltonian deformation of the repetition code. We prove that the error threshold to bit-flip decoherence is maximal throughout the entire ferromagnetic phase, even arbitrarily close to the quantum critical point, and that stabilizer measurements and majority voting can be used as an efficient optimal decoder.
\end{abstract}

\maketitle

\input{sections/intro.tex}

\newpage
\tableofcontents

\input{sections/background.tex}

\input{sections/overview.tex}

\input{sections/meas_decoding.tex}

\input{sections/pauli_meas_decoding.tex}

\input{sections/tfim.tex}

\input{sections/discussion.tex}

\begin{acknowledgments}
    We thank Ehud Altman, Ruihua Fan, Tim Hsieh, Michael Knap, Nathaniel Selub, and Simon Trebst for fruitful discussions, and Zac Tobias in particular for clarifying discussions on the worm algorithm. Z.W. acknowledges funding provided by the Institute for Quantum Information and Matter, an NSF Physics Frontiers Center (NSF Grant PHY-2317110). S.J.G. was supported by the Gordon \& Betty Moore Foundation, and in part by a Brown Investigator Award, a program of the Brown Institute for Basic Sciences at the California Institute of Technology. This research was done using services provided by the OSG Consortium~\cite{osg1,osg2,osg3,osg4}, which is supported by the National Science Foundation awards \#2030508 and \#2323298.

    \emph{Use of Artificial Intelligence.}---The authors acknowledge the use of artificial intelligence (AI) for finding additional references, assisting with proofs, analysis of numerical data, optimization of Monte Carlo parameters, and proofreading of this manuscript. No AI was used for writing code used in this work's Monte Carlo simulations, or in writing any part of this manuscript. 
\end{acknowledgments}

\newpage
\onecolumngrid
\appendix

\input{appendices/quantum_classical.tex}
\input{appendices/ite_overlaps.tex}

\input{appendices/decoding_fidelities.tex}
\input{appendices/imag_adiabatic.tex}

\input{appendices/pruning.tex}

\input{appendices/strong_symm.tex}

\input{appendices/coherent_decoding.tex}
\input{appendices/worm.tex}
\input{appendices/more_numerics.tex}

\input{appendices/tfim_numerics.tex}

\newpage
\twocolumngrid
\bibliography{refs}

\end{document}

%% file: sections/intro.tex
\section{Introduction}


Quantum error correction (QEC) is a concept of paramount importance in modern quantum physics. Besides its manifest practical importance for realizing large-scale fault-tolerant quantum computation \cite{kitaev_quantum_1997,kitaev_quantum_1997a,aharonov_faulttolerant_2008}, QEC has also emerged as an invaluable \emph{conceptual} framework for understanding a diverse array of problems across many-body physics. Indeed, the robust protection of quantum information in a degenerate ground-state manifold has long been considered a fundamental definition of topological quantum order \cite{kitaev_fault-tolerant_2003,nussinov_symmetry_2009,bravyi_topological_2010}. More recently, classic works on QEC in topological codes \cite{dennis_topological_2002,bombin_introduction_2013} have been reinterpreted through the lens of mixed-state topological order \cite{fan_diagnostics_2024,bao_mixedstate_2023}, prompting vigorous investigation into new mixed-state phases of matter \cite{leeQuantumCriticalityDecoherence2023,sangMixedStateQuantumPhases2024,chenSeparabilityTransitionsTopological2024,lessa2024strongtoweakspontaneoussymmetrybreaking,sala2024spontaneousstrongsymmetrybreaking,sang_stability_2025,wang_intrinsic_2025,sohal_noisy_2025,ellison_classification_2025,lessa_higherform_2025,weinstein_efficient_2025}. In addition to its central role in topologically ordered phases, QEC has provided new perspectives on diverse problems such as thermalization in unitary and monitored quantum dynamics \cite{haydenBlackHolesMirrors2007,hosurChaosQuantumChannels2016,brandao_quantum_2019,gullans_dynamical_2020,choi_quantum_2020,li_statistical_2021}, classical and quantum spin glasses \cite{rakovszky2024bottlenecksquantumchannelsfinite,placke2024topologicalquantumspinglass,placke2025expansioncreatesspinglassorder}, and the AdS-CFT correspondence \cite{almheiri_bulk_2015,pastawski_holographic_2015,yangBidirectionalHolographicCodes2016}.

\begin{figure}[!b]
    \includegraphics[width = \columnwidth]{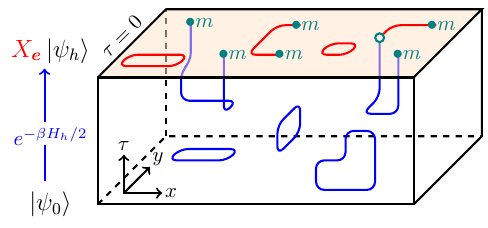}
    \caption{Schematic cartoon of bit-flip decoherence on the transverse-field toric code (TFTC). Each ground state $\ket{\psi_{h}}$ of the TFTC can be prepared with high fidelity via imaginary-time evolution from a reference stabilizer toric code state $\ket{\psi_{0}}$. The quantum fluctuations of $m$ anyons can be effectively captured through a sum over all possible spacetime histories (blue) during imaginary times $\tau < 0$. A random bit-flip error $X_{\ve}$ then acts on the state at imaginary time $\tau = 0$ (red), creating further $m$ anyons. A stabilizer measurement reveals the mod-two superposition of anyons from the ground-state fluctuations and from the bit-flip error.}
    \label{fig:cartoon}
\end{figure}

Although the protection of quantum information is widely believed to be an intrinsic property of topological phases \cite{nayak_nonabelian_2008}, QEC is typically understood in detail only for especially simple toy models \cite{dennis_topological_2002,wang_confinement-higgs_2003,katzgraber_error_2009,stace_thresholds_2009,bombin_strong_2012,chen_unconventional_2024,suTapestryDualitiesDecohered2024,lyons_understanding_2024,temkin_charge-informed_2025}. The most well-studied example is the toric code \cite{kitaev_fault-tolerant_2003}, where a theory of optimal decoding has been thoroughly developed in terms of the statistical physics of the Nishimori random-bond Ising model (RBIM) \cite{dennis_topological_2002,bombin_introduction_2013,fan_diagnostics_2024,wang_confinement-higgs_2003,lee_exact_2025}. 
%
%
In recent works~\cite{Bentsen2024approximatequantum,kim2024errorthresholdsykcodes,sang_approximate_2024,wang_fractional_2025,sommers_spectral_2025,lip}, the intrinsic error protection properties of certain analytically tractable nonstabilizer codes have been similarly investigated using statistical physics techniques. Unfortunately, although the existence of a decoder which recovers the protected information is guaranteed in these codes when the decoherence strength is below a critical threshold, it is unclear how to devise a conceptually simple or algorithmically efficient decoding algorithm which recovers the information. We are thus led to ask if it is ever possible to efficiently decode, i.e., carry out QEC, in these physically motivated nonstabilizer codes.

In this work, we develop decoding protocols for a class of topological quantum memories deformed away from their stabilizer limit. Our central focus is on the toric code perturbed by a transverse field \cite{wegnerDualityGeneralizedIsing1971,kogutIntroductionLatticeGauge1979,trebst_breakdown_2007,vidal_low-energy_2009,dusuel_robustness_2011,sachdevQuantumPhaseTransitions2011,fradkin_field_2013}, which provides quantum fluctuations to the $m$ anyons of the toric code. Below a critical field strength $h_{c}$, the transverse-field toric code (TFTC) retains a topologically ordered ground-state manifold which can encode two qubits of quantum information in a deformed\footnote{To avoid confusion, we emphasize that our code space is deformed by the Hamiltonian's transverse field and \emph{not} by ``wavefunction deformation'' as in Refs.~\cite{castelnovo_quantum_2008,zhu_gapless_2019,huxford_gaining_2023,sala_decoherence_2025}.} code space. Above $h_{c}$, the $m$ anyons condense and destroy the topological order, resulting in a unique gapped ground state incapable of encoding information.

Since the TFTC ground states themselves exhibit coherent quantum fluctuations of $m$ anyons, the `true' optimal recovery channel---defined so as to maximize the fidelity of recovered quantum information---is generally a highly nonlocal and exponentially complex quantum channel~\cite{schumacherQuantumDataProcessing1996,schumacher_approximate_2002}; in particular, its implementation requires the precise ground-state wavefunctions of the non-solvable TFTC Hamiltonian. Remarkably, we will nevertheless demonstrate that ground states of the TFTC under sufficiently weak bit-flip and phase-flip decoherence can be reliably decoded using projective measurements of the ``bare'' stabilizers of the unperturbed toric code, even when the transverse field is tuned close to criticality. As we will elaborate, the success of this simple-minded decoding strategy relies crucially on a special property of the TFTC Hamiltonian: namely, the Hamiltonian is \emph{stoquastic} \cite{bravyi2007complexitystoquasticlocalhamiltonian}. Conceptually, stoquasticity allows for a physical representation of the TFTC ground states using the classical statistical physics of lattice gauge theories~\cite{wegnerDualityGeneralizedIsing1971,fradkinPhaseDiagramsLattice1979,kogutIntroductionLatticeGauge1979}. More concretely, stoquasticity leads to a basis of approximate ground states which are strictly ``sign-free'', leading to several technical simplifications in the construction of optimal decoders for the postmeasurement states.

As a first step, we investigate the preservation of quantum information in pure TFTC ground states under projective stabilizer measurements. One motivation for this problem arises from previous investigations of the stabilizer toric code subjected to ``coherent errors'' \cite{bravyi_correcting_2018,venn_coherent_2023,behrends_surface_2024,behrends2024statisticalmechanicalmappingmaximumlikelihood,bao2024phasesdecodabilitysurfacecode,cheng_emergent_2025,yan_non-linear_2026}, and in particular the work of Refs.~\cite{lavasani_stability_2025,liu_detecting_2025}. Specifically, Refs.~\cite{bravyi_correcting_2018,venn_coherent_2023,behrends_surface_2024,behrends2024statisticalmechanicalmappingmaximumlikelihood,bao2024phasesdecodabilitysurfacecode,cheng_emergent_2025,yan_non-linear_2026} considered projective stabilizer measurements of toric code states deformed by a low-depth unitary circuit, and found that certain low-depth circuits (for example, uniform single-qubit rotations) followed by stabilizer measurements were sufficient to irreversibly destroy the encoded information. In contrast, Ref.~\cite{lavasani_stability_2025} studied quantum information encoded in the ground-state manifold of the Castelnovo-Chamon model \cite{castelnovo_quantum_2008,huxford_gaining_2023}, an analytically tractable (albeit non-generic) deformation of the toric code Hamiltonian. There it was shown that the encoded information was protected from stabilizer measurements throughout the entire topological phase of the model.
%
%
Ref.~\cite{liu_detecting_2025} extended these results in several directions, 
%
%
and drew connections between QEC and emergent higher-form symmetries. Since these Hamiltonian deformations correspond to unitary circuits whose depth diverges as the phase boundary is approached \cite{chen_local_2010}, it is natural to ask why this class of coherent errors can be effectively corrected via stabilizer measurements, while generic shallow-depth coherent errors cannot. 

\begin{figure}[t]
    \includegraphics[width=\columnwidth]{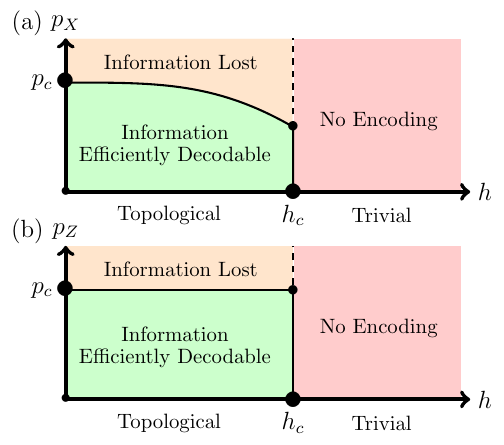}
    \caption{Phase diagrams of optimal postmeasurement decoding in the TFTC under independent bit-flip and phase-flip decoherence, with strengths $p_X$ and $p_Z$ respectively. (a) Even when the transverse field $h$ is tuned infinitesimally close to the quantum critical point $h_c$, quantum information is protected and efficiently decodable via stabilizer measurements up to a finite bit-flip strength $p_X$. The value of the bit-flip error threshold remains finite as $h$ approaches $h_c$; see Fig.~\ref{fig:lgt_bitflip_decoding}(c) for quantitative error thresholds in a closely related spacetime-isotropic model. (b) In contrast, the error threshold for phase-flip decoherence remains exactly constant and equal to its stabilizer value throughout the topological phase.}
    \label{fig:phase_diagrams}
\end{figure}

Here we extend the results of Refs.~\cite{lavasani_stability_2025,liu_detecting_2025} to the TFTC, a more generic Hamiltonian deformation of the toric code, showing once more that the encoded information is protected from stabilizer measurements throughout its entire topological phase. Furthermore, we explain the preservation of information following syndrome measurements in all of these models as arising from a common mechanism, namely, the stoquasticity of the Hamiltonian deformations in question. We first prove (under very mild physical assumptions) that quantum information is preserved under syndrome measurements of the TFTC throughout its topological phase. Then, we explicitly construct and numerically investigate the optimal decoder for the postmeasurement states. This decoder is described by a three-dimensional lattice gauge theory with a constraint along a single imaginary time slice arising from syndrome measurements. We can efficiently simulate this model, and thereby sample syndromes and implement the optimal postmeasurement decoder, using the \emph{worm algorithm} (WA) \cite{prokofev_worm_1998,prokofev_worm_2001,deng_dynamic_2007,collevecchio_worm_2016,tobias_optimal_2026}. From our simulations, we explicitly verify our theoretical prediction that the fidelity of recovered information is exponentially close to unity throughout the topological phase.

Our second motivation for studying stabilizer measurements of TFTC ground states is the construction of an efficient decoder for the TFTC under Pauli noise. In a companion work \cite{lip}, we argued that a broad class of Hamiltonian-deformed toric codes exhibit a finite \emph{intrinsic} error threshold to Pauli noise as a critical point is approached. This intrinsic threshold is defined by the behavior of the coherent information \cite{schumacherQuantumDataProcessing1996,schumacher_approximate_2002,nielsenQuantumComputationQuantum2010}, which guarantees the \emph{existence} of a decoder below threshold. Inspired by the preceding result in the undecohered code states, we investigate the efficacy of stabilizer measurements for performing QEC on the decohered code states. Specifically, we construct and numerically implement the optimal postmeasurement decoder for the TFTC under independent bit-flip and phase-flip decoherence and subsequent stabilizer measurements. The optimal postmeasurement decoder in the presence of decoherence is described by a somewhat more complex statistical physics model, involving a three-dimensional gauge theory coupled to a two-dimensional independent-spin model by a constraint, which can again be simulated efficiently via the WA. We demonstrate\footnote{As a technical point of caution: whereas the TFTC is quantitatively described by the extremely anisotropic (continuous-time) limit of a lattice gauge theory~\cite{fradkin_order_1978,kogutIntroductionLatticeGauge1979}, our numerics are performed on a \emph{spacetime-isotropic} lattice gauge theory for technical simplicity. We expect all \emph{universal} conclusions of our numerical results to carry over to the TFTC, but precise quantitative values of error thresholds are unreliable.} that this decoder exhibits a finite error threshold all the way to the TFTC critical point [see Fig.~\ref{fig:phase_diagrams}(a)], thereby providing direct numerical support for the results of Ref.~\cite{lip}. 

While the decoders we construct are optimal given that stabilizer measurements have been performed, a natural question is whether stabilizer measurements themselves are optimal---that is, whether the decoding threshold using stabilizer measurements agrees with the `intrinsic' threshold set by the `true' optimal recovery channel. Although we are unable to construct a fully rigorous proof for the case of bit-flip decoherence (where the transverse field and decoherence ``work together'' to create $m$ anyons), we provide analytical arguments suggesting that this is indeed the case. For phase-flip decoherence, we prove that the error threshold of the optimal postmeasurement decoder is entirely independent of the transverse field strength---i.e., it is equal to its value in the stabilizer toric code
%
%
($p^{\text{Nishi}}_{c} \approx 0.109$) throughout the topological phase of the TFTC [see Figure~\ref{fig:phase_diagrams}(b)]. Roughly speaking, this behavior arises because the TFTC ground states contain no $e$ anyons themselves, and so the only $e$ anyons present are those created by phase-flip decoherence.


To further illustrate the power of our approach, we also investigate the $d$-dimensional transverse-field Ising model (TFIM) \cite{fradkin_field_2013,sachdevQuantumPhaseTransitions2011} in its ferromagnetic phase as a Hamiltonian deformation of the repetition code \cite{lavasani_stability_2025}. Such a code can be viewed as a ``symmetry-protected'' quantum code, in the sense that it exhibits excellent protection against decoherence which respects the Ising symmetry, but no protection whatsoever against decoherence which breaks the symmetry. As we will show explicitly, the TFIM code exhibits a maximal threshold to bit-flip decoherence throughout the ferromagnetic phase, and QEC can be performed efficiently by measuring the repetition code stabilizers and utilizing a majority-vote decoder. In short, this decoder succeeds with high fidelity because each symmetry-broken code state exhibits a nonzero magnetization density with vanishingly small fluctuations, for all non-maximal decoherence strengths; consequently, the domain walls observed via stabilizer measurement can be unambiguously annihilated to recover a repetition code state.

This work is organized as follows. In Sec.~\ref{sec:background}, we review the necessary background on the TFTC, and particularly its connection to $\mathbb{Z}_{2}$ lattice gauge theory. In Sec.~\ref{sec:overview} we broadly set up the QEC problem of interest and lay out the primary observables and technical tools we will utilize. In Sec.~\ref{sec:meas_decoding} we demonstrate that stabilizer measurements of the TFTC ground states preserve their encoded information with high fidelity. In Sec.~\ref{sec:pauli_meas_decoding} we incorporate Pauli decoherence on these states, where we demonstrate that both bit-flip and phase-flip decoherence can be efficiently decoded throughout the topological phase of the TFTC. In Sec.~\ref{sec:tfim} we briefly discuss QEC in the TFIM, which admits a maximal threshold to bit-flip decoherence throughout the ferromagnetic phase. We conclude with a discussion in Sec.~\ref{sec:discussion}, followed by several pedagogical and technical appendices.

%% file: sections/background.tex
\section{Background}
\label{sec:background}
%
We begin by fixing our notation and conventions on the stabilizer toric code, and by reviewing the transverse-field toric code (TFTC) and its relation to $\mathbb{Z}_{2}$ lattice gauge theory \cite{wegnerDualityGeneralizedIsing1971,fradkinPhaseDiagramsLattice1979,kogutIntroductionLatticeGauge1979}. We additionally review the quantum-classical mapping for this model \cite{fradkin_order_1978,kogutIntroductionLatticeGauge1979,sachdevQuantumPhaseTransitions2011,fradkin_field_2013}, which will provide significant physical intuition for many results throughout this work.

\subsection{Stabilizer Toric Code}
\label{subsec:stabilizer_tc}
\begin{figure}
    \includegraphics[width = \columnwidth]{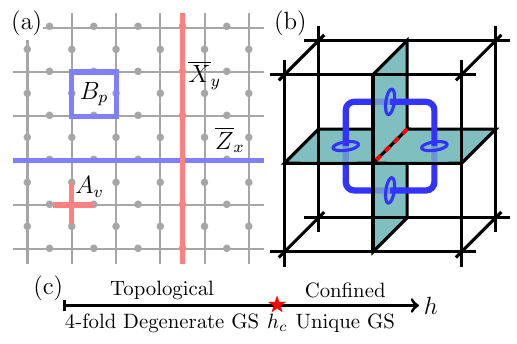}
    \caption{(a) Stabilizers $A_{v}, B_{p}$ and two representative logical operators $\overline{Z}_{x}, \overline{X}_{y}$ of the toric code. Blue (red) denotes Pauli-$Z$ (Pauli-$X$) operators. (b) Four unit cells of the 3$d$ classical $\mathbb{Z}_{2}$ LGT [see Eq.~\eqref{eq:lgt_classical}], obtained from the TFTC via the quantum-classical mapping. Classical Ising spins $U_{ij}$ are placed on the edges of a spacetime lattice with vertices $i,j,k,l$. One flipped spin $U_{ij} = -1$ (dashed red line) leads to four violated plaquettes (teal), with energetic costs $2\varepsilon$ and $2J$ for spacelike and timelike plaquettes respectively. Since every cube must have an even number of violated plaquettes, each configuration can be visualized as a closed-loop flux worldline in the dual lattice (blue) passing through violated plaquettes. (c) Phase diagram of the TFTC, as a function of the transverse field $h$. For $h < h_{c}$ the model exhibits a topological phase with a four-dimensional ground-state manifold $\mathcal{C}_{h}$, capable of encoding two qubits of logical information. For $h > h_{c}$ the $m$ anyons condense, leading to a confined phase with a unique ground state.}
    \label{fig:background}
\end{figure}

We consider an $L \times L$ square lattice on the torus, i.e., with periodic boundary conditions in both directions. Each link $\ell$ of the lattice is equipped with a qubit, on which the local Pauli operators $X_{\ell}$ and $Z_{\ell}$ act. The familiar stabilizer toric code is defined by the Hamiltonian~\cite{kitaev_fault-tolerant_2003}
\begin{equation}
    \label{eq:stabilizer_tc}
    H_{0} \equiv - \sum_{v} A_{v} - \sum_{p} B_{p}, 
\end{equation}
where the vertex stabilizers $A_{v}$ and the plaquette stabilizers $B_{p}$ are defined respectively as [see Fig.~\ref{fig:background}(a)]
\begin{equation}
    A_{v} \equiv \prod_{\ell \ni v} X_{\ell}, \quad B_{p} \equiv \prod_{\ell \in p} Z_{\ell} .
\end{equation}
The former product is taken over the four links $\ell$ containing the vertex $v$, while the latter product is taken over the four links $\ell$ belonging to a plaquette $p$. The code space $\mathcal{C}_{0}$ of the stabilizer toric code is the ground-state manifold of $H_{0}$, i.e., the simultaneous $+1$ eigenspace of the mutually commuting stabilizers $A_{v}$ and $B_{p}$. This subspace is fourfold degenerate on the torus. Excitations with $A_{v} = -1$ and $B_{p} = -1$ are the $e$ and $m$ anyons respectively. 

The logical operators of the stabilizer toric code generate two microscopic $\mathbb{Z}_{2}$ 1-form symmetries \cite{wen_emergent_2019,mcgreevy_generalized_2023}. They are given by the product of Pauli-$Z$ (Pauli-$X$) operators around non-contractible cycles of the torus:
\begin{equation}
    \label{eq:logicals}
    \overline{Z}_{\mu} \equiv \prod_{\ell \in C_{\mu}} Z_{\ell}, \quad \overline{X}_{\mu} \equiv \prod_{\ell \in \tilde{C}_{\mu}} X_{\ell} ,
\end{equation}
where $C_{\mu}$ ($\tilde{C}_{\mu}$) is a non-contractible closed loop through the direct (dual) lattice about the $\mu$th cycle of the torus [Fig.~\ref{fig:background}(a)], with $\mu = x, y$. While $\overline{Z}_{\mu}$ and $\overline{X}_{\mu}$ commute with each of the stabilizers $A_{v}$ and $B_{p}$, they satisfy the (anti)commutation relations $\acomm{\overline{X}_{x}}{\overline{Z}_{y}} = \acomm{\overline{X}_{y}}{\overline{Z}_{x}} = 0$ and $\comm{\overline{X}_{x}}{\overline{Z}_{x}} = \comm{\overline{X}_{y}}{\overline{Z}_{y}} = 0$. Notably, the action of $\overline{Z}_{\mu}$ and $\overline{X}_{\mu}$ on the code space $\mathcal{C}_{0}$ depends only on the \emph{homology}\footnote{Throughout this work, we say that a pair of Pauli-$X$ (Pauli-$Z$) operators are \emph{homologous} if they differ by products of $A_{v}$ ($B_{p}$) stabilizers. Topologically, two such homologous operators differ by \emph{contractible} closed loops of $X_{\ell}$ ($Z_{\ell}$) operators in the dual (direct) lattice.} of the loops $C_{\mu}$ and $\tilde{C}_{\mu}$, since logical operators defined on homologous closed loops differ by products of stabilizers. As a direct consequence\footnote{Explicitly, as long as $O$ is compactly supported, we can find realizations of the logical operators $\overline{X}_{\mu}$ and $\overline{Z}_{\mu}$ which commute with $O$. Then, $\bra{\psi^{\bs}_{0}} O \ket{\psi^{\bs}_{0}} = \bra*{\psi^{\bs'}_{0}} O \ket*{\psi^{\bs'}_{0}}$ follows from inserting $1 = \overline{X}_{\mu}^{2}$ into the expectation value, while $\bra{\psi^{\bs}_{0}} O \ket*{\psi^{\bs'}_{0}} = 0$ follows from inserting $1 = \overline{Z}_{\mu}^{2}$ into the expectation value.}, ground states in $\mathcal{C}_{0}$ obey a \emph{local indistinguishability} property: for any two states $\ket{\psi}, \ket{\varphi} \in \mathcal{C}_{0}$, and any compactly supported operator $O$, we have
\begin{equation}
    \label{eq:local_indistinguishability}
    \bra{\psi} O \ket{\varphi} = c_{O} \bra{\psi} \ket{\varphi} ,
\end{equation}
where $c_{O}$ is a constant which depends only on $O$. This property, which is closely related to the Knill-Laflamme condition for quantum error-correcting codes \cite{knill_theory_1997,knill_theory_2000,nielsenQuantumComputationQuantum2010}, underlies the stability of the toric code's topological order to arbitrary local perturbations \cite{bravyi_topological_2010,hastings_quasiadiabatic_2005,bravyi_short_2011,li_perturbative_2025}.

Throughout this work, we let $\qty{\ket{\psi^{\bs}_{0}}}$ denote the logical-$Z$ eigenstates within $\mathcal{C}_{0}$, with $\bs \equiv (\sigma_{x}, \sigma_{y})$ and $\sigma_{\mu} = 0, 1$ ($\mu = x, y$) labeling the logical-$Z$ eigenvalues. These states are defined by the relations
\begin{subequations}
\label{eq:logical_Z}
    \begin{align}
        A_{v} \ket{\psi^{\bs}_{0}} &= B_{p} \ket{\psi^{\bs}_{0}} = + \ket{\psi^{\bs}_{0}}, \\
        \overline{Z}_{\mu} \ket{\psi^{\bs}_{0}} &= (-1)^{\sigma_{\mu}} \ket{\psi^{\bs}_{0}} .
    \end{align}
\end{subequations}
Each state $\ket{\psi^{\bs}_{0}}$ can be regarded as a macroscopic superposition of closed loop states; in the computational (i.e., Pauli-$Z$) basis, these states are collections of closed loops of $Z_{\ell} = -1$ in the dual lattice. The logical-$Z$ operators $\overline{Z}_{\mu}$ count the number of loops mod two which are cut by the non-contractible cycles $C_{\mu}$. For example, the state $\ket*{\psi^{(0,0)}_{0}}$ contains only contractible closed-loop configurations, while $\ket*{\psi^{(1,0)}_{0}}$ with $\overline{Z}_{x} = -1$ contains only configurations with a non-contractible loop about the $y$ cycle.

Finally, it will occasionally be convenient to utilize an alternate notation $\overline{X}^{\bs}$ for the logical-$X$ operators, defined such that
\begin{equation}
    \label{eq:logical_x}
    \overline{X}^{\bs} \ket*{\psi^{\bs'}_{0}} = \ket*{\psi^{\bs \oplus \bs'}_{0}} ,
\end{equation}
where $\bs \oplus \bs' \equiv (\sigma_{x} \oplus \sigma_{x}', \sigma_{y} \oplus \sigma_{y}')$, with $\oplus$ denoting mod-2 addition. In other words, $\overline{X}^{(0,0)} = 1$, $\overline{X}^{(1, 0)} = \overline{X}_{y}$, etc.

\subsection{Transverse-Field Toric Code}
\label{subsec:tftc}
In this work, we will study QEC protocols for quantum information stored in the ground-state manifold of the transverse-field toric code (TFTC)  \cite{wegnerDualityGeneralizedIsing1971,kogutIntroductionLatticeGauge1979,trebst_breakdown_2007,vidal_low-energy_2009,dusuel_robustness_2011,sachdevQuantumPhaseTransitions2011,fradkin_field_2013}, a physically natural and well-studied deformation of the toric code. The Hamiltonian is
\begin{equation}
    \label{eq:tftc}
    H_{h} \equiv - \sum_{v} A_{v} - \sum_{p} B_{p} - h \sum_{\ell} X_{\ell} ,
\end{equation}
where we assume $h \geq 0$. So long as $h$ is not too large, $H_{h}$ remains topologically ordered: 
the ground-state manifold $\mathcal{C}_{h}$ of $H_{h}$ remains fourfold degenerate with an energetic splitting exponentially small in $L$, with an $L$-independent gap separating these energy eigenstates from the rest of the spectrum \cite{hastings_quasiadiabatic_2005,bravyi_topological_2010,bravyi_short_2011}. Furthermore, although the logical-$Z$ 1-form symmetry is explicitly broken by the transverse field, one can construct a ``fattened'' logical-$Z$ operator via quasiadiabatic continuation \cite{hastings_quasiadiabatic_2005}, and the ground states in $\mathcal{C}_{h}$ continue to satisfy a local indistinguishability criterion of the form \eqref{eq:local_indistinguishability} up to corrections exponentially small in $L$. As a consequence, $\mathcal{C}_{h}$ is expected to serve as a robust error-correcting code with a nonzero threshold throughout the topologically ordered phase of $H_{h}$.

Physically, the effect of the transverse field is to provide dynamics and quantum fluctuations to the $m$ anyons. At a critical field strength $h_{c} \approx 0.328$ \cite{wu_phase_2012}, the gap to creating $m$ anyons closes, and these anyons condense within the ground state for $h > h_{c}$ [Fig.~\ref{fig:background}(c)]. The resulting phase is a ``confined'' phase\footnote{This terminology refers to the fact that the $e$ anyons are confined, i.e., the creation of two far-separated $e$ anyons costs an energy linear in their separation.}, with a unique gapped ground state incapable of encoding any quantum information. The critical point at $h_{c}$ is described by the $3d$ Ising* universality class\footnote{The terminology Ising* means that the Ising order parameter is \emph{gauged} \cite{schuler_universal_2016,serna_worldsheet_2024}, and does not appear as a scaling operator in the critical theory.}; this follows from a well-known duality to the (2+1)$d$ transverse-field Ising model (TFIM)~\cite{wegnerDualityGeneralizedIsing1971,kogutIntroductionLatticeGauge1979,sachdevQuantumPhaseTransitions2011,fradkin_field_2013}.

Note that the vertex stabilizer $A_{v}$ commutes with $H_{h}$ for all values of $h$. The ground-state manifold $\mathcal{C}_{h}$ therefore exhibits $A_{v} = +1$, and the ground-state physics of $H_{h}$ is equivalent to that of a (2+1)$d$ $\mathbb{Z}_{2}$ quantum lattice gauge theory (LGT) \cite{wegnerDualityGeneralizedIsing1971,kogutIntroductionLatticeGauge1979}, defined by the Hamiltonian
\begin{equation}
    \label{eq:lgt}
    H_{\LGT} \equiv - \sum_{p} B_{p} - h \sum_{\ell} X_{\ell}
\end{equation}
together with the \emph{Hilbert space constraint} $A_{v} \stackrel{!}{=} +1$. In this model, $Z_{\ell}$ and $X_{\ell}$ are to be interpreted as the $\mathbb{Z}_{2}$-valued vector potential and electric field respectively; the $m$ anyon excitations $B_{p} = -1$ are therefore pointlike magnetic vortices, while ``Gauss's law'' $A_{v} = 1$ enforces that electric field lines $X_{\ell} = -1$ form closed loops in the absence of electric charges (the $e$ anyons). The $m$ anyons are gapped in the topological phase and condense within the confined phase, while the electric field lines are condensed within the topological phase and exhibit finite line tension in the confined phase.

\subsection{Quantum-Classical Mapping}
\label{subsec:quantum_classical}
Significant physical intuition for the QEC protocols we investigate will come from the quantum-classical mapping \cite{fradkin_order_1978,kogutIntroductionLatticeGauge1979,sachdevQuantumPhaseTransitions2011,fradkin_field_2013}. As reviewed in Appendix~\ref{app:quantum_classical}, matrix elements of the Gibbs operator $e^{-\beta H_{h}}$ can be expressed as partition functions for an extremely anisotropic \emph{classical} $\mathbb{Z}_{2}$ LGT in three dimensions; the third direction is viewed as the ``imaginary time'' direction, denoted $\tau$. The classical action is given by\footnote{We leave the boundary conditions of $S_{\LGT}$ unspecified for now; we will always utilize periodic boundary conditions in the $x$ and $y$ directions, but different matrix elements of $e^{-\beta H_{h}}$ will lead to different boundary conditions in the $\tau$ direction. The boundary conditions of $S_{\LGT}$ determine which topological class of flux loop configurations $\mathcal{L}$ are contained in the partition function sum~\eqref{eq:lgt_partition_fn}.}
\begin{align}
        S_{\LGT}[U] &\equiv - \varepsilon \sum_{[ijkl]_{\perp}} U_{ij} U_{jk} U_{kl} U_{li} - J \sum_{[ijkl]_{\parallel}} U_{ij} U_{jk} U_{kl} U_{li}  \notag \\
        &= - \sum_{[ijkl]} J_{ijkl} U_{ij} U_{jk} U_{kl} U_{li} , \label{eq:lgt_classical}
\end{align}
where $U_{ij} = \pm 1$ are Ising variables defined on the nearest-neighbor links $\expval{ij}$ of a three-dimensional spacetime lattice with sites $i \equiv (v, \tau)$, with $\tau = 0, \pm \varepsilon, \pm 2\varepsilon, \ldots$ discretized into $L_{\tau}$ time slices of width $\varepsilon \equiv \frac{\beta}{L_{\tau}}$. Each term in the action is the product of four links in a spacetime plaquette $[ijkl]$; the plaquettes within a single imaginary time slice (denoted $[ijkl]_{\perp}$) are proportional to $\varepsilon$, while the plaquettes across imaginary time slices (denoted $[ijkl]_{\parallel}$) are proportional to $J \equiv - \frac{1}{2} \log \tanh(\varepsilon h)$. In the second expression, we have defined $J_{ijkl} = \varepsilon$ ($J_{ijkl} = J$) on spacelike (timelike) plaquettes $[ijkl]$.

The statistical physics model \eqref{eq:lgt_classical} can be regarded as a model of one-dimensional Ising magnetic flux loops $\mathcal{L}$ through the dual spacetime lattice. Each violated plaquette $U_{ij} U_{jk} U_{kl} U_{li} = -1$ corresponds to a unit segment of flux through the plaquette [see Fig.~\ref{fig:background}(b)], with energy $2\varepsilon$ ($2J$) per plaquette in the ``time'' (``space'') direction; since each elementary cube of the lattice must contain an even number of violated plaquettes, these flux segments necessarily form closed loops. The partition function of the classical LGT can therefore be expressed as
\begin{equation}
\label{eq:lgt_partition_fn}
    \mathcal{Z}_{\LGT} \equiv \sum_{\qty{U}} e^{-S_{\LGT}[U]} \propto \sum_{\mathcal{L}} e^{-E[\mathcal{L}]} ,
\end{equation}
where $\sum_{\mathcal{L}}$ refers to a sum over closed-loop configurations $\mathcal{L}$ in the dual lattice, and $E[\mathcal{L}] \equiv 2\varepsilon \abs{\mathcal{L}}_{\parallel} + 2J \abs{\mathcal{L}}_{\perp}$ is the microscopic energy of $\mathcal{L}$, with $\abs{\mathcal{L}}_{\parallel}$ and $\abs{\mathcal{L}}_{\perp}$ respectively denoting the length of $\mathcal{L}$ in the time and space directions. These loops exhibit a finite coarse-grained line tension throughout the topological phase $h < h_{c}$, and a vanishing coarse-grained line tension throughout the trivial phase $h > h_{c}$. In analogy to the (2+1)$d$ quantum model, the $3d$ classical LGT is Wegner-dual to the $3d$ Ising model~\cite{wegnerDualityGeneralizedIsing1971}, and the phase transition is therefore once again in the $3d$ Ising* universality class.

\begin{figure}[t!]
    \includegraphics[width = \columnwidth]{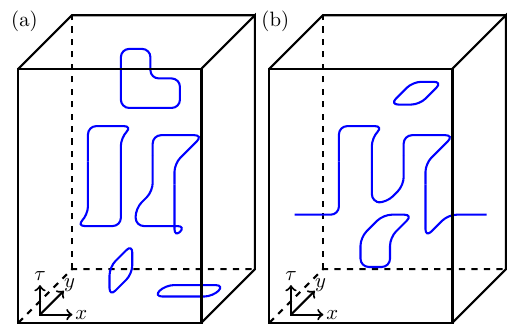}
    \caption{(a) Schematic depiction of an anyon worldline configuration contributing to the partition function $\mathcal{Z}_{h} \equiv \bra{\psi^{\bs}_{0}} e^{-\beta H_{h}} \ket{\psi^{\bs}_{0}}$. Notably, the global worldline configuration must be \emph{contractible}, so as to not implement a nontrivial logical-$X$ operator. (b) An anyon worldline configuration contributing to a ``twisted'' partition function $\mathcal{Z}^{\bs}_{h}$ in Eq.~\eqref{eq:Z_defn_new} for $\bs \neq (0,0)$; here $\bs = (0,1)$, since a non-contractible anyon worldline about the $x$ cycle of the torus modifies the value of $\overline{Z}_{y}$. Since such non-contractible worldlines exhibit an $\mathcal{O}(L)$ free energy within the topological phase $h < h_{c}$, $\mathcal{Z}_{h}^{\bs}$ for $\bs \neq (0,0)$ is exponentially suppressed relative to $\mathcal{Z}_{h} \equiv \mathcal{Z}_{h}^{(0,0)}$.}
    \label{fig:worldlines}
\end{figure}

The classical action in Eq.~\eqref{eq:lgt_classical} can be regarded as a path-integral description of the TFTC, where the flux loops precisely correspond to the imaginary-time worldlines of $m$ anyons which are conserved mod two. In particular, spacetime processes where anyon worldlines traverse the periodic boundaries of the system are associated with logical-$X$ operations. As explained in Appendix~\ref{app:quantum_classical}, the matrix element
\begin{equation}
\label{eq:Z_defn_new}
    \mathcal{Z}^{\bs}_h \equiv \bra*{\psi^{\bs \oplus \bs'}_0} e^{-\beta H_h} \ket*{\psi^{\bs'}_0}
\end{equation}
can be expressed as a LGT partition function over the imaginary-time interval $\tau \in [- \frac{\beta}{2}, \frac{\beta}{2}]$, with ``flux-free'' boundary conditions at the initial and final imaginary times $\tau = \pm \frac{\beta}{2}$, and which sums over only the spacetime processes for which the anyon worldlines implement the logical-$X$ operator $\overline{X}^{\bs}$. In particular, $\mathcal{Z}_h \equiv \mathcal{Z}_h^{(0,0)}$ includes only \emph{contractible} worldline configurations [Fig.~\ref{fig:worldlines}(a)], whereas $\mathcal{Z}^{\bs}_h$ for $\bs \neq (0,0)$ includes only (a subset of) \emph{non-contractible} worldline configurations [Fig.~\ref{fig:worldlines}(b)]. Note that Eq.~\eqref{eq:Z_defn_new} is independent of the choice of $\bs'$, since $H_h$ commutes with the logical-$X$ operators. 

Throughout the topological phase $h < h_{c}$, since worldlines exhibit finite line tensions, the free energy cost of system-spanning non-contractible worldline configurations grows with the linear system size $L$, and therefore $\mathcal{Z}^{\bs}_{h}$ is exponentially suppressed relative to $\mathcal{Z}_{h}$:
\begin{equation}
    \label{eq:Z_defn_cost}
    \frac{\mathcal{Z}^{\bs}_{h}}{\mathcal{Z}_{h}} = \begin{cases}
        1, & \bs = (0,0) \\
        \mathcal{O}(e^{-L / \xi}), & \bs \neq (0,0)
    \end{cases} \quad (h < h_{c}) .
\end{equation}
Although we are unaware of a rigorous proof of Eq.~\eqref{eq:Z_defn_cost} for all $h < h_{c}$, the analogous statement for an \emph{isotropic} 3$d$ classical LGT can be rigorously proven to hold at sufficiently low temperatures using the Peierls argument \cite{friedli_statistical_2017}. Furthermore, essentially identical observables have been numerically studied extensively in the literature~\cite{winter_geometric_2008,huang_wormalgorithmtype_2020}. In Appendix~\ref{app:winding_number}, we also numerically verify this statement in an isotropic 3$d$ LGT. We will therefore take Eq.~\eqref{eq:Z_defn_cost} to hold for all $h < h_c$ as our fundamental assumption throughout the remainder of this work.

%% file: sections/overview.tex
\section{Setup}
\label{sec:overview}
The primary focus of this work is to construct and study QEC protocols for the TFTC under Pauli decoherence. In Sec.~\ref{subsec:setup}, we broadly define the QEC problem we will investigate. In Sec.~\ref{subsec:ite} we introduce imaginary-time evolution (ITE) as a technique for constructing a useful approximate representation of TFTC code states, and in Sec.~\ref{subsec:modified_protocol} we describe a reformulation of our QEC protocol in terms of the code states constructed via ITE.


\subsection{Definition of QEC Protocol}
\label{subsec:setup}
Suppose two qubits of quantum information are encoded in the TFTC ground state $\ket{\psi_{h}} \in \mathcal{C}_{h}$, which is subsequently subjected to the Pauli decoherence channel
\begin{equation}
    \label{eq:decoherence}
    \mathcal{E}_{\ell}(\rho) = (1 - p) \rho + p_{X} X_{\ell} \rho X_{\ell} + p_{Y} Y_{\ell} \rho Y_{\ell} + p_{Z} Z_{\ell} \rho Z_{\ell}
\end{equation}
on each link $\ell$, where\footnote{Note that the variable $p$ is overloaded to stand for both plaquettes (for example, in the stabilizer $B_{p}$), and for the strength of decoherence. We hope that the meaning is clear from context.} $p \equiv p_{X} + p_{Y} + p_{Z}$. The resulting state is denoted $\rho_{h} \equiv \mathcal{E}(\dyad{\psi_{h}})$, where $\mathcal{E} \equiv \prod_{\ell} \mathcal{E}_{\ell}$; to avoid notational clutter, we suppress explicit dependence on the decoherence strengths $(p_{X}, p_{Y}, p_{Z})$. In the special case where only $p_{X}$ or $p_{Z}$ is nonzero, we refer to $\mathcal{E}_{\ell}$ as a ``bit-flip'' or ``phase-flip'' channel respectively. Throughout this work we will make the simplifying assumption $p_{Y} = p_{X} p_{Z} / (1 - p)$, so that $\mathcal{E}_{\ell} = \mathcal{E}_{\ell}^{X} \circ \mathcal{E}_{\ell}^{Z}$ describes independent bit-flip and phase-flip channels. 

As described in Sec.~\ref{subsec:tftc}, we expect the code space $\mathcal{C}_{h}$ to robustly protect quantum information against weak Pauli decoherence for all $h < h_{c}$. In a companion work~\cite{lip}, we showed that a \emph{finite} strength of Pauli decoherence is required to irreversibly destroy information encoded in $\mathcal{C}_{h}$, even when $h$ is infinitesimally close to $h_{c}$. Nevertheless, even if quantum information is \emph{in principle} protected below some intrinsic decoherence threshold, a central question is whether this encoded information can be \emph{practically} recovered---that is, whether there exists an efficient and conceptually simple QEC protocol which recovers the encoded information. 

Motivated by the results of Refs.~\cite{lavasani_stability_2025,liu_detecting_2025}, we will investigate the efficacy of stabilizer measurements for performing QEC on the decohered TFTC states. Starting from the decohered state $\rho_{h}$, we projectively measure each stabilizer $A_{v}, B_{p}$ and obtain the postmeasurement state
\begin{equation}
    \label{eq:pauli_postmeas}
    \rho_{h}(\vs) \equiv \frac{\hat{\Pi}_{\vs} \rho_{h} \hat{\Pi}_{\vs}}{P_{\vs}}
\end{equation}
with probability $P_{\vs} \equiv \tr [ \hat{\Pi}_{\vs} \rho_{h} ]$, where
\begin{equation}
    \label{eq:projector}
    \hat{\Pi}_{\vs} \equiv \prod_{v} \qty( \frac{1 + (-1)^{s_{v}} A_{v}}{2} ) \prod_{p} \qty( \frac{1 + (-1)^{s_{p}} B_{p}}{2} ) .
\end{equation}
is the measurement projector. The \emph{syndrome} $\vs \equiv \qty{s_{v}, s_{p}}$ with $s_{v}, s_{p} = 0, 1$ encodes the measurement outcome, and indicates the presence of an $e$ anyon ($m$ anyon) on each vertex (plaquette) for which $s_{v} = 1$ ($s_{p} = 1$).

Given the syndrome $\vs$, we ask whether there exists a recovery unitary $V_{\vs}$ which recovers $\ket{\psi_{h}}$ from $\rho_{h}(\vs)$ with high fidelity $\mathcal{F}_{\vs}(\ket{\psi_{h}})$, given by
\begin{equation}
    \label{eq:decoding_fidelity}
    \mathcal{F}_{\vs}(\ket{\psi_{h}}) \equiv \bra{\psi_{h}} V_{\vs} \rho_{h}(\vs) V_{\vs}^{\dag} \ket{\psi_{h}} .
\end{equation}
This question can be addressed both for individual syndromes $\vs$, or for the statistical ensemble of syndromes; here we are particularly interested in whether $\ket{\psi_{h}}$ can be recovered from $\rho_{h}(\vs)$ with high fidelity, with high probability over the ensemble of syndromes. Our metric for success will be the measurement-averaged decoding fidelity,
\begin{equation}
    \label{eq:avg_decoding_fidelity}
    \begin{split}
        \overline{\mathcal{F}}(\ket{\psi_{h}}) &\equiv \sum_{\vs} P_{\vs} \mathcal{F}_{\vs}(\ket{\psi_{h}}) \\
        &= \sum_{\vs} \bra{\psi_{h}} V_{\vs} \hat{\Pi}_{\vs} \rho_{h} \hat{\Pi}_{\vs} V_{\vs}^{\dag} \ket{\psi_{h}} .
    \end{split}
\end{equation}
Note that $\overline{\mathcal{F}}$ is close to unity if and only if $\mathcal{F}_{\vs}$ is close to unity with high probability: to wit, Markov's inequality~\cite{ross_first_2020}
yields
\begin{equation}
    \mathbb{P}[ \mathcal{F}_{\vs} \leq 1 - \varepsilon ] \leq \frac{1 - \overline{\mathcal{F}}}{\varepsilon}
\end{equation}
for any $\varepsilon > 0$, where $\mathbb{P}[\cdot]$ is the total probability of event $(\cdot)$.

In QEC algorithms for the stabilizer toric code under Pauli decoherence \cite{dennis_topological_2002,wang_confinement-higgs_2003}, one seeks to pairwise annihilate the observed anyons to recover a stabilizer state $\ket{\psi_{0}} \in \mathcal{C}_{0}$. Throughout this work we will consider an analogous two-step QEC protocol: we first pairwise annihilate the observed anyons to obtain a stabilizer state in $\mathcal{C}_{0}$, and then we adiabatically re-encode into the deformed code space $\mathcal{C}_{h}$. 
Such recovery unitaries $V_{\vs}$ 
have the form\footnote{In principle, one can also consider more general recoveries which perform coherent logical rotations on the stabilizer code space; as we will see in Secs.~\ref{sec:meas_decoding} and~\ref{sec:pauli_meas_decoding}, stoquasticity will ensure that such recoveries are strictly unnecessary in the present context.\label{fn:coherent}}
\begin{equation}
    \label{eq:recovery_twostep}
    \begin{split}
        V_{\vs} &= \mathcal{U}_{h} X_{\vs} Z_{\vs} , \\ 
        \mathcal{U}_{h} &\equiv \mathcal{T} \exp \qty{ -i \int_{0}^{T} \dd{t} H_{h(t)} }.
    \end{split}
\end{equation}
Here\footnote{We mention conventional adiabatic evolution only for conceptual simplicity; strictly speaking, $\mathcal{U}_{h}$ as defined above incurs finite-$T$ errors for any $T < \infty$, and leads to unwanted dynamical phases between nearly-degenerate ground states when $T \gtrsim e^{L/\xi}$. Although these errors can be made small for $T \gtrsim \poly(L, \Delta^{-1})$, to be technically precise, one should replace $\mathcal{U}_{h}$ with a \emph{quasiadiabatic} unitary to avoid finite-$T$ errors~\cite{hastings_quasiadiabatic_2005,bachmann_automorphic_2012}. In any case, we will soon dispense with $\mathcal{U}_h$ altogether.} $\mathcal{U}_{h}$ adiabatically evolves $\mathcal{C}_{0}$ to $\mathcal{C}_{h}$ over a time $T$ which is very large compared to the inverse gap of $H_{h}$, while $X_{\vs}$ ($Z_{\vs}$) is a product of Pauli-$X$ (Pauli-$Z$) operators which pairwise annihilates the $m$ anyons ($e$ anyons) observed in the syndrome $\vs$, i.e., $X_{\vs}$ and $Z_{\vs}$ are required to satisfy
\begin{equation}
    \label{eq:anyon_pairings}
    X_{\vs} B_{p} = (-1)^{s_{p}} B_{p} X_{\vs}, \quad Z_{\vs} A_{v} = (-1)^{s_{v}} A_{v} Z_{\vs} .
\end{equation}
Such an error correction protocol will succeed with high fidelity if $X_{\vs} Z_{\vs} \rho_h(\vs) Z_{\vs} X_{\vs}$ is approximately equal to the pure stabilizer code state $\ket{\psi_0} = \mathcal{U}_h^{\dag} \ket{\psi_h}$. A full cycle of successful QEC is therefore depicted in the following diagram:
\begin{equation}
    \label{eq:flow_diagram}
    \begin{tikzcd}[column sep = huge, row sep = huge]
        \ket{\psi_h} \arrow[r, "\mathcal{E}"]  & \rho_h \arrow[d, "\hat{\Pi}_{\vs}"] \\
        \ket{\psi_0} \arrow[u, "\mathcal{U}_h"] & \rho_h(\vs) \arrow[l, "X_{\vs} Z_{\vs}"]
    \end{tikzcd}
\end{equation}
Throughout this work, we will assume that no additional errors arise during the course of stabilizer measurement, anyon pairing, or adiabatic re-encoding: we assume that each of these operations can be performed perfectly, and focus our attention on determining how to best pair the anyons (i.e., how to choose $X_{\vs}$ and $Z_{\vs}$).


There is significant freedom in choosing the operators $X_{\vs}$ and $Z_{\vs}$; \emph{homologous} choices differ by stabilizers and act identically on the postmeasurement states, while \emph{non-homologous} choices differ by logical operators and lead to distinct recovery unitaries. Our central task is therefore to determine the optimal \emph{homology classes} for the anyon-pairing unitaries $X_{\vs}$ and $Z_{\vs}$ for each $\vs$. An algorithm for choosing these homology classes for each observed syndrome $\vs$ is called a \emph{decoder}. We will refer to an algorithm which maximizes the decoding fidelity \eqref{eq:decoding_fidelity} for each $\vs$ as an \emph{optimal postmeasurement decoder}, to distinguish from the true optimal recovery channel which may not involve measurements of the ``bare'' stabilizers $A_{v}, B_{p}$. The latter is generally highly nonlocal and exponentially complex \cite{schumacherQuantumDataProcessing1996,schumacher_approximate_2002}, due to the nontrivial quantum fluctuations within the code space $\mathcal{C}_{h}$.

It is useful to briefly mention an alternative interpretation of the QEC problem depicted in the diagram \eqref{eq:flow_diagram}. Rather than viewing $\mathcal{C}_{h}$ as our quantum code and $\mathcal{U}_{h}$ as a step in our recovery protocol, we can instead view the stabilizer space $\mathcal{C}_{0}$ as our code, and regard $\mathcal{U}_{h}$ as a \emph{coherent error} acting on $\mathcal{C}_{0}$. This coherent error is then followed by the incoherent error $\mathcal{E}$, and then by stabilizer measurements. Most previous works on coherent errors \cite{bravyi_correcting_2018,venn_coherent_2023,behrends_surface_2024,behrends2024statisticalmechanicalmappingmaximumlikelihood,bao2024phasesdecodabilitysurfacecode,cheng_emergent_2025} have focused on finite-depth (typically depth-one) unitary circuits; in contrast, $\mathcal{U}_{h}$ generally constitutes a deep unitary circuit, whose depth diverges especially as $h \to h_{c}$. From this perspective, one might naively guess that stabilizer measurements could be highly destructive to the initially encoded information, even without additional Pauli decoherence (i.e., with $p = 0$). 


The central result of this work is to demonstrate that stabilizer measurements can in fact be utilized as part of an effective decoding protocol, even in close vicinity to the quantum critical point at $h_{c}$. In Section~\ref{sec:meas_decoding}, we will first prove that in the limiting case of no decoherence ($p = 0$), stabilizer measurements and an (asymptotically) optimal postmeasurement decoder achieve a decoding fidelity exponentially close to unity throughout the topological phase $h < h_c$. Then, in Section~\ref{sec:pauli_meas_decoding}, we will numerically demonstrate that the same result holds for decoherence strengths below a nonzero critical threshold, even in close proximity to the quantum critical point.

\subsection{Imaginary Time Evolution}
\label{subsec:ite}
To analyze the quantum code $\mathcal{C}_{h}$, our essential technical strategy is to use \emph{imaginary time evolution} (ITE) to construct an excellent approximate basis for $\mathcal{C}_{h}$. As we will see, the utility of the ITE approach arises from the quantum-classical mapping (see Sec.~\ref{subsec:quantum_classical}), which will provide significant technical and conceptual insights in constructing our QEC protocols.

Let $\ket{\psi_{h}} \in \mathcal{C}_{h}$ be a code state in the TFTC. An exact description of $\ket{\psi_{h}}$ would require the complete wavefunctions for the four lowest eigenstates of $H_{h}$, which are difficult to describe analytically beyond the perturbative regime. Instead, to excellent approximation, we can generate $\ket{\psi_{h}}$ using ITE from a reference stabilizer code state $\ket{\psi_{0}} \in \mathcal{C}_{0}$:
\begin{equation}
    \label{eq:imag_time_evo}
    \ket{\psi_{h}} \approx \ket{\varphi_{h}} \equiv \frac{e^{-\beta H_{h} / 2} \ket{\psi_{0}}}{\sqrt{\bra{\psi_{0}} e^{-\beta H_{h}} \ket{\psi_{0}}}} .
\end{equation}
Here $\beta$ is taken sufficiently large [say, $\beta \sim \text{poly}(L)$] so that $e^{-\beta H_{h} / 2}$ acts approximately as a projector onto $\mathcal{C}_{h}$, but not so large $[\beta \ll \exp(L)]$ that it can resolve the exponentially small energetic splitting within $\mathcal{C}_{h}$. It is then a straightforward exercise to show\footnote{Here we require a very mild physical assumption: namely, that the norm of the projection of $\ket{\psi_{0}}$ onto $\mathcal{C}_{h}$ scales with $L$ no worse than $\mathcal{O}(e^{-\text{poly}(L)})$. A common expectation is that this norm scales as $\mathcal{O}(e^{-\alpha_{h} L^{2}})$ for some $\alpha_{h} > 0$, for all $h < h_{c}$ \cite{gu_fidelity_2010}.\label{fn:ite_projection}} that $\ket{\varphi_{h}}$ lies within $\mathcal{C}_{h}$ up to exponentially small corrections in $L$, as long as $h < h_{c}$ and the gap of $H_{h}$ remains open; see Appendix~\ref{app:ite_truecodestates} for details. We will suppress explicit $\beta$ dependence of states and observables throughout, taking $\beta$ to be some fixed function of $L$. 


Throughout the topological phase\footnote{Throughout this work, we always take the thermodynamic limit $L \to \infty$ \emph{before} tuning $h$ towards $h_{c}$; that is, we always work in the regime where the correlation length $\xi$ is much smaller than $L$. Of course, this assumption necessarily breaks down in our numerical simulations close to the critical point.} $h < h_{c}$, we claim that the ITE in Eq.~\eqref{eq:imag_time_evo} acts approximately \emph{unitarily} on $\mathcal{C}_{0}$, up to exponentially small corrections. To see this, first let
\begin{equation}
    \label{eq:logicalZ_ITE}
    \ket{\varphi^{\bs}_{h}} \equiv \frac{1}{\sqrt{\mathcal{Z}_{h}}} e^{-\beta H_{h} / 2} \ket{\psi^{\bs}_{0}}
\end{equation}
denote the logical-$Z$ stabilizer states evolved under ITE, where $\mathcal{Z}_{h} \equiv \mathcal{Z}_{h}^{(0,0)}$ was defined in Eq.~\eqref{eq:Z_defn_new}. Using Eq.~\eqref{eq:Z_defn_cost}, these four states are approximately orthonormal for all $h < h_{c}$:
\begin{equation}
    \label{eq:ortho_deformed_basis}
    \bra{\varphi^{\bs}_{h}} \ket*{\varphi^{\bs'}_{h}} = \frac{\mathcal{Z}_{h}^{\bs \oplus \bs'}}{\mathcal{Z}_{h}} = \begin{cases}
        1, & \bs = \bs' \\
        \mathcal{O}(e^{-L/\xi}), & \bs \neq \bs'
    \end{cases} .
\end{equation}
In other words, ITE preserves the inner products between the logical-$Z$ states up to exponentially small corrections. This observation can be understood simply via the quantum-classical mapping: the only spacetime processes which spoil the orthogonality of the states $\ket{\varphi^{\bs}_{h}}$ are those for which $m$ anyon worldlines nontrivially wrap the torus [see Fig.~\ref{fig:worldlines}(b)], and these processes are exponentially suppressed throughout the topological phase.

Eq.~\eqref{eq:ortho_deformed_basis}, combined with the logical-$X$ symmetry of $H_{h}$, implies that ITE acts approximately unitarily on all of $\mathcal{C}_{0}$. Explicitly,
%
%
let $\hat{\Pi}_{0}$ denote the projector onto $\mathcal{C}_{0}$, i.e., the measurement operator in Eq.~\eqref{eq:projector} with trivial syndrome $s_{v} , s_{p} = 0$. We perform a polar decomposition of the operator $e^{-\beta H_{h} / 2} \hat{\Pi}_{0}$ to obtain
\begin{equation}
    \frac{1}{\sqrt{\mathcal{Z}_{h}}} e^{-\beta H_{h} / 2} \hat{\Pi}_{0} = \mathcal{V}_{h} \mathcal{I}_{h}^{1/2} ,
\end{equation}
where the Hermitian operator $\mathcal{I}_{h}$ is defined as
\begin{equation}
    \label{eq:almost_identity_0}
    \mathcal{I}_{h} \equiv \frac{\hat{\Pi}_{0} e^{-\beta H_{h}} \hat{\Pi}_{0}}{\mathcal{Z}_{h}} ,
\end{equation}
and the isometry $\mathcal{V}_{h}$ is correspondingly given by\footnote{Note that $\bra{\psi_{0}} e^{-\beta H_{h}} \ket{\psi_{0}} > 0$ for all $\ket{\psi_{0}} \in \mathcal{C}_{0}$, and so the operator $\mathcal{I}_{h}$ can be inverted on $\mathcal{C}_{0}$.}
\begin{equation}
    \label{eq:ite_isometry}
    \mathcal{V}_{h} \equiv \frac{1}{\sqrt{\mathcal{Z}_{h}}} e^{-\beta H_{h} / 2} \hat{\Pi}_{0} \mathcal{I}_{h}^{-1/2} \hat{\Pi}_{0} ,
\end{equation}
which satisfies $\mathcal{V}_{h}^{\dag} \mathcal{V}_{h} = \hat{\Pi}_{0}$ exactly. $\mathcal{V}_{h}$ therefore serves as an exactly unitary implementation of ITE in Eq.~\eqref{eq:imag_time_evo}. Notably, $\mathcal{V}_{h}$ is approximately identical to ITE when $\mathcal{I}_{h}$ is close to the identity operator within $\mathcal{C}_{0}$. Since $\mathcal{I}_{h}$ acts entirely within $\mathcal{C}_{0}$ and commutes with the logical-$X$ operators, it admits the general form
\begin{equation}
\label{eq:almost_identity}
        \mathcal{I}_{h} = \hat{\Pi}_{0} \qty[ 1 +  \sum_{\bs \neq (0, 0)} \frac{\mathcal{Z}^{\bs}_{h}}{\mathcal{Z}_{h}} \overline{X}^{\bs} ] \hat{\Pi}_{0} ,
\end{equation}
where the coefficients can be verified by taking matrix elements of Eq.~\eqref{eq:almost_identity_0} using the logical-$Z$ states $\ket{\psi^{\bs}_{0}}$. Since the amplitudes $\mathcal{Z}^{\bs}_{h} / \mathcal{Z}_{h}$ for nontrivial logical operations are exponentially small in $L$ for all $h < h_{c}$, the operator $\mathcal{I}_{h}$ is indeed approximately equal to $\hat{\Pi}_{0}$. As a result, the unitarily evolved states $\mathcal{V}_{h} \ket{\psi_{0}}$ are exponentially close to the ITE states $\ket{\varphi_{h}}$ in Eq.~\eqref{eq:imag_time_evo}, and therefore ITE acts approximately unitarily on $\mathcal{C}_{0}$.


Throughout the remainder of this work, we will utilize the ITE states $\ket{\varphi_{h}}$ as an analytically tractable proxy for the true code states $\ket{\psi_{h}}$. Since each $\ket{\varphi_{h}}$ approximately lies within the ``true'' code space $\mathcal{C}_{h}$, the four-dimensional space spanned by the ITE states must approximately coincide with $\mathcal{C}_{h}$; that is, for any $\ket{\psi_{h}} \in \mathcal{C}_{h}$, there exists an ITE state $\ket{\varphi_{h}}$ for which $\bra{\varphi_{h}} \ket{\psi_{h}} \geq 1 - \mathcal{O}(e^{-L/\xi})$, and vice versa. This implies that any QEC protocol will succeed on $\mathcal{C}_{h}$ with high average fidelity if and only if it succeeds with high average fidelity on the ITE states. Explicitly, in Appendix~\ref{app:decoding_fidelities} we prove the following inequality on the difference of average decoding fidelities:
\begin{equation}
\label{eq:fidelity_bound}
    \abs{\overline{\mathcal{F}}(\ket{\psi_{h}}) - \overline{\mathcal{F}}(\ket{\varphi_{h}})} \leq 4 \sqrt{ 1 - \abs{\bra{\varphi_{h}} \ket{\psi_{h}}}^{2} } . 
\end{equation}
Furthermore, in Appendix~\ref{app:imag_time_adiabatic} we argue that the ITE state in Eq.~\eqref{eq:imag_time_evo} for each $\ket{\psi_0} \in \mathcal{C}_0$ is approximately equal to the state $\mathcal{U}_h \ket{\psi_0}$ obtained from adiabatic unitary evolution, up to an overall dynamical phase and exponentially small corrections. 

We briefly mention an interesting alternative method of promoting ITE to a bona-fide unitary encoding, which we call ``pruning'' and describe in Appendix~\ref{app:pruning}. In short, since the nearly orthogonal states $\ket{\varphi^{\bs}_{h}}$ are non-negative in the computational basis (see Sec.~\ref{subsec:stoquasticity}), they must also exhibit nearly disjoint support in the computational basis. We can therefore ``prune'' these four code states to construct a basis of exactly orthonormal and strictly non-negative code states, which once again approximately span $\mathcal{C}_{h}$. This ``pruning procedure'' can itself be regarded as a form of error correction; more precisely, it can be regarded as a postselected measurement onto a perfectly correctable subspace, which succeeds with probability exponentially close to unity for all $h < h_{c}$.

\subsection{Modified QEC Protocol}
\label{subsec:modified_protocol}
The discussion of the preceding subsection motivates the following slight modification of the QEC problem described in Sec.~\ref{subsec:setup}. First, instead of starting from the TFTC state $\ket{\psi_{h}} = \mathcal{U}_{h} \ket{\psi_{0}} \in \mathcal{C}_{h}$, we will instead begin with the stabilizer code state $\ket{\psi_{0}} \in \mathcal{C}_{0}$ and view the corresponding ITE state $\ket{\varphi_{h}} \propto e^{-\beta H_{h} / 2} \ket{\psi_{0}}$ as arising from a (nonlinear) coherent error on $\ket{\psi_{0}}$. We then subject these states to incoherent Pauli decoherence to obtain $\rho_{h} \equiv \mathcal{E}(\dyad{\varphi_{h}})$. Next, we attempt to decode by measuring the stabilizers $A_{v}, B_{p}$, resulting in the postmeasurement states $\rho_{h}(\vs)$ as 
in Eq.~\eqref{eq:pauli_postmeas}. Finally, we attempt to recover $\ket{\psi_{0}}$ by pairwise annihilating the observed $m$ ($e$) anyons using $X_{\vs}$ ($Z_{\vs}$). The flow diagram \eqref{eq:flow_diagram} for a complete round of successful QEC is modified in this protocol to
\begin{equation}
    \label{eq:flow_diagram_2}
    \begin{tikzcd}[column sep = huge, row sep = huge]
        \ket{\varphi_h} \arrow[r, "\mathcal{E}"]  & \rho_h \arrow[d, "\hat{\Pi}_{\vs}"] \\
        \ket{\psi_0} \arrow[u, "e^{-\beta H_{h} / 2}"] & \rho_h(\vs) \arrow[l, "X_{\vs} Z_{\vs}"]
    \end{tikzcd}
\end{equation}
So long as $h < h_{c}$, Eq.~\eqref{eq:fidelity_bound} guarantees that the average recovery fidelity of this protocol,
\begin{subequations}
    \begin{align}
        \overline{\mathcal{F}}^{\ITE}(\ket{\psi_0}) &\equiv \sum_{\vs} P_{\vs} \mathcal{F}^{\ITE}_{\vs}(\ket{\psi_0}), \label{eq:avg_ite_fidelity} \\
        \mathcal{F}_{\vs}^{\ITE}(\ket{\psi_{0}}) &\equiv \bra{\psi_{0}} X_{\vs} Z_{\vs} \rho_{h}(\vs) Z_{\vs} X_{\vs} \ket{\psi_{0}} , \label{eq:ite_fidelity}
    \end{align}
\end{subequations}
is equal to the average recovery fidelity of the original protocol up to exponentially small corrections in $L$; note that we label $\overline{\mathcal{F}}^{\ITE}$ and $\mathcal{F}_{\vs}^{\ITE}$ by the \emph{stabilizer} code state $\ket{\psi_{0}}$, rather than the corresponding TFTC code state $\ket{\psi_{h}} = \mathcal{U}_{h} \ket{\psi_{0}}$ as in Eq.~\eqref{eq:decoding_fidelity}. In turn, the optimal postmeasurement decoders we will construct for the ITE code states are guaranteed to be \emph{asymptotically} optimal for the original protocol of Sec.~\ref{subsec:setup}, in the sense that no other decoder can achieve a larger fidelity on the true code states by more than an $\mathcal{O}(e^{-L/\xi})$ amount. For the remainder of this work, we will therefore focus our attention on constructing the optimal postmeasurement decoder which maximizes the ITE decoding fidelity \eqref{eq:ite_fidelity}. 

As we will see in Sec.~\ref{subsec:phase_flip}, it will be sufficient to construct optimal postmeasurement decoders for bit-flip and phase-flip decoherence separately. We will therefore set $p_{Z} = 0$ and $Z_{\vs} = 1$ for the majority of this work, and this will lead to yet another simplification: rather than attempting to decode to an \emph{arbitrary} code state $\ket{\psi_{0}}$, we can focus our attention on decoding the \emph{specific} logical-$Z$ code state $\ket*{\psi^{(0,0)}_{0}}$. This is because the entire protocol in the diagram~\eqref{eq:flow_diagram_2} exhibits a \emph{strong} logical-$X$ symmetry: every component of the diagram, including every Kraus operator of the decoherence channel $\mathcal{E}$, commutes with the logical-$X$ operators. As demonstrated in Appendix~\ref{app:strong_symm}, this fact implies that the average ITE fidelity~\eqref{eq:avg_ite_fidelity} for $\ket{\psi_{0}} = \ket*{\psi^{(0,0)}_{0}}$ can be used to lower-bound the worst-case average ITE fidelity for \emph{any} code state. In other words, our QEC protocol will succeed with high fidelity on an arbitrary code state if and only if it succeeds with high fidelity on $\ket*{\psi^{(0,0)}_{0}}$. This observation will help to simplify our construction of the optimal postmeasurement decoders in subsequent sections.

%% file: sections/meas_decoding.tex
\section{Stabilizer Measurements of Transverse-Field Toric Code States}
\label{sec:meas_decoding}

In this section, we will demonstrate that projective measurements of the ``bare'' toric code stabilizers preserve information encoded within the TFTC code space with high fidelity. This section can be regarded as the $p = 0$ edge case of the more general decoherence problem defined in the previous section (which will be discussed in Sec.~\ref{sec:pauli_meas_decoding}), and serves as a useful first benchmark for the efficacy of stabilizer measurements in performing QEC on the TFTC states.

Starting from an arbitrary ITE code state $\ket{\varphi_{h}} \propto e^{-\beta H_{h} / 2} \ket{\psi_{0}}$ [see Eq.~\eqref{eq:imag_time_evo}], we perform a projective measurement of the stabilizers $B_{p}$ and obtain the postmeasurement state
\begin{equation}
    \label{eq:ite_postmeas_0}
    \ket{\varphi_{h}(\vs)} \equiv \frac{1}{\sqrt{P_{\vs}}} \hat{\Pi}_{\vs} \ket{\varphi_{h}}
\end{equation}
with probability
\begin{equation}
    \label{eq:ite_prob}
    P_{\vs} \equiv \bra{\varphi_{h}} \hat{\Pi}_{\vs} \ket{\varphi_{h}} ,
\end{equation}
where the projector $\hat{\Pi}_{\vs}$ is defined in Eq.~\eqref{eq:projector}. Given the syndrome $\vs$, we would like to determine an $m$ anyon pairing $X_{\vs}$ which recovers $\ket{\psi_{0}}$ from $\ket{\varphi_{h}(\vs)}$ with high fidelity. Note that $A_{v} = 1$ in the absence of phase-flip decoherence, and so we can set the $e$ anyon pairing $Z_{\vs}$ to unity in this section.


In Appendix~\ref{app:coherent_decoding}, we prove that quantum information is perfectly preserved by stabilizer measurement with syndrome $\vs$ (i.e., a perfect recovery unitary $V_{\vs}$ exists) if and only if the following two criteria are satisfied:
\begin{enumerate}
    \item The measurement probabilities $P^{\bs}_{\vs} \equiv \bra{\varphi^{\bs}_{h}} \hat{\Pi}_{\vs} \ket{\varphi^{\bs}_{h}}$ are identical for each basis state, i.e., $P^{\bs}_{\vs}$ is independent of $\bs$. 
    \item The basis ITE states remain orthonormal after measurement, i.e., $\bra*{\varphi^{\bs}_{h}(\vs)} \ket*{\varphi^{\bs'}_{h}(\vs)} = \delta_{\bs \bs'}$.
\end{enumerate}
Similarly, a high-fidelity approximate recovery unitary exists if these two criteria are approximately satisfied. Intuitively, the first condition guarantees that the measurement does not reveal any information about the underlying code state, while the second condition guarantees that superpositions of code states are not collapsed by syndrome measurement\footnote{Note that this intuition is basis-dependent; revealing information in the logical-$Z$ basis is equivalent to collapsing superpositions in the logical-$X$ basis, and vice versa.}. Together, they imply that the measurement operator $\hat{\Pi}_{\vs}$ acts \emph{unitarily} on the code space.


For the logical-$Z$ ITE states $\ket{\varphi^{\bs}_{h}}$ defined in Eq.~\eqref{eq:logicalZ_ITE}, the first criterion above holds exactly for all $h$ due to logical-$X$ symmetry. In Sec.~\ref{subsec:stoquasticity}, we will explicitly demonstrate that the second criterion is satisfied throughout the topological phase up to exponentially small corrections, with high probability in the syndrome measurement, and subsequently in Sec.~\ref{subsec:meas_decoding} we determine the optimal anyon pairing $X_{\vs}$ for each syndrome $\vs$. In Sec.~\ref{subsec:meas_decoding_numerics}, we numerically implement this optimal postmeasurement decoder and demonstrate its efficacy throughout the topological phase.

\subsection{Decodability via Stoquasticity}
\label{subsec:stoquasticity}
Although $H_{h}$ in Eq.~\eqref{eq:tftc} represents a rather natural perturbation to the stabilizer toric code, it also possesses a special property called \emph{stoquasticity}. A Hamiltonian is called \emph{stoquastic} in a given basis if all off-diagonal matrix elements in this basis are real and non-positive \cite{bravyi2007complexitystoquasticlocalhamiltonian}. By this definition, $H_{h}$ is stoquastic in the Pauli-$Z$ basis, the Pauli-$X$ basis, and the stabilizer basis. Here we focus on stoquasticity in the computational (i.e., Pauli-$Z$) basis, whose basis states we denote by $\ket{\vb{z}}$ with $\vb{z} \equiv \qty{z_{\ell}}$ and $z_{\ell} = 0, 1$, such that $Z_{\ell} \ket{\vb{z}} = (-1)^{z_{\ell}} \ket{\vb{z}}$. An immediate consequence of stoquasticity is that matrix elements of the Gibbs operator $e^{-\beta H_{h}}$ in this basis are strictly non-negative\footnote{A simple proof of this fact is to define $E_{\text{max}}$ as the largest eigenvalue of $H_{h}$, so that $M \equiv \beta (E_{\text{max}} \mathds{1} - H_{h})$ is a matrix of non-negative entries. We can then rewrite the left-hand side of Eq.~\eqref{eq:stoquastic} as $e^{-\beta E_{\text{max}}} \bra{\vb{z}} e^{M} \ket{\vb{z}'}$, which is manifestly non-negative upon Taylor expanding $e^{M}$.}:
\begin{equation}
    \label{eq:stoquastic}
    \bra{\vb{z}} e^{-\beta H_{h}} \ket{\vb{z}'} \geq 0 \quad \forall \vb{z}, \vb{z}'.
\end{equation}
Note that this property was already implicitly used in Sec.~\ref{subsec:quantum_classical} to construct the quantum-classical mapping for~$H_{h}$. 

We now show that stoquasticity directly implies that stabilizer measurements preserve the approximate orthogonality of the logical-$Z$ ITE states $\ket{\varphi^{\bs}_{h}}$, with high probability in the measurement outcome. Under a stabilizer measurement, each such state collapses to 
\begin{equation}
    \label{eq:ite_basis_measured}
    \begin{split}
        \ket{\varphi^{\bs}_{h}(\vs)} &\equiv \frac{1}{\sqrt{P_{\vs}}} \hat{\Pi}_{\vs} \ket{\varphi^{\bs}_{h}} \\
        &= \frac{1}{\sqrt{\mathcal{Z}_{h}(\vs)}} \hat{\Pi}_{\vs} e^{-\beta H_{h} / 2} \ket{\psi^{\bs}_{0}}
    \end{split}
\end{equation}
with probability
\begin{equation}
    P_{\vs} = \bra{\varphi^{\bs}_{h}} \hat{\Pi}_{\vs} \ket{\varphi^{\bs}_{h}} = \frac{\mathcal{Z}_{h}(\vs)}{\mathcal{Z}_{h}} ,
\end{equation}
where we have defined the postmeasurement partition function
\begin{equation}
    \label{eq:postmeas_partition_fn}
    \mathcal{Z}_{h}(\vs) \equiv \bra{\psi^{\bs}_{0}} e^{-\beta H_{h} / 2} \hat{\Pi}_{\vs} e^{-\beta H_{h} / 2} \ket{\psi^{\bs}_{0}} .
\end{equation}
Note that since both $H_{h}$ and $\hat{\Pi}_{\vs}$ commute with the logical-$X$ operators $\overline{X}_{\mu}$, $\mathcal{Z}_{h}(\vs)$ and $P_{\vs}$ are independent of $\bs$. Crucially, stoquasticity implies that the ITE states $\ket{\varphi^{\bs}_{h}}$ and their postmeasurement analogs $\ket{\varphi^{\bs}_{h}(\vs)}$ are all strictly non-negative in the computational basis; consequently, each postmeasurement overlap $\bra{\varphi^{\bs}_{h}(\vs)} \ket*{\varphi^{\bs'}_{h}(\vs)}$ is strictly non-negative. From this observation, we can compute the average modulus of the overlap between the postmeasurement states as follows:
\begin{equation}
    \begin{split}
        \sum_{\vs} P_{\vs} \abs{ \bra*{\varphi^{\bs}_{h}(\vs)} \ket*{\varphi^{\bs'}_{h}(\vs)} } &= \sum_{\vs} P_{\vs} \qty[ \frac{\bra{\varphi^{\bs}_{h}} \hat{\Pi}_{\vs} \ket*{\varphi^{\bs'}_{h}}}{P_{\vs}} ] \\
        &= \bra{\varphi^{\bs}_{h}} \ket*{\varphi^{\bs'}_{h}} ,
    \end{split}
\end{equation}
where we have employed the resolution of identity $\sum_{\vs} \hat{\Pi}_{\vs} = 1$. Finally using Markov's inequality \cite{ross_first_2020}, we obtain the result
\begin{equation}
    \label{eq:logical_markov}
    \mathbb{P} \qty[ \abs{\bra*{\varphi^{\bs}_{h}(\vs)} \ket*{\varphi^{\bs'}_{h}(\vs)}} > \varepsilon ] \leq \frac{\bra*{\varphi^{\bs}_{h}} \ket*{\varphi^{\bs'}_{h}}}{\varepsilon} .
\end{equation}
Since the right-hand side is $\mathcal{O}(e^{-L/\xi})$ for $\bs \neq \bs'$ [see Eq.~\eqref{eq:ortho_deformed_basis}], by choosing $\varepsilon = \mathcal{O}(e^{-L / \xi'})$ for any $\xi' > \xi$, we learn that the postmeasurement overlaps are also exponentially small in $L$, with probability exponentially close to unity. 

\begin{figure}
    \includegraphics[width = \columnwidth]{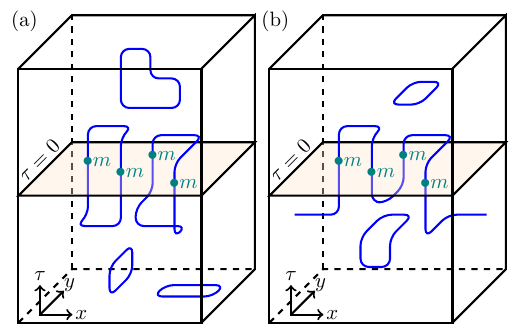}
    \caption{(a) Schematic depiction of an anyon worldline configuration contributing to the postmeasurement partition function $\mathcal{Z}_{h}(\vs)$ in Eq.~\eqref{eq:postmeas_partition_fn}. Worldlines are required to pierce the $\tau = 0$ surface at precisely the locations of the observed $m$ anyons, and to form a globally contractible closed-loop configuration. (b) An anyon worldline configuration contributing to the ``twisted'' postmeasurement partition function $\mathcal{Z}^{\bs}_{h}(\vs)$ in Eq.~\eqref{eq:twisted_partition_postmeas}, for $\bs \neq (0, 0)$. The worldlines are again consistent with the observed anyon positions, but now wrap the torus to implement a nontrivial logical-$X$ operator. For typical measurement syndromes $\vs$ (i.e., the observed locations of $m$ anyons), $\mathcal{Z}_{h}^{\bs}(\vs)$ is exponentially suppressed relative to $\mathcal{Z}_{h}(\vs)$.}
    \label{fig:worldlines_postmeas}
\end{figure}

\begin{figure*}[t!]
    \centering
    \includegraphics[width = \textwidth]{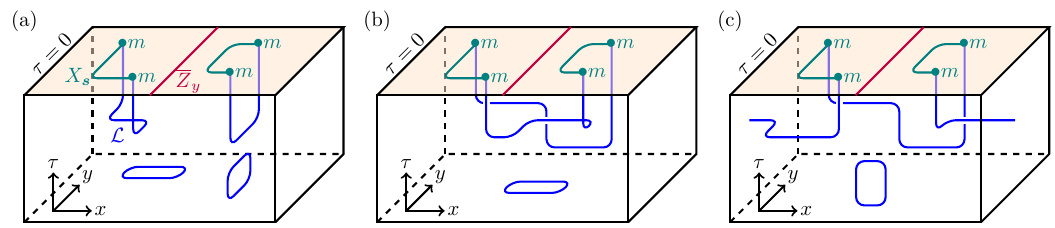}
    \caption{Schematic depiction of the anyon worldline configurations contributing to the postmeasurement ITE states $\ket{\varphi_{h}(\vs)}$. (a) Each syndrome $\vs$ (i.e., the collection of observed $m$ anyons) can be visualized as the endpoints of anyon worldlines (blue) in imaginary spacetime, with imaginary time $\tau \in [-\frac{\beta}{2}, 0]$. For a given worldline trajectory, the underlying stabilizer code state $\ket{\psi_{0}}$ is recovered without logical error when the worldlines and the anyon pairing $X_{\vs}$ form contractible closed loops. (b) Another anyon worldline trajectory which is \emph{homologous} to that of (a). Both worldlines induce the same value of the logical-$Z$ operator relative to $\ket{\psi_{0}}$, and can be corrected by the same pairing. (c) Another worldline trajectory which is \emph{non-homologous} to the preceding two. The net effect of this spacetime process, and the pairing $X_{\vs}$, is to implement a logical-$X$ operator on $\ket{\psi_{0}}$. Assuming the spacetime homology class of (a, b) is the most probable, the class of worldlines homologous to (c) contributes an exponentially small amplitude to $\ket{\varphi_{h}(\vs)}$, and consequently $X_{\vs} \ket{\varphi_{h}(\vs)}$ is equal to $\ket{\psi_{0}}$ up to exponentially small corrections.}
    \label{fig:meas_decoding_cartoon}
\end{figure*}

The preceding result has a simple physical interpretation within the quantum-classical mapping. Recall from Sec.~\ref{subsec:ite} that the overlaps $\bra*{\varphi^{\bs}_{h}} \ket*{\varphi^{\bs'}_{h}}$ are closely related to the twisted LGT partition functions $\mathcal{Z}^{\bs}_{h}$ [Eq.~\eqref{eq:Z_defn_new}], which demand that $m$ anyon worldlines nontrivially wrap the torus for $\bs \neq (0,0)$. These processes are exponentially suppressed throughout the topological phase $h < h_{c}$. Similarly, the postmeasurement overlaps $\bra{\varphi^{\bs}_{h}(\vs)} \ket*{\varphi^{\bs'}_{h}(\vs)}$ can be expressed in terms of the \emph{conditional} twisted partition functions
\begin{equation}
    \label{eq:twisted_partition_postmeas}
    \mathcal{Z}^{\bs}_{h}(\vs) = \bra*{\psi^{\bs' \oplus \bs}_{0}} e^{-\beta H_{h} / 2} \hat{\Pi}_{\vs} e^{-\beta H_{h} / 2} \ket*{\psi^{\bs'}_{0}} ,
\end{equation}
which are independent of $\bs'$ due to logical-$X$ symmetry. Using the quantum-classical mapping, these partition functions can be expressed as sums over $m$ anyon worldline configurations which are \emph{constrained} at the midpoint of the ITE, which we denote as $\tau = 0$. Specifically, the anyon worldlines are required to pierce the $\tau = 0$ imaginary-time surface at precisely the plaquettes $p$ where an $m$ anyon was observed, i.e., where $s_{p} = 1$; see Fig.~\ref{fig:worldlines_postmeas}. Eq.~\eqref{eq:logical_markov} therefore expresses the simple statement that, so long as system-spanning worldlines are exponentially rare in the unconditional ensemble, they must also be exponentially rare after conditioning on $\vs$, for typical syndromes $\vs$. In other words,
\begin{equation}
    \begin{split}
        \bra*{\varphi^{\bs}_{h}(\vs)} \ket*{\varphi^{\bs'}_{h}(\vs)} &= \frac{\mathcal{Z}^{\bs \oplus \bs'}_{h}(\vs)}{\mathcal{Z}_{h}(\vs)} \\
        &= \begin{cases}
        1, & \bs = \bs' \\
        \mathcal{O}(e^{-L / \xi'}), & \bs \neq \bs'
    \end{cases} ,
    \end{split}
\end{equation}
where the latter expression holds with high probability.

In summary, we have demonstrated that both criteria 1 and 2 from the preceding subsection are satisfied with high probability, and therefore stabilizer measurements act approximately unitarily on an arbitrary ITE code state $\ket{\varphi_{h}} = \sum_{\bs} b_{\bs} \ket{\varphi^{\bs}_{h}}$. In particular, the probability $P_{\vs}$ of observing the syndrome $\vs$ is independent of the code state, up to exponentially small relative corrections:
\begin{equation}
    \bra{\varphi_{h}} \hat{\Pi}_{\vs} \ket{\varphi_{h}} = \frac{\mathcal{Z}_{h}(\vs)}{\mathcal{Z}_{h}} \qty[ 1 + \mathcal{O}(e^{-L / \xi'}) ] ,
\end{equation}
and the corresponding postmeasurement state is
\begin{equation}
    \ket{\varphi_{h}(\vs)} = \sum_{\bs} b_{\bs} \ket{\varphi^{\bs}_{h}(\vs)} + \mathcal{O}(e^{-L / \xi'}) .
\end{equation}
Our remaining goal is to determine an anyon pairing $X_{\vs}$ which maps each $\ket{\varphi^{\bs}_{h}(\vs)}$ to $\ket{\psi^{\bs}_{0}}$ with high fidelity. If such an anyon pairing can be found, then we will have successfully achieved high-fidelity QEC of the measured ITE states.


\subsection{Optimal Postmeasurement Decoder}
\label{subsec:meas_decoding}
Having shown that stabilizer measurements preserve the encoded information with high fidelity for $h < h_c$, we will now use stoquasticity once more to determine the optimal postmeasurement anyon pairing $X_{\vs}$. Recall from Sec.~\ref{subsec:modified_protocol} (as well as Appendix~\ref{app:strong_symm}) that, due to the logical-$X$ symmetry of the entire QEC protocol, it is sufficient to decode the specific postmeasurement code state $\ket*{\varphi^{(0,0)}_{h}(\vs)}$. This state is an eigenstate of all stabilizers $A_{v}, B_{p}$ with eigenvalues $+1$ and $(-1)^{s_{p}}$ respectively, and therefore can be written without loss of generality as follows:
\begin{equation}
    \ket*{\varphi^{(0,0)}_{h}(\vs)} = \mathcal{X}_{\vs} \sum_{\bs} c_{\bs}(\vs) \ket{\psi^{\bs}_{0}} .
\end{equation}
Here $\mathcal{X}_{\vs}$ is an arbitrarily chosen product of Pauli-$X$ operators which creates the syndrome $\vs$ on top of the stabilizer states $\ket{\psi^{\bs}_{0}}$, i.e., such that $\mathcal{X}_{\vs} B_{p} = (-1)^{s_{p}} B_{p} \mathcal{X}_{\vs}$. As with the decoding unitary $X_{\vs}$, there is significant freedom in the choice of $\mathcal{X}_{\vs}$; homologous choices differ by stabilizers $A_v$ and act identically on the code states $\ket{\psi^{\bs}_0}$, while non-homologous choices differ by logical-$X$ operators and permute the four coefficients $c_{\bs}(\vs)$. Without loss of generality, we choose the homology of $\mathcal{X}_{\vs}$ such that $c_{(0,0)}(\vs)$ is the largest of the four coefficients. From Eq.~\eqref{eq:ite_fidelity}, it is then clear that $X_{\vs} = \mathcal{X}_{\vs}$ is the optimal postmeasurement decoder, and achieves a decoding fidelity
\begin{equation}
    \begin{split}
        \mathcal{F}^{\text{ITE}}_{\vs} &= \abs{ \bra*{\psi^{(0,0)}_{0}} X_{\vs} \ket*{\varphi^{(0,0)}_{h}(\vs)} }^{2} = [c_{(0,0)}(\vs)]^2 \\
        &= \max_{\bs} [c_{\bs}(\vs)]^2 .
    \end{split}
\end{equation}

We now demonstrate that this optimal postmeasurement decoder achieves a high decoding fidelity with high probability. To this end, we write the four coefficients $c_{\bs}(\vs)$ more explicitly as follows:
\begin{equation}
\label{eq:c_op}
    \begin{split}
        c_{\bs}(\vs) &= \bra{\psi^{\bs}_{0}} X_{\vs} \ket*{\varphi^{(0,0)}_{h}(\vs)} \\
        &= \frac{1}{\sqrt{\mathcal{Z}_{h}(\vs)}} \bra{\psi^{\bs}_{0}} X_{\vs} e^{-\beta H_{h} / 2} \ket*{\psi^{(0,0)}_{0}} .
    \end{split}
\end{equation}
Notably, these coefficients are strictly non-negative due to stoquasticity. Furthermore, using logical-$X$ symmetry, we have the identity
\begin{equation}
    \label{eq:postmeas_overlap}
    \begin{split}
        \bra{\varphi^{\bs}_{h}(\vs)} \ket*{\varphi^{(0,0)}_{h}(\vs)} &= \bra*{\varphi^{(0,0)}_{h}(\vs)} \overline{X}^{\bs} \ket*{\varphi^{(0,0)}_{h}(\vs)} \\
        &= \sum_{\bs'} c_{\bs'}(\vs) c_{\bs \oplus \bs'}(\vs) .
    \end{split}
\end{equation}
Finally, the results of the preceding section imply that the left-hand side of Eq.~\eqref{eq:postmeas_overlap} is exponentially small for $\bs \neq (0,0)$, with high probability. We therefore find
\begin{equation}
    \sum_{\bs'} c_{\bs'}(\vs) c_{\bs \oplus \bs'}(\vs) = \begin{cases}
        1, & \bs = (0,0) \\
        \mathcal{O}(e^{-L / \xi'}), & \bs \neq (0,0)
    \end{cases} .
\end{equation}
Since each $c_{\bs}(\vs) \geq 0$, the only solution to these equations is that the largest coefficient $c_{(0,0)}(\vs)$ is exponentially close to unity, while the other three are exponentially close to zero. We therefore find that the optimal anyon pairing $X_{\vs}$ achieves a decoding fidelity
\begin{equation}
        \mathcal{F}^{\text{ITE}}_{\vs} = [c_{(0,0)}(\vs)]^2 = 1 - \mathcal{O}(e^{-L / \xi'}) 
\end{equation}
with high probability.

We can once again construct a simple physical interpretation of this result through the quantum-classical mapping. By trotterizing the ITE in Eq.~\eqref{eq:c_op}, we can express each amplitude $c_{\bs}(\vs)$ as a sum over anyon worldlines in the imaginary-time half-space $\tau \in [- \frac{\beta}{2}, 0]$, such that the worldlines pierce the $\tau = 0$ surface at precisely the plaquettes $p$ for which $s_{p} = 1$; see Figure~\ref{fig:meas_decoding_cartoon}. The collection of all such worldline configurations $\mathcal{L}$ can be divided into four \emph{spacetime homology classes}; two worldline configurations $\mathcal{L}, \mathcal{L}'$ are considered homologous if their mod-two superposition forms contractible closed loops in the half-space, and are considered non-homologous otherwise. Since non-homologous worldline trajectories result in quantum states differing by logical-$X$ operators, we can precisely identify each coefficient $c_{\bs}(\vs)$ as the amplitude of a single spacetime homology class.
%
%
The preceding calculation implies that, throughout the topological phase $h < h_{c}$, only one of four homology classes contributes a non-negligible amplitude to each syndrome $\vs$ with high probability. The optimal anyon pairing $X_{\vs}$ is obtained by demanding that the operator $X_{\vs}$, visualized along the $\tau = 0$ surface, forms contractible closed loops with worldlines $\mathcal{L}$ in the most likely class.

In practice, to implement the optimal postmeasurement decoder, we must first numerically compute the amplitudes $c_{\bs}(\vs)$ for each homology class, and choose the anyon pairing $X_{\vs}$ to correct the most likely class. In the following section, we will demonstrate that these coefficients can be computed \emph{efficiently}\footnote{We emphasize that our claim of numerical efficiency is only empirical, and not (for example) a rigorous proof of polynomial runtime.} using Monte Carlo simulations of a constrained LGT. To understand how each $c_{\bs}(\vs)$ is computed in principle, recall first from Eq.~\eqref{eq:lgt_partition_fn} that the partition function $\mathcal{Z}_h$ can be represented as a statistical sum over $m$ anyon worldlines $\mathcal{L}$. Similarly, as discussed in Sec.~\ref{subsec:stoquasticity}, the conditional partition function $\mathcal{Z}_h(\vs)$ in Eq.~\eqref{eq:postmeas_partition_fn} can be expressed as a statistical sum of anyon worldlines which are \emph{constrained} by the observed syndrome $\vs$:
\begin{equation}
    \mathcal{Z}_h(\vs) \propto \sum_{\mathcal{L} | \vs} e^{-E[\mathcal{L}]} .
\end{equation}
Here the sum is performed over all anyon worldlines $\mathcal{L}$ which are consistent with the observed syndrome $\vs$, i.e., which pierce the $\tau = 0$ surface at precisely the plaquettes $p$ for which $s_p = 1$. The boundary conditions are such that each $\mathcal{L}$ cannot pass through the imaginary-time boundaries at $\tau = \pm \frac{\beta}{2}$, and the entire configuration is required to be contractible so as not to implement a logical-$X$ operator. This collection of worldline configurations can be further subdivided by their homology classes in the $\tau \leq 0$ space, and each coefficient $[c_{\bs}(\vs)]^2$ is the total probability of all worldline configurations in a single such homology class. Therefore, by sampling configurations $\mathcal{L}$ from the partition function $\mathcal{Z}_h(\vs)$ and tracking their homologies, we can compute the probabilities of each class and determine the optimal decoder.

\subsection{Numerical Implementation of the Optimal Postmeasurement Decoder}
\label{subsec:meas_decoding_numerics}
We now demonstrate that the optimal postmeasurement decoder described in the preceding section can be efficiently implemented using Monte Carlo numerical simulations. Rather than sampling from the extremely anisotropic partition functions $\mathcal{Z}_h$ and $\mathcal{Z}_h(\vs)$, for technical simplicity, our numerics are performed on the \emph{isotropic} $\mathbb{Z}_{2}$ LGT defined by the classical action
\begin{equation}
    \label{eq:lgt_classical_isotropic}
    \widetilde{S}_{\LGT}[U] \equiv - \frac{1}{\lambda} \sum_{[ijkl]} U_{ij} U_{jk} U_{kl} U_{li} ,
\end{equation}
and the (unconditional) partition function
\begin{equation}
    \label{eq:lgt_classical_isotropic_partition}
    \begin{split}
        \widetilde{\mathcal{Z}}_{\lambda} &\equiv \sum_{\qty{U}} e^{- \widetilde{S}_{\LGT}[U]} \\
        &\propto \sum_{\mathcal{L}} e^{- 2 \abs{\mathcal{L}} / \lambda} ,
    \end{split}
\end{equation}
where in the second line, each $\mathcal{L}$ represents a collection of $\mathbb{Z}_{2}$-valued flux loops through the dual lattice. To match the boundary conditions in the quantum matrix elements $\mathcal{Z}_{h}$ and $\mathcal{Z}_{h}(\vs)$, we impose periodic boundary conditions in the $x$ and $y$ directions, and fixed boundary conditions in the $\tau$ direction, so that $U_{ij} U_{jk} U_{kl} U_{li} = +1$ along the boundaries $\tau = \pm L_{\tau} / 2$, where $L_{\tau}$ is the total extent of the system in the $\tau$ direction. In other words, no anyon worldlines can pass through the imaginary-time boundaries at $\tau = \pm L_{\tau} / 2$. Moreover, we demand that the flux loops are globally contractible: concretely, this is equivalent to imposing that $\prod_{\expval{ij} \in \gamma} U_{ij} = 1$ for any closed loop $\gamma$ on the $\tau = \pm \frac{\beta}{2}$ surfaces.

The isotropic $\mathbb{Z}_{2}$ LGT hosts a ``deconfined'' topological phase for temperatures $\lambda$ below $\lambda_{c} \approx 1.313$, obtained from the 3$d$ Ising critical temperature $\beta_{c}^{\Ising} \approx 0.221$ \cite{ferrenberg_pushing_2018} via the Wegner duality relation $\tanh \beta_{c}^{\Ising} = e^{-2/\lambda_{c}}$ \cite{wegnerDualityGeneralizedIsing1971}. While an extremely anisotropic limit (i.e., a continuous-time limit) is necessary for an exact description of the ITE states $\ket{\varphi_{h}}$ in Eq.~\eqref{eq:imag_time_evo}, we expect that all of the universal properties near the critical point, as well as the qualitative features within the two phases, should be completely identical to the isotropic case\footnote{If desired, one could alternatively regard our numerical results as describing decoding from a space of code states $\ket{\widetilde{\varphi}_{\lambda}} \propto T_{\lambda}^{L_{\tau}/2} \ket{\psi_{0}}$, where $T_{\lambda}$ is the transfer matrix of the isotropic LGT.}.

To simulate the LGT, we utilize the \emph{worm algorithm} (WA) \cite{prokofev_worm_1998,prokofev_worm_2001,deng_dynamic_2007,collevecchio_worm_2016}, which we describe in detail in Appendix~\ref{app:worm_algo}. The WA is a highly efficient nonlocal Monte Carlo update for statistical physics systems whose configuration space consists of closed loops, and has recently been utilized in other QEC contexts~\cite{temkin_charge-informed_2025,vijay2026holographicallyemergentgaugetheory,vijay2026informationcriticalphasesdecoherence,tobias_optimal_2026}. A crucial fact about the WA is that it naturally and efficiently tunnels between all four homology classes of worldlines, and can therefore be used to estimate their relative probabilities~\cite{tobias_optimal_2026} from their observed relative frequencies.

Our numerical protocol first simulates the unconstrained 3$d$ LGT \eqref{eq:lgt_classical_isotropic} in an $L \times L \times L_{\tau}$ space using the WA, which samples flux loop configurations $\mathcal{L}$ with relative probabilities $e^{-2 \abs{\mathcal{L}} / \lambda}$. By sampling such a configuration $\mathcal{L}$, we obtain a syndrome $\vs$ from the configuration of fluxes along the $\tau = 0$ slice, leading to a conditional ensemble described by the conditional partition function
\begin{equation}
    \widetilde{\mathcal{Z}}_{\lambda}(\vs) \propto \sum_{\mathcal{L} | \vs} e^{-2 \abs{\mathcal{L}} / \lambda} ,
\end{equation}
where $\mathcal{L} | \vs$ indicates that only worldline configurations $\mathcal{L}$ consistent with the observed syndrome $\vs$ are retained in the conditional ensemble. To sample further flux loops from this ensemble, we perform further WA updates constrained to only add closed loops to the $\tau < 0$ and $\tau > 0$ spaces respectively---that is, additional flux loops added by the WA can no longer cross the timelike boundary at $\tau = 0$---so that the flux configuration at $\tau = 0$ cannot be changed.

\begin{figure}[t]
    \centering
    \includegraphics[width = \columnwidth]{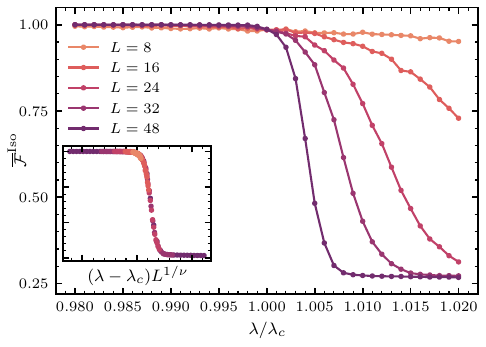}
    \caption{Average optimal postmeasurement decoding fidelity $\overline{\mathcal{F}}^{\text{Iso}}$ for the spacetime-isotropic TFTC states under stabilizer measurements, computed via WA Monte Carlo simulations of a classical $\mathbb{Z}_{2}$ LGT [see Eq.~\eqref{eq:lgt_classical_isotropic}], for several linear system sizes $L$. For all $\lambda < \lambda_{c}$, $\overline{\mathcal{F}}^{\text{Iso}}$ is exponentially close to unity. Inset: finite-size scaling collapse using the 3$d$ Ising correlation length exponent $\nu \approx 0.630$~\cite{ferrenberg_pushing_2018}.}
    \label{fig:meas_decoding_numerics}
\end{figure}

As described in the preceding section, the flux configurations in the conditional ensemble for each $\vs$ can be sorted into four homology classes by their relative homologies in the $\tau \leq 0$ half-space (see Fig.~\ref{fig:meas_decoding_cartoon}). As we sample further flux loops from the conditional ensemble, we track their homology classes and obtain statistics for the relative frequency of each class. The per-syndrome decoding fidelity $\mathcal{F}^{\text{Iso}}_{\vs}$ is then estimated from
\begin{equation}
\label{eq:F_iso}
    \mathcal{F}^{\text{Iso}}_{\vs} \equiv \frac{\max_{\bs} n_{\bs}(\vs)}{\sum_{\bs} n_{\bs}(\vs)} ,
\end{equation}
where $n_{\bs}(\vs)$ is the number of sampled configurations in the homology class $\bs$; note that $n_{\bs}(\vs) / \sum_{\bs'} n_{\bs'}(\vs)$ is the analog of $[c_{\bs}(\vs)]^2$ from the preceding section.
%
%
Finally, we average the result over many syndrome samples to obtain an estimate for the average decoding fidelity $\overline{\mathcal{F}}^{\text{Iso}}$.

Figure~\ref{fig:meas_decoding_numerics} depicts the average decoding fidelity $\overline{\mathcal{F}}^{\text{Iso}}$ as a function of $\lambda$, averaged over many sampled syndromes $\vs$. As expected from the preceding discussion, we find that the decoding fidelity is nearly perfect for all $\lambda < \lambda_{c}$. Furthermore, $\overline{\mathcal{F}}^{\text{Iso}}$ asymptotes to its minimal value of $0.25$ for all $\lambda > \lambda_{c}$; in this regime, ITE fails to produce nearly orthonormal basis states $\ket{\varphi^{\bs}_{h}}$ in the first place, and our construction fails completely. Note that the maximization over homology classes leads to a noticeable finite sample size bias in the asymptotic behavior of $\overline{\mathcal{F}}^{\text{Iso}}$ at large $\lambda$, i.e., $\overline{\mathcal{F}}^{\text{Iso}}$ actually asymptotes to a value slightly larger than 0.25. This bias can be mitigated by increasing the sample size from each observed syndrome, or by considering a \emph{probabilistic likelihood} decoder~\cite{yan_non-linear_2026,lavasani_stability_2025} in place of our ``maximum likelihood'' decoder.

Near the critical point, we find that the onset of decoding failures is well-described by the finite-size scaling ansatz
\begin{equation}
\label{eq:fss}
    \overline{\mathcal{F}}^{\text{Iso}} = \Phi((\lambda - \lambda_{c}) L^{1/\nu}) ,
\end{equation}
where $\nu \approx 0.630$ is the bulk correlation length exponent of the 3$d$ Ising model (and therefore of the 3$d$ LGT) \cite{ferrenberg_pushing_2018}, and $\Phi$ is a universal scaling function. In this way, we obtain an excellent scaling collapse with no fitting parameters whatsoever; see Fig.~\ref{fig:meas_decoding_numerics} inset.

In practice, any implementation of the optimal postmeasurement decoder crucially relies on knowledge of the transverse field strength $h$ (or the temperature $\lambda$ in our isotropic numerical simulation). A natural question is whether a simpler model-agnostic decoder, such as minimal-weight perfect matching (MWPM) \cite{dennis_topological_2002,wang_confinement-higgs_2003}, can successfully recover the encoded information at least for sufficiently small $h$. In fact, we show in Appendix~\ref{app:mwpm} that MWPM \emph{also} appears to achieve high-fidelity decoding for all $h < h_{c}$, to the best of our numerical resolution [see Fig.~\ref{fig:mwpm_1}(a)]. We are not aware of any theoretical reason that MWPM should perform as well as the optimal postmeasurement decoder in this $p = 0$ setting. In Sec.~\ref{subsec:bitflip_meas_numerics}, where we introduce bit-flip decoherence on top of the TFTC ground states, we will see numerically that MWPM exhibits a slightly smaller bit-flip decoding threshold [$p_{c}^{\text{MWPM}}(\lambda) < p_{c}^{\text{opt}}(\lambda)$] as in the stabilizer toric code.

In summary, we have shown that quantum information in the ground-state manifold of the TFTC is preserved under stabilizer measurements with high fidelity, with high probability in the syndrome measurement. Motivated by this result, we now investigate the utility of stabilizer measurements in decoding TFTC states subjected to Pauli decoherence.

%% file: sections/pauli_meas_decoding.tex
\section{QEC of Pauli Decoherence via Stabilizer Measurements}
\label{sec:pauli_meas_decoding}
Let us again suppose that two qubits of quantum information are encoded in the ITE state $\ket{\varphi_{h}} \propto e^{-\beta H_{h} / 2} \ket{\psi_{0}}$. We now subject each $\ell$th qubit to the Pauli decoherence channel~\eqref{eq:decoherence}, resulting in the decohered state $\rho_{h} \equiv \mathcal{E}(\dyad{\varphi_{h}})$. As discussed in Sec.~\ref{subsec:modified_protocol}, we will attempt to recover the information encoded originally in $\ket{\varphi_{h}}$ by measuring the stabilizers $A_{v}, B_{p}$, resulting in the postmeasurement state
\begin{equation}
    \rho_{h}(\vs) \equiv \frac{\hat{\Pi}_{\vs} \rho_{h} \hat{\Pi}_{\vs}}{P_{\vs}}
\end{equation}
with probability
\begin{equation}
    \label{eq:mixed_syndrome_prob}
    P_{\vs} \equiv \tr[\hat{\Pi}_{\vs} \rho_{h}] .
\end{equation}
Similar to Sec.~\ref{sec:meas_decoding}, we seek to determine a Pauli correction $X_{\vs} Z_{\vs}$ for each syndrome $\vs$ which recovers $\ket{\psi_{0}}$ from $\rho_{h}(\vs)$ with high fidelity.

As a first hint that stabilizer measurements can serve as an effective QEC strategy, we will show in Sec.~\ref{subsec:pauli_meas_probs} that the probability $P_{\vs}$ of observing the syndrome $\vs$ is once again independent of the initial state $\ket{\psi_{0}}$ up to exponentially small relative corrections; that is, the measurement reveals negligibly little information about the underlying encoded state. This is a necessary (but not sufficient) condition for performing high-fidelity QEC using stabilizer measurements. In Sec.~\ref{subsec:bitflip_meas_decoding}, we derive the optimal postmeasurement decoder for the bit-flip decohered ITE states. We then numerically implement (a spacetime isotropic version of) this decoder in Sec.~\ref{subsec:bitflip_meas_numerics}, demonstrating explicitly that this decoder exhibits a nonzero threshold even as the TFTC's critical point is approached. In Sec.~\ref{subsec:phase_flip} we reinstate phase-flip decoherence, where we prove that the phase-flip threshold of the TFTC is identical to that of the stabilizer toric code. Finally, in Sec.~\ref{subsec:optimality} we will argue (without a fully rigorous proof) that stabilizer measurements are asymptotically optimal for decoding bit-flip decoherence---that is, the decoding threshold for the optimal postmeasurement decoder coincides with the ``intrinsic'' error threshold for bit-flip decoherence.

\subsection{Stabilizer Measurements Cannot Reveal Encoded Information}
\label{subsec:pauli_meas_probs}

To demonstrate that the syndrome probabilities~\eqref{eq:mixed_syndrome_prob} are approximately independent of $\ket{\varphi_{h}}$, we first consider the case in which $\ket{\varphi_{h}} = \ket{\varphi^{\bs}_{h}}$ is a logical-$Z$ ITE state. In this case, the probability of observing syndrome $\vs$ is given by
\begin{equation}
    P_{\vs}^{\bs} \equiv \tr[ \hat{\Pi}_{\vs} \mathcal{E}(\dyad{\varphi^{\bs}_{h}}{\varphi^{\bs}_{h}})] = \frac{\mathcal{Z}_{h,p}(\vs)}{\mathcal{Z}_{h}} .
\end{equation}
Similar to Sec.~\ref{subsec:stoquasticity}, we have defined the postmeasurement partition function
\begin{equation}
    \label{eq:pauli_postmeas_partition}
    \mathcal{Z}_{h,p}(\vs) \equiv \bra{\psi^{\bs}_{0}} e^{-\beta H_{h} / 2} \mathcal{E}(\hat{\Pi}_{\vs}) e^{-\beta H_{h} / 2} \ket{\psi^{\bs}_{0}} ,
\end{equation}
where we have used the fact that the channel $\mathcal{E}$ is self-adjoint (i.e., its Kraus operators are Hermitian) to move the channel onto the projector. As in the preceding section, we immediately find that $P^{\bs}_{\vs}$ is independent of $\bs$ due to logical-$X$ symmetry.

For a more general state $\ket{\varphi_{h}} = \sum_{\bs} b_{\bs} \ket{\varphi^{\bs}_{h}}$, the syndrome probability $P_{\vs}$ contains additional off-diagonal contributions of the form
\begin{align}
    \tr[ \hat{\Pi}_{\vs} \mathcal{E}(\dyad*{\varphi^{\bs}_{h}}{\varphi^{\bs'}_{h}})] &= \frac{\bra*{\psi^{\bs'}_{0}} e^{-\beta H_{h} / 2} \mathcal{E}(\hat{\Pi}_{\vs}) e^{-\beta H_{h} / 2} \ket{\psi^{\bs}_{0}}}{\mathcal{Z}_{h}} \notag \\
    &= \frac{\mathcal{Z}^{\bs \oplus \bs'}_{h,p}(\vs)}{\mathcal{Z}_{h}} , \label{eq:pauli_meas_offdiag}
\end{align}
where we have defined the twisted partition functions
\begin{equation}
    \label{eq:pauli_postmeas_partition_twisted}
    \mathcal{Z}^{\bs}_{h, p}(\vs) \equiv \bra*{\psi^{\bs \oplus \bs'}_{0}} e^{-\beta H_{h} / 2} \mathcal{E}(\hat{\Pi}_{\vs}) e^{-\beta H_{h} / 2} \ket*{\psi^{\bs'}_{0}} 
\end{equation}
which are independent of $\bs'$. Notably, each $\mathcal{Z}^{\bs}_{h,p}(\vs)$ is non-negative due to stoquasticity, and satisfies
\begin{equation}
    \begin{split}
        \sum_{\vs} \mathcal{Z}^{\bs}_{h,p}(\vs) &= \bra*{\psi^{\bs \oplus \bs'}_{0}} e^{-\beta H_{h} / 2} \mathcal{E}(1) e^{-\beta H_{h} / 2} \ket*{\psi^{\bs'}_{0}} \\
        &= \mathcal{Z}^{\bs}_{h} , 
    \end{split}
\end{equation}
where $\mathcal{Z}^{\bs}_{h}$ is the unconditional twisted partition function defined in Eq.~\eqref{eq:Z_defn_new}. We therefore find
\begin{equation}
    \begin{split}
        \sum_{\vs} P^{\bs}_{\vs} \abs{ \frac{\mathcal{Z}^{\bs}_{h,p}(\vs)}{\mathcal{Z}_{h,p}(\vs)} } &= \sum_{\vs} \frac{\mathcal{Z}_{h,p}(\vs)}{\mathcal{Z}_{h}} \frac{\mathcal{Z}^{\bs}_{h,p}(\vs)}{\mathcal{Z}_{h,p}(\vs)} \\
        &= \frac{\mathcal{Z}^{\bs}_{h}}{\mathcal{Z}_{h}} .
    \end{split}
\end{equation}
Recall from Eq.~\eqref{eq:Z_defn_cost} that the final expression is exponentially small for $\bs \neq (0,0)$. Markov's inequality therefore implies that
\begin{equation}
    \label{eq:pauli_twisted_small}
    \frac{\mathcal{Z}^{\bs}_{h,p}(\vs)}{\mathcal{Z}_{h,p}(\vs)} = \begin{cases}
        1, & \bs = (0,0) \\
        \mathcal{O}(e^{-L / \xi'}), & \bs \neq (0,0)
    \end{cases}
\end{equation}
with probability exponentially close to unity. Therefore, for an arbitrary ITE state $\ket{\varphi_{h}}$, the off-diagonal contributions~\eqref{eq:pauli_meas_offdiag} to the syndrome probability $P_{\vs}$ are vanishingly small \emph{relative} to the diagonal components. Altogether, we find
\begin{equation}
    P_{\vs} = \frac{\mathcal{Z}_{h,p}(\vs)}{\mathcal{Z}_{h}} \qty[ 1 + \mathcal{O}(e^{-L / \xi'}) ]
\end{equation}
with high probability. In other words, stabilizer measurements reveal vanishingly little information encoded in the decohered states $\rho_{h}$.

\begin{figure}[t]
    \includegraphics[width=\columnwidth]{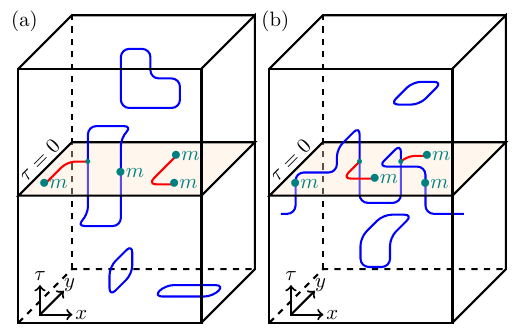}
    \caption{(a) Schematic depiction of an anyon worldline and bit-flip error configuration contributing to the postmeasurement partition function $\mathcal{Z}_{h,p}(\vs)$ in Eq.~\eqref{eq:pauli_postmeas_partition}. Contractible anyon worldlines (blue) piercing the $\tau = 0$ surface combine with bit-flip errors (red) to yield the observed syndrome $\vs$. (b) A worldline/error configuration contributing to the twisted partition function $\mathcal{Z}^{\bs}_{h,p}(\vs)$ in Eq.~\eqref{eq:pauli_postmeas_partition_twisted}, for $\bs \neq (0,0)$. The worldlines and errors are again together consistent with the observed syndrome, but now the worldlines implement a nontrivial logical-$X$ operator. For typical syndromes $\vs$, $\mathcal{Z}^{\bs}_{h,p}(\vs)$ is exponentially suppressed relative to $\mathcal{Z}_{h,p}(\vs)$.}
    \label{fig:worldlines_postmeas_pauli}
\end{figure}

Physically, we can once again regard each $\mathcal{Z}^{\bs}_{h,p}(\vs)$ as the partition function of a classical LGT, which sums over all anyon worldlines which wrap the torus so as to implement the logical operation $\overline{X}^{\bs}$. Unlike $\mathcal{Z}^{\bs}_{h}(\vs)$ in Eq.~\eqref{eq:twisted_partition_postmeas}, where the worldlines were constrained to rigidly agree with the observed syndrome $\vs$ along the $\tau = 0$ slice, here the observed syndrome is also influenced by the decoherence channel $\mathcal{E}$; see Figure~\ref{fig:worldlines_postmeas_pauli} for the case with $p_{Z} = 0$. Nevertheless, since the probability of a system-spanning worldline is exponentially suppressed in the unconditional ensemble, it must once again remain exponentially suppressed in the conditional ensemble for typical syndromes $\vs$, leading to Eq.~\eqref{eq:pauli_twisted_small}.

\subsection{Bit-Flip Decoherence: Optimal Decoding of Stabilizer Measurements}
\label{subsec:bitflip_meas_decoding}
We now move onto the construction of the optimal postmeasurement decoder for the ITE states. We begin with the case of bit-flip decoherence, where the transverse field of $H_{h}$ and the decoherence cooperate to spread $m$ anyons around the system. We will see in Sec.~\ref{subsec:phase_flip} that it is sufficient to decode bit-flips and phase-flips independently, and so here we set $p_Z = 0$ with no loss of generality. 

To construct the optimal postmeasurement decoder in this setting, it will be convenient to unravel $\rho_{h} \equiv \mathcal{E}(\dyad{\varphi_{h}})$ into a sum over all possible error strings $\ve$ as follows:
\begin{equation}
\label{eq:bitflip_expansion}
    \rho_{h} = \mathcal{E}(\dyad{\varphi_{h}}) = \sum_{\ve} p_{\ve} X_{\ve} \dyad{\varphi_{h}} X_{\ve} .
\end{equation}
Here $\ve \equiv \qty{e_{\ell}}$, with $e_{\ell} = 0, 1$, specifies a collection of bit-flip errors at each $\ell$th qubit for which $e_{\ell} = 1$. The full error is $X_{\ve} \equiv \prod_{\ell} X_{\ell}^{e_{\ell}}$, and $p_{\ve} \equiv (1 - p)^{2L^{2} - \abs{\ve}} p^{\abs{\ve}}$ is the corresponding probability of the full error. 

Notably, the bit-flip channel $\mathcal{E}$ exhibits a \emph{strong} $\mathbb{Z}_{2}$ symmetry with respect to each logical-$X$ operator $\overline{X}_{\mu}$, i.e., each error $X_{\ve}$ commutes with $\overline{X}_{\mu}$. The measurement operator $\hat{\Pi}_{\vs}$ and the $m$ anyon pairing $X_{\vs}$ similarly commute with $\overline{X}_{\mu}$. As mentioned in Sec.~\ref{subsec:modified_protocol} and explained in Appendix~\ref{app:strong_symm}, this observation implies that it is sufficient to focus on decoding the specific stabilizer code state $\ket*{\psi^{(0,0)}_{0}}$. In particular, our decoders achieve high average decoding fidelity on an arbitrary code state if and only if they do so on $\ket*{\psi^{(0,0)}_{0}}$. To simplify the notation in the remainder of this section, we will neglect the $(0,0)$ superscript and write $\ket{\varphi_{h}} \equiv \ket*{\varphi^{(0,0)}_{h}}$ and $\ket{\psi_{0}} \equiv \ket*{\psi^{(0,0)}_{0}}$.

\begin{figure*}[t]
    \includegraphics[width = \textwidth]{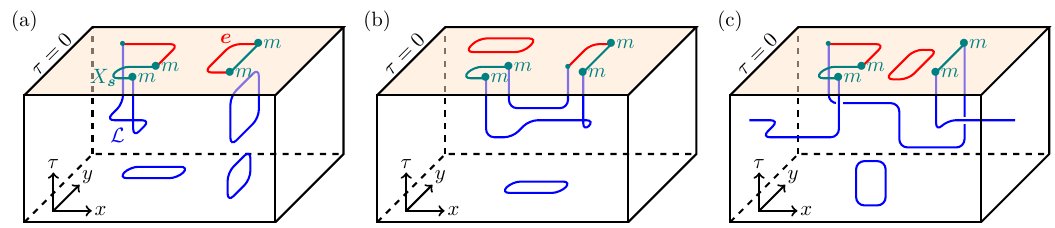}
    \caption{Schematic depiction of worldline and error configurations contributing to the postmeasurement decohered ITE states $\rho_h(\vs)$, restricted to the $\tau \leq 0$ half-space for visual simplicity. (a) Anyon worldlines $\mathcal{L}$ (blue) in the $\tau < 0$ bulk combine with bit-flip errors $\ve$ (red) to yield a syndrome $\vs$, (i.e., the collection of observed $m$ anyons). For a given $(\mathcal{L}, \ve)$ pair, the underlying stabilizer code state $\ket{\psi_0}$ is recovered without logical error when the anyon pairing $X_{\vs}$ forms contractible closed loops with $\mathcal{L}$ and $\ve$ together. (b) Another $(\mathcal{L}, \ve)$ configuration which is homologous to that of (a). Both configurations can be corrected by the same anyon pairing. (c) Another $(\mathcal{L}, \ve)$ configuration which is non-homologous to the preceding two. The net effect of the spacetime process $\mathcal{L}$, bit-flip error $\ve$, and anyon pairing $X_{\vs}$, is to implement a logical-$X$ operator on $\ket{\psi_0}$. Optimal decoding succeeds with high fidelity when the vast majority of probable $(\mathcal{L}, \ve)$ configurations fall into the same homology class.}
    \label{fig:pauli_meas_decoding_cartoon}
\end{figure*}

Using Eq.~\eqref{eq:logicalZ_ITE}, the decohered state $\rho_{h} = \mathcal{E}(\dyad{\varphi_{h}})$ is given by
\begin{equation}
    \begin{split}
        \rho_{h} &= \frac{1}{\mathcal{Z}_{h}} \mathcal{E}( e^{-\beta H_{h} / 2} \dyad{\psi_{0}} e^{-\beta H_{h} / 2} ) \\
        &= \frac{1}{\mathcal{Z}_{h}} \sum_{\ve} p_{\ve} X_{\ve} e^{-\beta H_{h} / 2} \dyad{\psi_{0}} e^{-\beta H_{h} / 2} X_{\ve} .
    \end{split}
\end{equation}
We now measure the stabilizers $B_p$ and obtain a syndrome $\vs$; as in the pure-state case of Sec.~\ref{sec:meas_decoding}, $A_{v} = +1$ in the absence of phase-flip decoherence, and so we can once again ignore $e$ anyons entirely in this section. The postmeasurement state $\rho_{h}(\vs)$ is an eigenstate of each stabilizer $A_{v}, B_{p}$ with eigenvalues $+1$ and $(-1)^{s_{p}}$ respectively, and can therefore be generally represented by a $4 \times 4$ matrix $M_{\bs \bs'}(\vs)$ as follows:
\begin{equation}
    \begin{split}
        \rho_{h}(\vs) &= \frac{\hat{\Pi}_{\vs} \rho_{h} \hat{\Pi}_{\vs}}{P_{\vs}} \\
        &= \mathcal{X}_{\vs} \qty( \sum_{\bs \bs'} M_{\bs \bs'}(\vs) \dyad*{\psi^{\bs}_{0}}{\psi^{\bs'}_{0}} ) \mathcal{X}_{\vs} .
    \end{split}
\end{equation}
As in Sec.~\ref{subsec:meas_decoding}, $\mathcal{X}_{\vs}$ is an arbitrarily chosen Pauli-$X$ operator which creates the syndrome $\vs$ on top of the stabilizer code states $\ket{\psi^{\bs}_{0}}$. The matrix elements $M_{\bs \bs'}(\vs)$ of $\rho_{h}(\vs)$ can be expressed, after straightforward algebra, as
\begin{equation}
    \label{eq:m_defn_pauli}
    \begin{split}
        M_{\bs \bs'}(\vs) &\equiv \bra{\psi^{\bs}_{0}} \mathcal{X}_{\vs} \rho_{h}(\vs) \mathcal{X}_{\vs} \ket*{\psi^{\bs'}_{0}} \\
        & \hspace{-1.2cm}= \frac{\bra{\psi_{0}} e^{-\beta H_{h} / 2} \mathcal{E}(\mathcal{X}_{\vs} \dyad*{\psi^{\bs'}_{0}}{\psi^{\bs}_{0}} \mathcal{X}_{\vs}) e^{-\beta H_{h} / 2} \ket{\psi_{0}}}{\mathcal{Z}_{h,p}(\vs)} .
    \end{split}
\end{equation}
Using $\sum_{\bs} \mathcal{X}_{\vs} \dyad{\psi^{\bs}_{0}} \mathcal{X}_{\vs} = \hat{\Pi}_{\vs}$, we immediately find that these matrix elements further satisfy the relations 
\begin{equation}
    \label{eq:M_sum}
    \sum_{\bs'} M_{\bs', \bs \oplus \bs'}(\vs) = \frac{\mathcal{Z}^{\bs}_{h,p}(\vs)}{\mathcal{Z}_{h,p}(\vs)} ,
\end{equation}
where we have defined both the postmeasurement partition function $\mathcal{Z}_{h,p}(\vs)$ and its twisted analog $\mathcal{Z}_{h,p}^{\bs}(\vs)$ in Eqs.~\eqref{eq:pauli_postmeas_partition} and~\eqref{eq:pauli_postmeas_partition_twisted} respectively. Recalling from Eq.~\eqref{eq:pauli_twisted_small} that the right-hand side is exponentially small for $\bs \neq (0,0)$, we arrive at the results
\begin{subequations}
    \begin{align}
        \sum_{\bs} M_{\bs \bs}(\vs) &= 1, \\
        M_{\bs \bs'}(\vs) &= \mathcal{O}(e^{-L / \xi'}) \quad (\bs \neq \bs') .
    \end{align}
\end{subequations}
The latter equation, which holds with high probability, follows from noting once more that $M_{\bs \bs'}(\vs) \geq 0$ due to stoquasticity, so if the sum in Eq.~\eqref{eq:M_sum} for $\bs \neq (0,0)$ is exponentially small, each term in the sum must individually be exponentially small.

An immediate useful observation from the preceding discussion is that, whereas $\rho_{h}$ exhibits both incoherent (diagonal) and coherent (off-diagonal) fluctuations in the stabilizer basis, the postmeasurement state $\rho_{h}(\vs)$ is effectively incoherent; i.e., $\rho_{h}(\vs)$ could have been obtained to excellent approximation from $\ket{\psi_{0}}$ by an incoherent Pauli channel of the form
\begin{equation}
    \label{eq:M_channel}
    \mathcal{M}_{\vs}(\rho) = \sum_{\bs} M_{\bs \bs}(\vs) \mathcal{X}_{\vs} \overline{X}^{\bs} \rho \overline{X}^{\bs} \mathcal{X}_{\vs} ,
\end{equation}
where $M_{\bs \bs}(\vs)$ is interpreted as the probability of the error $\mathcal{X}_{\vs} \overline{X}^{\bs}$. This observation is a direct result of stoquasticity, which has guaranteed that the off-diagonal elements $M_{\bs \bs'}(\vs)$ are exponentially small. By a suitable choice of $\mathcal{X}_{\vs}$, we can arrange that $M_{\bs \bs}(\vs)$ is largest for $\bs = (0, 0)$, after which it is clear that $X_{\vs} = \mathcal{X}_{\vs}$ serves as the optimal anyon pairing with decoding fidelity
\begin{equation}
    \begin{split}
        \mathcal{F}^{\text{ITE}}_{\vs}(\ket{\psi_{0}}) &= \bra{\psi_{0}} X_{\vs} \rho_{h}(\vs) X_{\vs} \ket{\psi_{0}} \\
        &= M_{(0,0), (0,0)}(\vs) .
    \end{split}
\end{equation}
On the other hand, unlike in the $p = 0$ setting of Sec.~\ref{sec:meas_decoding}, stoquasticity has \emph{not} guaranteed that $\ket{\psi_{0}}$ is recoverable from $\rho_{h}(\vs)$. Indeed, the discussion thus far has made no mention of $p$, which may be below or above the decoding threshold at this stage. High-fidelity QEC is possible whenever the single diagonal matrix element $M_{(0,0),(0,0)}(\vs)$ is large while the other three are small.

We can once again construct a physical interpretation for the diagonal coefficients $M_{\bs \bs}(\vs)$. First consider the conditional partition function $\mathcal{Z}_{h,p}(\vs)$. In the absence of the measurement projector $\hat{\Pi}_{\vs}$ (i.e., in the unconditional partition function $\sum_{\vs} \mathcal{Z}_{h,p}(\vs) = \mathcal{Z}_h$), we have an independent sum over all possible $m$ anyon worldlines $\mathcal{L}$ in spacetime and all possible bit-flip errors $\ve$; the former are weighted by the relative probabilities $e^{-E[\mathcal{L}]}$ [see Eq.~\eqref{eq:lgt_partition_fn}], while the latter are weighted by the probabilities $p_{\ve} \propto [p / (1 - p)]^{\abs{\ve}}$. The insertion of the projector $\hat{\Pi}_{\vs}$ inside $\mathcal{Z}_{h,p}(\vs)$ then couples the anyon worldlines to the bit-flip errors by a \emph{constraint}: specifically, the only $(\mathcal{L}, \ve)$ configurations contributing to $\mathcal{Z}_{h,p}(\vs)$ are those for which the worldline syndrome $\vs_{\mathcal{L}}$ along the $\tau = 0$ slice and the error syndrome $\vs_{\ve} \equiv \partial \ve$ combine to yield the observed syndrome $\vs$:
\begin{equation}
    \begin{split}
        \mathcal{Z}_{h,p}(\vs) &\propto \sum_{\mathcal{L}, \ve} \, p_{\ve} \, e^{-E[\mathcal{L}]} \, \delta(\vs_{\mathcal{L}} \oplus \vs_{\ve} = \vs) \\
        &= \sum_{\mathcal{L}, \ve | \vs} p_{\ve} \, e^{-E[\mathcal{L}]} ,
    \end{split}
\end{equation}
where in the second line, the sum is performed over only the pairs $(\mathcal{L}, \ve)$ which are together consistent with the syndrome $\vs$. Since the error syndrome combines with the worldline syndrome at $\tau = 0$, it is convenient to visualize the bit-flip errors as additional ``spacelike'' anyon worldlines constrained to the $\tau = 0$ time slice (see Fig.~\ref{fig:worldlines_postmeas_pauli}).

As in Sec.~\ref{subsec:meas_decoding}, the collection of all anyon worldlines $\mathcal{L}$ consistent with a given syndrome $\vs_{\mathcal{L}}$ can be divided into four \emph{spacetime} homology classes by considering their behavior in the half-space $\tau \in [- \frac{\beta}{2}, 0]$. Similarly, the collection of all bit-flip errors $\ve$ can be divided into four \emph{error} homology classes: two errors $\ve, \ve'$ belong to the same class if $X_{\ve}$ and $X_{\ve'}$ differ by products of $A_v$ stabilizers. Finally, the collection of all \emph{pairs} $(\mathcal{L}, \ve)$ of worldlines and errors can be divided into four \emph{joint} homology classes: two pairs $(\mathcal{L}, \ve)$ and $(\mathcal{L}', \ve')$ with the same syndrome are called homologous if they lead to the same logical state at the $\tau = 0$ time slice; see Figure~\ref{fig:pauli_meas_decoding_cartoon}. By inspecting the second line of Eq.~\eqref{eq:m_defn_pauli}, we see that the diagonal matrix elements $M_{\bs \bs}(\vs)$ are precisely the relative probabilities of these four joint homology classes.

To implement the optimal postmeasurement decoder in practice, we must numerically compute each diagonal coefficient $M_{\bs \bs}(\vs)$ and choose an anyon pairing which corrects in the most likely homology class. Analogously to Sec.~\ref{subsec:meas_decoding_numerics}, these coefficients can be computed efficiently using Monte Carlo simulations of the constrained statistical physics model described by the partition function $\mathcal{Z}_{h,p}(\vs)$, as we will now show.

\subsection{Numerical Implementation of Optimal Postmeasurement Decoder for Bit-Flip Decoherence}
\label{subsec:bitflip_meas_numerics}
We now numerically investigate the optimal postmeasurement decoder derived in the preceding section. We will once again utilize an \emph{isotropic} $\mathbb{Z}_2$ LGT for technical simplicity [see Eqs.~\eqref{eq:lgt_classical_isotropic} and \eqref{eq:lgt_classical_isotropic_partition}]; we emphasize once more that we expect all universal properties of our isotropic protocol will be identical to those of the extremely anisotropic limit necessary for an exact description of the ITE states $\ket{\varphi_h}$.

\begin{figure*}[t]
    \includegraphics[width = \textwidth]{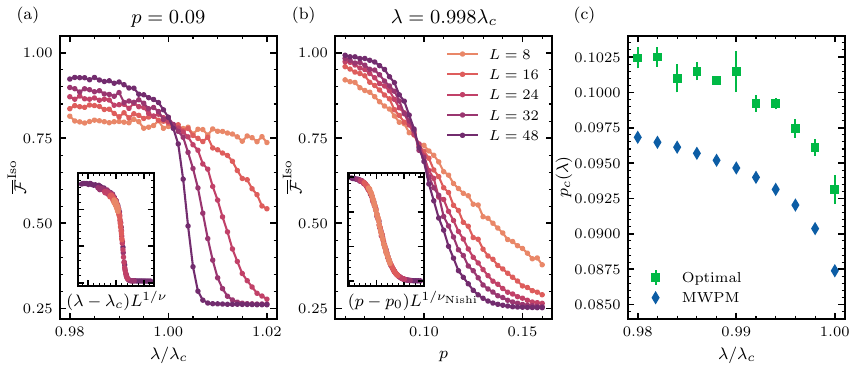}
    \caption{(a) Average optimal postmeasurement decoding fidelity $\overline{\mathcal{F}}^{\text{Iso}}$ for the spacetime-isotropic TFTC under bit-flip decoherence at strength $p = 0.09$, as a function of the LGT temperature $\lambda$ [see Eq.~\eqref{eq:lgt_classical_isotropic}], for several linear system sizes $L$. Inset: finite-size scaling collapse using the 3$d$ Ising correlation length exponent $\nu \approx 0.630$~\cite{ferrenberg_pushing_2018}. (b) Decoding fidelity $\overline{\mathcal{F}}^{\text{Iso}}$ at the fixed LGT temperature $\lambda = 0.998 \lambda_c$, as a function of the bit-flip decoherence strength $p$. Inset: finite-size scaling collapse using the critical threshold $p_0 \equiv p_c^{\text{opt}}(\lambda = 0.998\lambda_c) \approx 0.0961$ and the Nishimori RBIM exponent $\nu_{\text{Nishi}} \approx 1.534$~\cite{wan_revisiting_2025}. (c) Optimal and MWPM postmeasurement decoding thresholds, $p_c^{\text{opt}}(\lambda)$ and $p_c^{\text{MWPM}}(\lambda)$ respectively, as a function of $\lambda$. Error bars in the MWPM thresholds are too small to be shown. We see that both thresholds approach a finite value as $\lambda$ approaches $\lambda_c$ from below, and remain finite at $\lambda = \lambda_c$.}
    \label{fig:lgt_bitflip_decoding}
\end{figure*}

To sample a syndrome $\vs$, we first independently sample a flux loop configuration $\mathcal{L}$ with relative probability $e^{-2 \abs{\mathcal{L}} / \lambda}$ and a bit-flip error $\ve$ with probability $p_{\ve} = (1 - p)^{2L^2 - \abs{\ve}} p^{\abs{\ve}}$. These yield an initial worldline syndrome $\vs_{\mathcal{L}}$ and an error syndrome $\vs_{\ve}$ respectively, which together correctly sample the syndrome $\vs = \vs_{\mathcal{L}} \oplus \vs_{\ve}$ with probability
\begin{equation}
    \widetilde{P}_{\vs} \equiv \frac{\widetilde{\mathcal{Z}}_{\lambda,p}(\vs)}{\widetilde{\mathcal{Z}}_{\lambda}}, \quad \widetilde{\mathcal{Z}}_{\lambda,p}(\vs) \propto \sum_{\mathcal{L}, \ve | \vs} p_{\ve} e^{-2 \abs{\mathcal{L}} / \lambda} .
\end{equation}
To sample further pairs $(\mathcal{L}, \ve)$ from the conditional ensemble defined by $\widetilde{Z}_{\lambda, p}(\vs)$, we combine three distinct Monte Carlo updates:
\begin{enumerate}
    \item First, we update $\mathcal{L}$ in the same manner as in Sec.~\ref{subsec:meas_decoding_numerics}; i.e., we perform the WA restricted to the half-spaces $\tau > 0$ and $\tau < 0$. This update preserves the worldline syndrome $\vs_{\mathcal{L}}$, and thereby preserves $\vs$.
    \item Second, we update $\ve$ using a similar WA update, which modifies $\ve$ by a closed loop; see Appendix~\ref{app:worm_algo_stab} for explicit details. This update preserves the error syndrome $\vs_{\ve}$, and thereby preserves $\vs$.
    \item The preceding two updates are not ergodic by themselves, since they never modify $\vs_{\mathcal{L}}$ or $\vs_{\ve}$. To remedy this, our third update modifies both $\vs_{\mathcal{L}}$ and $\vs_{\ve}$ while preserving $\vs = \vs_{\mathcal{L}} \oplus \vs_{\ve}$: specifically, at each link $\ell$ of the imaginary time slice $\tau = 0$, we simultaneously add or remove an elementary flux loop encircling $\ell$ and add or remove a single bit-flip error at $\ell$. We accept this update with the usual Metropolis probability:
    \begin{equation}
        P_{\text{accept}} = \min \qty{ 1, \frac{p_{\ve'} e^{-2 \abs{\mathcal{L}'} / \lambda}}{p_{\ve} e^{-2 \abs{\mathcal{L}}/\lambda}} } ,
    \end{equation}
    where $(\mathcal{L}, \ve)$ and $(\mathcal{L}', \ve')$ are the joint configurations before and after the update, respectively.
\end{enumerate}
Finally, similar to Sec.~\ref{subsec:meas_decoding_numerics}, we obtain an estimate for the decoding fidelity $\mathcal{F}^{\text{Iso}}_{\vs}$ from the number $n_{\bs}(\vs)$ of configurations $(\mathcal{L}, \ve)$ sampled from each joint homology class $\bs$:
\begin{equation}
    \mathcal{F}^{\text{Iso}}_{\vs} = \frac{\max_{\bs} n_{\bs}(\vs)}{\sum_{\bs} n_{\bs}(\vs)} .
\end{equation}
We then average over many observed syndromes $\vs$ to obtain the average decoding fidelity $\overline{\mathcal{F}}^{\text{Iso}}$.

Figure~\ref{fig:lgt_bitflip_decoding} presents our main numerical results, which are obtained using the optimal postmeasurement decoder described above [except for the blue diamonds in Fig.~\ref{fig:lgt_bitflip_decoding}(c)]. Additional numerical plots for the optimal postmeasurement decoder can be found in Appendix~\ref{app:lgt_bitflip_moreplots}. Numerical plots for the minimal-weight perfect matching (MWPM) decoder can be found in Appendix~\ref{app:mwpm}.

Figure~\ref{fig:lgt_bitflip_decoding}(a) depicts the average decoding fidelity using the optimal postmeasurement decoder, at a fixed bit-flip decoherence strength $p = 0.09$, as a function of the LGT temperature $\lambda$. We see that for all $\lambda < \lambda_c$, $\overline{\mathcal{F}}^{\text{Iso}}$ systematically increases towards unity. We expect that $\overline{\mathcal{F}}^{\text{Iso}}$ will eventually relax towards unity exponentially quickly in the linear system size $L$, although the system sizes present are not large enough to verify this fact at this large value of $p$. The inset demonstrates a good finite-size scaling collapse using the bulk critical temperature $\lambda_c \approx 1.313$ (see Sec.~\ref{subsec:meas_decoding_numerics}) and the 3$d$ Ising correlation length exponent $\nu \approx 0.630$~\cite{ferrenberg_pushing_2018}, using a scaling form identical to Eq.~\eqref{eq:fss}; similarly good scaling collapses are obtained at smaller values of $p$ (see Appendix~\ref{app:lgt_bitflip_moreplots}). Note that the curves for different system sizes do not perfectly cross, and in fact appear to cross at $\lambda$ slightly beyond $\lambda_c$; however, the location of \emph{pairwise} crossings between adjacent system sizes appears to drift towards smaller values of $\lambda_c$. We expect that for sufficiently large system sizes, these pairwise crossings will asymptote to $\lambda_c$, yielding a clean crossing.

Figure~\ref{fig:lgt_bitflip_decoding}(b) depicts the average decoding fidelity at the fixed LGT temperature $\lambda = 0.998 \lambda_c$, just barely within the topological phase. We see clear evidence of an error threshold at $p_0 \approx 0.0961$. This threshold is noticeably below, but not dramatically reduced from, the well-known error threshold in the stabilizer limit $p_c^{\text{Nishi}} \approx 0.109$~\cite{honecker_universality_2001,dennis_topological_2002,wan_revisiting_2025}. We also achieve an excellent scaling collapse using the correlation length exponent $\nu_{\text{Nishi}} \approx 1.534$ obtained from numerical studies of the Nishimori random-bond Ising model (RBIM)~\cite{wan_revisiting_2025}. Similar scaling collapses are achieved for other values of $\lambda$ below $\lambda_c$ (see Appendix~\ref{app:lgt_bitflip_moreplots}). This numerical evidence is consistent with the expectation that the entire phase boundary $p_c^{\text{opt}}(\lambda)$ for $\lambda < \lambda_c$, which separates regimes where encoded information is recovered or destroyed under bit-flip decoherence and stabilizer measurements [see Fig.~\ref{fig:phase_diagrams}(a)], exhibits a renormalization group flow towards the optimal decoding threshold in the $\lambda = 0$ stabilizer toric code (which is well-known to be described by the Nishimori RBIM~\cite{dennis_topological_2002}).

Finally, Figure~\ref{fig:lgt_bitflip_decoding}(c) depicts the $\lambda$-dependent decoding threshold $p_c^{\text{opt}}(\lambda)$ of our optimal postmeasurement decoder, as well as the decoding threshold $p_c^{\text{MWPM}}(\lambda)$ of a MWPM decoder, both obtained from crossings and scaling collapses similar to Fig.~\ref{fig:lgt_bitflip_decoding}(b). Up to statistical fluctuations in the optimal thresholds, we find that the thresholds of both decoders monotonically decrease with increasing $\lambda$ as expected. Remarkably, we observe that both decoders exhibit a \emph{finite} decoding threshold as $\lambda$ approaches $\lambda_c$ from below. In fact, although our theoretical discussion thus far pertains only to the topological phase $\lambda < \lambda_c$ and not directly to the critical point, we find that both the optimal and MWPM numerical protocols exhibit finite decoding thresholds at $\lambda_c$. Since both $p_c^{\text{opt}}$ and $p_c^{\text{MWPM}}$ lower-bound the ``intrinsic'' error threshold (see Sec.~\ref{subsec:optimality}), the latter must also remain finite as $\lambda$ approaches $\lambda_c$. Our numerical results therefore confirm Ref.~\cite{lip}'s theoretical prediction of a finite intrinsic error threshold as the TFTC's quantum critical point is approached.

\subsection{Phase-Flip Decoherence}
\label{subsec:phase_flip}
Thus far, we have focused on decoding bit-flip decoherence in the TFTC. In that setting, both the transverse field and bit-flip decoherence cooperate to spread $m$ anyons around the system, while $e$ anyons remain entirely non-fluctuating. Here we demonstrate that phase-flip decoherence, which incoherently proliferates $e$ anyons, can also be efficiently corrected via stabilizer measurements. In fact, we prove that the critical threshold for phase-flip decoherence is entirely independent of the transverse field $h$, i.e., it is equal to its value in the stabilizer toric code (obtained from the Nishimori RBIM \cite{dennis_topological_2002}). 

For simplicity, let us first set $p_{X} = 0$ and focus on phase-flip decoherence alone. Since no $e$ anyons are present in the initial pure state $\ket{\varphi_{h}}$, the probability distribution of $e$ anyons in the decohered state $\rho_{h} \equiv \mathcal{E}(\dyad{\varphi_{h}})$ is entirely determined by the decoherence channel, and is independent of $h$. A naive decoding algorithm is therefore to projectively measure each $A_{v}$ and employ the optimal decoder of the stabilizer toric code. Unfortunately this approach fails immediately due to the presence of $m$ anyons: unlike the stabilizer toric code, different homologous pairings $Z_{\vs}$ of the $e$ anyons exhibit inequivalent action on the ITE code states $\ket{\varphi_{h}}$. The effect of performing this naive decoding strategy is to convert the phase-flip channel to a channel which applies a random Wilson loop $\prod_{\ell \in C} Z_{\ell}$, which nontrivially dephases $\ket{\varphi_{h}}$ for $h > 0$. 

The solution is to first measure the $m$ anyons. As discussed in Sec.~\ref{sec:meas_decoding}, measuring each $B_{p}$ in the state $\ket{\varphi_{h}}$ preserves the encoded information with high fidelity, with high probability in the syndrome measurement; in this case, the net effect of the $B_{p}$ stabilizer measurement and the subsequent $m$ anyon pairing $X_{\vs}$ is to \emph{unitarily} map $\ket{\varphi_{h}}$ to a corresponding stabilizer state $\ket{\psi_{0}}$, up to small corrections. Since the $B_{p}$ measurement commutes with the phase-flip channel, we have
\begin{equation}
    \begin{split}
        \frac{\hat{\Pi}^{B}_{\vs} \rho_{h} \hat{\Pi}^{B}_{\vs}}{\tr[\hat{\Pi}^{B}_{\vs} \rho_{h}]} &= \frac{\mathcal{E}(  \hat{\Pi}^{B}_{\vs} \dyad{\varphi_{h}} \hat{\Pi}^{B}_{\vs} )}{\bra{\varphi_{h}} \hat{\Pi}^B_{\vs} \ket{\varphi_{h}}} \\
        &= \mathcal{E}( X_{\vs} \dyad{\psi_{0}} X_{\vs}) + \mathcal{O}(e^{-L/\xi'}) \\
        &= X_{\vs} \mathcal{E}(  \dyad{\psi_{0}} ) X_{\vs} + \mathcal{O}(e^{-L/\xi'}) ,
    \end{split}
\end{equation}
where $\hat{\Pi}^{B}_{\vs}$ here denotes the $B_{p}$ component of the measurement projector, and the middle equality holds with probability exponentially close to unity. After correcting the $m$ anyons via $X_{\vs}$, we can subsequently measure each $A_{v}$ and perform optimal decoding of the stabilizer code~\cite{dennis_topological_2002}, which recovers $\ket{\psi_{0}}$ from $\mathcal{E}(\dyad{\psi_{0}})$ with fidelity exponentially close to unity for all $p_{Z} < p_{c}^{\text{Nishi}} \approx 0.109$. We have therefore proven that the TFTC exhibits a threshold to phase-flip decoherence given by $p_{c}^{\text{Nishi}}$, independent of $h$, for all $h < h_{c}$.

More generally, suppose $\mathcal{E}$ describes independent bit-flip and phase-flip decoherence [see Eq.~\eqref{eq:decoherence}, with $p_{Y} = p_{X} p_{Z} / (1 - p)$]. We saw in Sec.~\ref{subsec:bitflip_meas_decoding} that after $B_{p}$ stabilizer measurements, the combined effect of bit-flips and coherent fluctuations within $\ket{\varphi_{h}}$ is described by an approximately \emph{incoherent} bit-flip channel $\mathcal{M}_{\vs}$ for each syndrome $\vs$ [see Eq.~\eqref{eq:M_channel}]. Since the original bit-flip channel is independent of the phase-flip channel, we have
\begin{equation}
    \frac{\hat{\Pi}^{B}_{\vs} \rho_{h} \hat{\Pi}^{B}_{\vs}}{\tr[\hat{\Pi}^{B}_{\vs} \rho_{h}]} = \mathcal{E}^{Z} \circ \mathcal{M}_{\vs} (\dyad{\psi_{0}}) + \mathcal{O}(e^{-L/\xi'})
\end{equation}
with high probability, where $\mathcal{E}^{Z}$ is the total phase-flip channel. We now see that bit-flips and phase-flips can be corrected independently, and that the threshold for phase-flip decoherence is generally given by $p_{c}^{\text{Nishi}}$, independent of the bit-flip decoherence strength.

Finally, since the $m$ anyon correction $X_{\vs}$ commutes with the $e$ anyon stabilizers $A_{v}$, it is sufficient to measure all of the $e$ and $m$ anyons at once. We then pairwise annihilate the $m$ anyons using the correction $X_{\vs}$ discussed in Sec.~\ref{subsec:bitflip_meas_decoding}, and pairwise annihilate the $e$ anyons using $Z_{\vs}$ obtained from the optimal decoder of the stabilizer toric code. In summary, we have shown that bit-flip and phase-flip decoherence can be decoded independently, and that the threshold for phase-flip decoherence is identical to that of the stabilizer toric code for all $h < h_c$.

From this result, it is tempting to conclude that stabilizer measurements themselves serve as an optimal decoder for phase-flip decoherence. Strictly speaking, this conclusion does not follow from the preceding result: in principle, it is logically possible that the TFTC with nonzero $h$ admits a higher-threshold decoder for phase-flip decoherence than the stabilizer toric code. We believe that this scenario is unlikely, but leave a proof for future analysis.

\subsection{Optimality of Stabilizer Measurements for Decoding Bit-Flip Decoherence}
\label{subsec:optimality}
We have seen in Sec.~\ref{sec:meas_decoding} that stabilizer measurements effectively act unitarily on the TFTC code space for all $h < h_{c}$, and in Sec.~\ref{subsec:bitflip_meas_numerics} that the decoding threshold $p_{c}^{\text{opt}}(h)$ for bit-flip decoherence using stabilizer measurements remains quite large even as $h \to h_{c}$. A natural question is whether stabilizer measurements themselves constitute an (asymptotically) optimal decoding strategy, i.e., whether $p^{\text{opt}}_{c}(h)$ obtained using stabilizer measurements agrees with the \emph{intrinsic} error threshold for the TFTC under bit-flip decoherence. Here we precisely define this intrinsic error threshold, and provide theoretical arguments (falling short of a rigorous proof) which suggest that stabilizer measurements indeed constitute an asymptotically optimal decoding strategy. 

For our purposes, the intrinsic threshold $p^{\text{int}}_{c}(h)$ will be most conveniently defined via \emph{decoupling}~\cite{schumacherQuantumDataProcessing1996,schumacher_approximate_2002,preskillLectureNotesPhysics2018}. Let us denote our principal TFTC system as $Q$, and introduce an auxiliary two-qubit reference system as $R$. We begin with a maximally entangled state $\ket{\Psi_{QR}}$ between $R$ and the code space of $Q$:
\begin{equation}
    \ket{\Psi_{QR}} \equiv \frac{1}{\sqrt{4}} \sum_{\bs} \ket{\psi^{\bs}_{h}}_{Q} \otimes \ket{\bs}_{R} ,
\end{equation}
where $\ket{\psi^{\bs}_{h}}_{Q} \equiv \mathcal{U}_{h} \ket{\psi^{\bs}_{0}}_{Q}$ denotes an orthonormal basis of $\mathcal{C}_{h}$, while $\ket{\bs}_{R}$ is an orthonormal basis for all of $R$. We will further represent the bit-flip channel $\mathcal{E}$ using a Stinespring dilation~\cite{nielsenQuantumComputationQuantum2010} by introducing an explicit environment subsystem $E$, which interacts unitarily with $Q$. The resulting state on $QRE$ is given by
\begin{equation}
    \ket{\Psi_{QRE}} \equiv \frac{1}{\sqrt{4}} \sum_{\bs, \ve} \sqrt{p_{\ve}} X_{\ve} \ket{\psi^{\bs}_{h}}_{Q} \otimes \ket{\bs}_{R} \otimes \ket{\ve}_{E} ,
\end{equation}
where $\ket{\ve}_{E}$ is a basis for $E$. Then, for a fixed value of the transverse field $h$, we say that the decoherence strength $p$ is below the intrinsic threshold $p^{\text{int}}_{c}(h)$ if subsystems $R$ and $E$ become decoupled in the thermodynamic limit $L \to \infty$. In other words\footnote{In many recent works \cite{fan_diagnostics_2024,lee_exact_2025,kim2024errorthresholdsykcodes,sang_approximate_2024,wang_fractional_2025,sommers_spectral_2025,lip}, the intrinsic threshold was defined by the behavior of the coherent information $I_{c}(R \rangle Q) \equiv S(\rho_{Q}) - S(\rho_{QR})$, where $S(\rho) \equiv - \tr[ \rho \log_{2} \rho]$ is the von Neumann entropy of $\rho$. Note that the quantum Pinsker inequality~\cite{schumacher_approximate_2002} implies that $I_{c}(R \rangle Q) \leq 2 - \frac{1}{2 \log 2} \norm{\rho_{RE} - \rho_{R} \otimes \rho_{E}}_{1}^{2}$, and so a near-maximal coherent information implies a near-vanishing trace distance.\label{fn:coherent_info}},
\begin{equation}
    p^{\text{int}}_{c}(h) \equiv \sup \qty{ p : \norm{\rho_{RE} - \rho_{R} \otimes \rho_{E}}_{1} \stackrel{L \to \infty}{\longrightarrow} 0 } ,
\end{equation}
where the supremum is taken over the range $p \in [0, \frac{1}{2}]$, $\rho_{S} \equiv \tr_{S^{c}} \dyad{\Psi_{QRE}}$ is the reduced density matrix on subsystem $S$, and $\norm{O}_{1} \equiv \tr[\sqrt{O^{\dag}O}]$ is the trace norm. If $\norm{\rho_{RE} - \rho_{R} \otimes \rho_{E}}_{1} \leq \varepsilon$, then there \emph{exists} a recovery channel $\mathcal{R}$ acting on $Q$ such that \cite{schumacher_approximate_2002}
\begin{equation}
    \mathcal{F}_{e} \equiv \bra{\Psi_{QR}} \mathcal{R}(\rho_{QR}) \ket{\Psi_{QR}} \geq 1 - \varepsilon .
\end{equation}
The quantity $\mathcal{F}_{e}$, called the \emph{entanglement fidelity}~\cite{schumacher_approximate_2002,nielsenQuantumComputationQuantum2010}, can in turn be used to lower-bound the recovery fidelity $\bra{\psi_{h}} \mathcal{R}(\rho_{h}) \ket{\psi_{h}}$ for any code state $\ket{\psi_{h}} \in \mathcal{C}_{h}$; see Appendix~\ref{app:strong_symm}. Therefore, $p^{\text{int}}_{c}(h)$ precisely quantifies the \emph{intrinsic} protection of quantum information, i.e., the capacity for information to be recovered by some (perhaps extraordinarily complicated) channel $\mathcal{R}$.

Now, suppose that a stabilizer measurement is performed on subsystem $Q$, yielding the state
\begin{equation}
    \ket{\Psi_{QRE}(\vs)} \equiv \sum_{\bs, \ve} \sqrt{\frac{p_{\ve}}{4 P_{\vs}}} \hat{\Pi}_{\vs} X_{\ve} \ket{\psi^{\bs}_{h}}_{Q} \otimes \ket{\bs}_{R} \otimes \ket{\ve}_{E}
\end{equation}
with probability $P_{\vs} \equiv \tr[\hat{\Pi}_{\vs} \rho_{Q}] = \frac{1}{4} \sum_{\bs} \bra{\psi^{\bs}_{h}} \mathcal{E}(\hat{\Pi}_{\vs}) \ket{\psi^{\bs}_{h}}$. We denote the postmeasurement reduced density matrices for each subsystem $S$ by $\rho_{S}(\vs)$. We would like to argue that, if $p < p^{\text{int}}_{c}(h)$, then with high probability in the syndrome $\vs$, there exists a recovery unitary $V_{\vs, Q}$ acting on $Q$ such that $V_{\vs, Q} \rho_{QR}(\vs) V_{\vs,Q}^{\dag}$ returns $\dyad{\Psi_{QR}}$ with high fidelity.

Since subsystem $Q$ is reduced to a four-dimensional effective Hilbert space following the measurement, the desired unitary $V_{\vs,Q}$ exists if and only if $Q$ and $R$ are nearly maximally entangled following the measurement. It is convenient to break this requirement into the following two distinct criteria:
\begin{enumerate}
    \item Subsystem $R$ remains nearly maximally entangled with $QE$, i.e., $\rho_{R}(\vs) \approx \frac{1}{4} \sum_{\bs} \dyad{\bs}_{R} = \frac{1}{4} \mathds{1}_R$ is nearly maximally mixed.
    \item Subsystems $R$ and $E$ remain nearly uncorrelated, i.e., $\norm{\rho_{RE}(\vs) - \rho_{R}(\vs) \otimes \rho_{E}(\vs)}_{1}$ remains small.
\end{enumerate}
Physically, the first criterion states that the syndrome measurement does not collapse or reveal any encoded information, similar to Sec.~\ref{sec:meas_decoding}. The second criterion states that the measurement does not teleport any information previously contained within $Q$ into $E$. Since any two purifications of $RE$ differ at most by a unitary on $Q$, they together imply that
\begin{equation}
    \ket{\Psi_{QRE}(\vs)} \approx (V_{\vs, Q}^{\dag} \otimes 1_{RE}) \ket{\Psi_{QR}} \otimes \ket{\psi_{E}(\vs)} ,
\end{equation}
i.e., the stabilizer measurement has purified subsystem $E$ and disentangled it from $QR$.

To analyze these two criteria in the TFTC, it is convenient as in preceding sections to work with ITE states $\ket{\varphi^{\bs}_{h}} \propto e^{-\beta H_{h} / 2} \ket{\psi^{\bs}_{0}}$ rather than the ``true'' code states $\ket{\psi^{\bs}_{h}}$. We therefore replace the maximally entangled state $\ket{\Psi_{QR}}$ with the modified state
\begin{equation}
    \begin{split}
        \ket{\Phi_{QR}} &\equiv \frac{1}{\sqrt{4}} \sum_{\bs} \ket{\varphi^{\bs}_{h}}_{Q} \otimes \ket{\bs}_{R} \\
        &= \frac{1}{\sqrt{4 \mathcal{Z}_{h}}} \sum_{\bs} e^{-\beta H_{h} / 2} \ket{\psi^{\bs}_{0}}_{Q} \otimes \ket{\bs}_{R} ,
    \end{split}
\end{equation}
which exhibits fidelity with $\ket{\Psi_{QR}}$ exponentially close to unity; in particular, $Q$ and $R$ are approximately maximally entangled in this state. It is easily seen that the probability $P_{\vs}$ of measuring syndrome $\vs$ in this state is precisely
\begin{equation}
    P_{\vs} = \frac{1}{4} \sum_{\bs} \bra{\varphi^{\bs}_h} \mathcal{E}( \hat{\Pi} )_{\vs} \ket{\varphi^{\bs}_h} = \frac{\mathcal{Z}_{h,p}(\vs)}{\mathcal{Z}_h} ,
\end{equation}
where $\mathcal{Z}_{h,p}(\vs)$ was defined in Eq.~\eqref{eq:pauli_postmeas_partition}. 
Furthermore, a short calculation shows that the postmeasurement reduced density matrix on $R$ is given by 
\begin{equation}
    \begin{split}
        \rho_R(\vs) 
        &= \frac{1}{4} \sum_{\bs \bs'} \frac{\mathcal{Z}_{h,p}^{\bs \oplus \bs'}(\vs)}{\mathcal{Z}_{h,p}(\vs)} \dyad{\bs}{\bs'} , 
    \end{split}
\end{equation}
with $\mathcal{Z}^{\bs}_{h,p}(\vs)$ defined in Eq.~\eqref{eq:pauli_postmeas_partition_twisted}. From Eq.~\eqref{eq:pauli_twisted_small}, we see that the $\bs \neq \bs'$ matrix elements are exponentially small, with probability exponentially close to unity. Therefore, we find immediately that stoquasticity and logical-$X$ symmetry have guaranteed that
\begin{equation}
\label{eq:rho_R_s_bound}
    \norm{\rho_R(\vs) - \frac{1}{4} \mathds{1}_R}_1 \leq \mathcal{O}(e^{-L / \xi'})
\end{equation}
with high probability, and therefore $R$ remains nearly maximally entangled with $QE$ after the measurement. Notice that this result holds for arbitrary $p$, even for $p > p^{\text{int}}_c(h)$, so long as $h < h_c$. Our remaining goal is therefore to argue that, whenever $p < p^{\text{int}}_c(h)$, criterion 2 is satisfied.

Unfortunately, we are unable to prove criterion 2 without further assumptions [i.e., besides our central assumption in Eq.~\eqref{eq:Z_defn_cost}]. Here we will discuss an additional physically motivated assumption, which we conjecture (but cannot prove) holds for all $p < p^{\text{int}}_c(h)$, and which implies that criterion 2 is satisfied with high probability.

To state our physical assumption, consider the four \emph{classical} probability distributions over bit-strings $\vb{z}$,
\begin{equation}
    Q^{\bs}_{\vb{z}} \equiv \sum_{\ve} p_{\ve} \abs{\bra{\vb{z}} X_{\ve} \ket{\varphi^{\bs}_h}}^2 ,
\end{equation}
which are simply the diagonal matrix elements of the decohered basis states $\rho^{\bs}_h \equiv \mathcal{E}(\dyad{\varphi^{\bs}_h})$. Our assumption is that these classical distributions are approximately distinguishable from each other. More precisely, let us suppose that their Bhattacharyya overlap is upper-bounded by
\begin{equation}
\label{eq:bhattacharyya}
    \sum_{\vb{z}} \sqrt{Q^{\bs}_{\vb{z}} Q^{\bs'}_{\vb{z}}} \leq \varepsilon \quad (\bs \neq \bs') ,
\end{equation}
where $\varepsilon \to 0$ as $L \to \infty$. In the $p = 0$ case, this inequality holds immediately due to our previous assumption in Eq.~\eqref{eq:Z_defn_cost}. As discussed below, it also holds in the stabilizer limit $h = 0$ if and only if $p < p^{\text{int}}_c(h = 0)$. In the more general case with $h$ and $p$ both nonzero, we will now show that this assumption implies that $p < p^{\text{int}}_c(h)$, and also that criterion 2 holds with high probability.

To see first that Eq.~\eqref{eq:bhattacharyya} implies that $p < p^{\text{int}}_c(h)$, we begin by using the triangle inequality to replace $\rho_R$ with the true maximally mixed state:
\begin{equation}
    \begin{split}
        &\norm{\rho_{RE} - \rho_R \otimes \rho_E}_1 \\
        & \quad \quad \leq \norm{\rho_{RE} - \frac{1}{4} \mathds{1}_R \otimes \rho_E}_1 + \norm{\rho_R - \frac{1}{4} \mathds{1}_R}_1 .
    \end{split}
\end{equation}
Noting that $\rho_R = \frac{1}{4} \sum_{\bs \bs'} \bra*{\varphi^{\bs'}_h} \ket{\varphi^{\bs}_h} \dyad{\bs}{\bs'}$, the second term is $\mathcal{O}(e^{-L/\xi'})$ for all $h < h_c$ by Eq.~\eqref{eq:ortho_deformed_basis}, and so we can focus on the first term. Using logical-$X$ symmetry, it is straightforward to show that
\begin{subequations}
    \begin{align}
        \rho_{RE} - \frac{1}{4} \mathds{1}_R \otimes \rho_E &= \sum_{\substack{\bs \neq \bs' \\ \ve, \ve'}} \Gamma^{\bs \bs'}_{\ve \ve'} \dyad{\bs}{\bs'} \otimes \dyad{\ve}{\ve'} , \\
        \Gamma^{\bs \bs'}_{\ve \ve'} &\equiv \frac{\sqrt{p_{\ve} p_{\ve'}}}{4} \bra*{\varphi^{\bs'}_h} X_{\ve'} X_{\ve} \ket{\varphi^{\bs}_h} ,
    \end{align}
\end{subequations}
and so
\begin{equation}
        \norm{\rho_{RE} - \frac{1}{4} \mathds{1}_R \otimes \rho_E}_1 \leq \sum_{\bs \neq \bs'} \norm{\sum_{\ve \ve'} \Gamma^{\bs \bs'}_{\ve \ve'} \dyad{\ve}{\ve'}}_1 .
\end{equation}
To bound the remaining trace norm, define for each $\vb{z}$ the state vector
\begin{equation}
    \ket{E^{\bs}_{\vb{z}}} = \sum_{\ve} \sqrt{p_{\ve}} \bra{\vb{z}} X_{\ve} \ket{\varphi^{\bs}_h} \ket{\ve} ,
\end{equation}
whose squared norm is $\bra{E^{\bs}_{\vb{z}}} \ket{E^{\bs}_{\vb{z}}} = Q^{\bs}_{\vb{z}}$. Then,  inserting a resolution of the identity $\mathds{1}_Q = \sum_{\vb{z}} \dyad{\vb{z}}$ and using the triangle inequality once more,
\begin{equation}
\label{eq:gamma_bound}
    \begin{split}
        \norm{ \sum_{\ve \ve'} \Gamma^{\bs \bs'}_{\ve \ve'} \dyad{\ve}{\ve'} }_1 &\leq \frac{1}{4} \sum_{\vb{z}} \norm{ \dyad*{E^{\bs}_{\vb{z}}}{E^{\bs'}_{\vb{z}}} }_1 \\
        & \hspace{-1cm} = \frac{1}{4} \sum_{\vb{z}} \sqrt{ \bra*{E^{\bs}_{\vb{z}}} \ket*{E^{\bs}_{\vb{z}}} \bra*{E^{\bs'}_{\vb{z}}} \ket*{E^{\bs'}_{\vb{z}}} } \\
        & \hspace{-1cm} = \frac{1}{4} \sum_{\vb{z}} \sqrt{ Q^{\bs}_{\vb{z}} Q^{\bs'}_{\vb{z}} } \\
        & \hspace{-1cm} \leq \frac{\varepsilon}{4} .
    \end{split}
\end{equation}
Altogether, we find that 
\begin{equation}
    \norm{\rho_{RE} - \rho_R \otimes \rho_E}_1 \leq 3 \varepsilon + \mathcal{O}(e^{-L/\xi}) ,
\end{equation}
and therefore $p$ is below the intrinsic threshold $p^{\text{int}}_c(h)$.

On the other hand, we can also show that Eq.~\eqref{eq:bhattacharyya} implies that criterion 2 is satisfied with high probability, i.e., any encoded information can be recovered with high fidelity from the postmeasurement states. To this end, define the \emph{conditional} probability distributions
\begin{equation}
    Q^{\bs}_{\vb{z}}(\vs) \equiv \frac{1}{P_{\vs}} \sum_{\ve} p_{\ve} \abs{\bra{\vb{z}} \hat{\Pi}_{\vs} X_{\ve} \ket{\varphi^{\bs}_h}}^2 ,
\end{equation}
which are simply the diagonal matrix elements of the postmeasurement decohered basis states $\rho^{\bs}_h(\vs) \propto \hat{\Pi}_{\vs} \rho^{\bs}_h \hat{\Pi}_{\vs}$. Crucially, these conditional probability distributions satisfy
\begin{equation}
    \sum_{\vs} P_{\vs} \sum_{\vb{z}} \sqrt{Q^{\bs}_{\vb{z}}(\vs) Q^{\bs'}_{\vb{z}}(\vs)} = \sum_{\vb{z}} \sqrt{Q^{\bs}_{\vb{z}} Q^{\bs'}_{\vb{z}}} \leq \varepsilon ,
\end{equation}
and so Markov's inequality implies
\begin{equation}
    \sum_{\vb{z}} \sqrt{Q^{\bs}_{\vb{z}}(\vs) Q^{\bs'}_{\vb{z}}(\vs)} \leq \delta
\end{equation}
with probability $1 - (\varepsilon / \delta)$. Since $\varepsilon \to 0$ as $L \to \infty$, we can choose $\delta$ to vanish as $L \to \infty$ slightly slower than $\varepsilon$, and so the four conditional distributions $Q^{\bs}_{\vb{z}}(\vs)$ are almost perfectly distinguishable from each other with high probability as $L \to \infty$.

We can now construct an identical bound on the trace distance $\norm{\rho_{RE}(\vs) - \rho_R(\vs) \otimes \rho_E(\vs)}_1$ as we just constructed for $\norm{\rho_{RE} - \rho_R \otimes \rho_E}_1$. In particular, we have
\begin{align}
        &\rho_{RE}(\vs) - \frac{1}{4} \mathds{1}_R \otimes \rho_E(\vs) = \sum_{\substack{\bs \neq \bs' \\ \ve, \ve'}} \Gamma^{\bs \bs'}_{\ve \ve'}(\vs) \dyad{\bs}{\bs'} \otimes \dyad{\ve}{\ve'}, \notag \\
        &\Gamma^{\bs \bs'}_{\ve \ve'}(\vs) \equiv \frac{\sqrt{p_{\ve} p_{\ve'}}}{4 P_{\vs}} \bra*{\varphi^{\bs'}_h} X_{\ve'} \hat{\Pi}_{\vs} X_{\ve} \ket{\varphi^{\bs}_h} ,
\end{align}
and so
\begin{equation}
    \begin{split}
        \norm{\rho_{RE}(\vs) - \frac{1}{4} \mathds{1}_R \otimes \rho_E(\vs)}_1 &\leq \sum_{\bs \neq \bs'} \norm{ \sum_{\ve \ve'} \Gamma^{\bs \bs'}_{\ve \ve'}(\vs) \dyad{\ve}{\ve'} }_1  \\
        &\leq 3 \delta ,
    \end{split}
\end{equation}
where the second bound is obtained via an identical calculation to that of Eq.~\eqref{eq:gamma_bound}. Combined with Eq.~\eqref{eq:rho_R_s_bound}, we find
\begin{equation}
    \norm{\rho_{RE}(\vs) - \rho_R(\vs) \otimes \rho_E(\vs)}_1 \leq 3 \delta + \mathcal{O}(e^{-L / \xi'})
\end{equation}
with high probability, and so criterion 2 is satisfied. 

In summary, we have shown that the physical assumption~\eqref{eq:bhattacharyya} implies that stabilizer measurements can be used as part of an asymptotically optimal decoding strategy for the TFTC under bit-flip decoherence. While we have proven that Eq.~\eqref{eq:bhattacharyya} implies that $p$ is below the intrinsic threshold, we are unable to prove the converse statement in full generality. Indeed, general-purpose inequalities give the bound
\begin{align}
    \sum_{\vb{z}} \sqrt{Q^{\bs}_{\vb{z}} Q^{\bs'}_{\vb{z}}} &\leq  \sum_{\ve \ve'} \sqrt{p_{\ve} p_{\ve'}} \sum_{\vb{z}} \bra{\varphi^{\bs}_h} X_{\ve} \ket{\vb{z}} \bra{\vb{z}} X_{\ve'} \ket*{\varphi^{\bs'}_h} \notag \\
    &= 4\sum_{\ve \ve'} \Gamma^{\bs \bs'}_{\ve \ve'} \notag \\
    &\leq 4\mathcal{D}_E \norm{ \sum_{\ve \ve'} \Gamma^{\bs \bs'}_{\ve \ve'} \dyad{\ve}{\ve'} }_1 , \label{eq:bhattacharyya_reverse_bound}
\end{align}
where stoquasticity has implied that each matrix element $\bra{\vb{z}} X_{\ve} \ket{\varphi^{\bs}_h}$ is non-negative, allowing for its square root to be taken. In the final line, we have used the inequality $\norm{A}_{\ell^1} \equiv \sum_{ij} \abs{A_{ij}} \leq \mathcal{D} \norm{A}_1$ for a $\mathcal{D} \times \mathcal{D}$ matrix $A$~\cite{JohnstonMO2015EntrywiseSchatten}, so that $\mathcal{D}_{E}$ is the Hilbert space dimension of subsystem $E$. Although the final trace norm can be upper-bounded by $\norm{\rho_{RE} - \rho_R \otimes \rho_E}_1$, we generally expect the latter to be upper-bounded as roughly $\mathcal{O}(e^{-L / \xi})$; unfortunately, $\mathcal{D}_E = 2^{2L^2}$ diverges too quickly to obtain a useful bound for the Bhattacharyya overlap. 

On the other hand, we expect on physical grounds that the inequality leading to the final line of Eq.~\eqref{eq:bhattacharyya_reverse_bound} is very loose. Whereas the inequality is saturated when $\Gamma^{\bs \bs'}_{\ve \ve'}$ is proportional to a matrix of $1$s, we in fact expect that $\Gamma^{\bs \bs'}_{\ve \ve'}$ is effectively \emph{sparse}\footnote{To be slightly more precise: since exponentially many errors $\ve$ lead to the same syndrome, many different pairs of errors $\ve, \ve'$ exhibit $\bra*{\psi^{\bs'}_0} X_{\ve'} X_{\ve} \ket{\psi^{\bs}_0} = 1$ exactly, and so $\Gamma^{\bs \bs'}_{\ve \ve'}$ is not sparse as constructed. This issue can be trivially alleviated by noting that we are working with a far larger $E$ Hilbert space than is strictly necessary. After grouping errors into homology classes $\boldsymbol{E} = [\ve]$, we can assign a single quantum state in $E$ to each homology class rather than to every individual error. We then expect the resulting matrix $\Gamma^{\bs \bs'}_{\boldsymbol{E} \boldsymbol{E}'}$ to be close to a sparse matrix for $p < p^{\text{int}}_c(h)$ and $h < h_c$.}. This is because the states $\ket{\varphi^{\bs}_h}$ are short-range correlated, and so only matrix elements $\bra*{\varphi^{\bs'}_h} X_{\ve'} X_{\ve} \ket{\varphi^{\bs}_h}$ where $X_{\ve}$ and $X_{\ve'}$ create sufficiently similar anyon excitations will be appreciably large. Indeed, in the stabilizer limit $h = 0$ this intuition can be made completely precise, and a dimension-independent version of the bound in Eq.~\eqref{eq:bhattacharyya_reverse_bound} can be constructed. It would be interesting to promote this intuition to a rigorous inequality in the $h > 0$ case.

Physically, we can think of the pure-state distributions $\abs{\bra{\vb{z}} \ket{\varphi^{\bs}_h}}^2$ as forming a \emph{classical} code, and the mixed-state distributions $Q^{\bs}_{\vb{z}}$ as arising from bit-flip errors on this classical code. Such errors are correctable within the classical code so long as the resulting distributions remain distinguishable, i.e., if Eq.~\eqref{eq:bhattacharyya} is satisfied. Our physical assumption is therefore that the \emph{quantum} code defined by the code states $\ket{\varphi^{\bs}_h}$ exhibits the same threshold as the classical code defined by their matrix elements in the Pauli-$Z$ basis. The plausibility of this statement rests on the fact that the code states $\ket{\varphi^{\bs}_h}$ are non-negative in the computational basis, and that bit-flip errors $X_{\ve}$ simply shuffle the computational basis states without additional phases. Another way of phrasing our assumption is that bit-flip errors are correctable if and only if the four decohered states $\mathcal{E}(\dyad{\varphi^{\bs}_h})$ exhibit nearly disjoint support in the computational basis.

%% file: sections/tfim.tex
\section{Transverse-Field Ising Model under Bit-Flip Decoherence and Stabilizer Measurements}
\label{sec:tfim}
Thus far, we have focused on the TFTC \eqref{eq:tftc} as a Hamiltonian deformation of the stabilizer toric code. However, our general formalism is widely applicable to a broad class of stoquastic Hamiltonian deformations of CSS stabilizer codes. As another concrete example, here we briefly investigate a simpler problem: the ($d$+1)-dimensional ferromagnetic transverse-field Ising model (TFIM), viewed as a deformation of the repetition code. Previous work \cite{lavasani_stability_2025} demonstrated that information encoded in pure (1+1)$d$ TFIM ground states is protected from stabilizer measurements throughout the ferromagnetic phase; here we build on this result by studying the efficacy of stabilizer measurements in correcting Pauli decoherence on the ($d$+1)-dimensional TFIM ground states. In short, we show that the TFIM exhibits a maximal threshold to bit-flip decoherence throughout the ferromagnetic phase, and QEC can be performed efficiently by measuring the repetition code stabilizers and utilizing a majority vote decoder.

One approach to analyzing the TFIM is to follow an identical strategy as in the TFTC; i.e., utilize stoquasticity to describe code states via ITE, and then construct the optimal decoder using the quantum-classical mapping. However, as we shall see, the presence of a spontaneously broken global (0-form) symmetry---as opposed to the broken 1-form symmetry in the TFTC---will lead to a dramatically simpler construction of an asymptotically optimal postmeasurement decoder.

\subsection{Setup}
The ($d$+1)-dimensional TFIM is defined by the Hamiltonian\footnote{To avoid cluttered notation, we will repurpose several notations from the TFTC for the TFIM. We hope that the meaning is clear from context.}
\begin{equation}
    \label{eq:tfim}
    H_{g} \equiv - \sum_{\expval{v v'}} Z_{v} Z_{v'} - g \sum_{v} X_{v} ,
\end{equation}
where $v$ labels the vertices of a $d$-dimensional square lattice with $N \equiv L^{d}$ sites, with periodic boundary conditions for convenience. At $g = 0$, the ground-state manifold $\mathcal{C}_{0}$ of $H_{g=0}$ is the simultaneous $+1$ eigenspace of the repetition code stabilizers $Z_{v} Z_{v'}$, i.e., linear combinations of the code states $\ket{\vb{0}_{0}} \equiv \ket{0}^{\otimes N}$ and $\ket{\vb{1}_{0}} \equiv \ket{1}^{\otimes N}$, and therefore can be used to encode a single logical qubit. Such a code can be viewed as a ``symmetry-protected'' quantum code, in the sense that it exhibits excellent protection against errors which commute with the $\mathbb{Z}_{2}$ symmetry $\hat{\eta} \equiv \prod_{v} X_{v}$, but no protection whatsoever against symmetry-breaking errors. 

For $g$ below a critical threshold $g_{c}$, $H_{g}$ exhibits a stable ferromagnetic phase, in which the $\mathbb{Z}_{2}$ symmetry is spontaneously broken. Consequently, $H_{g}$ continues to exhibit a two-dimensional nearly degenerate ground state manifold $\mathcal{C}_{g}$, with an energetic splitting exponentially small in $L^{d}$, and a gap to all other excited states. Furthermore, ground states in $\mathcal{C}_{g}$ satisfy a local indistinguishability property analogous to Eq.~\eqref{eq:local_indistinguishability} (up to exponentially small corrections) \cite{hastings_quasiadiabatic_2005}, but only for $\mathbb{Z}_{2}$-symmetric observables $O$. We therefore expect that information encoded in $\mathcal{C}_{g}$ will be robustly protected against $\mathbb{Z}_{2}$-symmetric decoherence for all $g < g_{c}$. 


In more detail, let $\ket{\vb{0}_{g}}$ and $\ket{\vb{1}_{g}}$ denote the two spin-polarized ground states\footnote{Similar to the TFTC, these states are not exact eigenstates of $H_{g}$. Instead, the state $\ket{\boldsymbol{+}_{g}} \equiv \frac{1}{\sqrt{2}} (\ket{\vb{0}_{g}} + \ket{\vb{1}_{g}})$ is the exact ground state of $H_{g}$, and $\ket{\boldsymbol{-}_{g}} \equiv \frac{1}{\sqrt{2}}(\ket{\vb{0}_{g}} - \ket{\vb{1}_{g}})$ exhibits an exponentially small gap $\sim e^{-(L/\xi)^{d}}$.} of $H_{g}$. These are short-range entangled states, satisfying $\hat{\eta} \ket{\vb{0}_{g}} = \ket{\vb{1}_{g}}$, and exhibiting nonzero magnetization densities of opposite signs:
\begin{equation}
    \label{eq:magnetization}
    \begin{split}
        m(g) &\equiv \bra{\vb{0}_{g}} \hat{m} \ket{\vb{0}_{g}} \equiv \frac{1}{N} \sum_{v} \bra{\vb{0}_{g}} Z_{v} \ket{\vb{0}_{g}} \\
        &\sim \mathcal{O}(1) ,
    \end{split}
\end{equation}
and $\bra{\vb{1}_g} \hat{m} \ket{\vb{1}_g} = - m(g)$. A general code state $\ket{\psi_{g}} \in \mathcal{C}_{g}$ is given by a superposition
\begin{equation}
    \ket{\psi_{g}} = \sum_{\bs = \vb{0}, \vb{1}} b_{\bs} \ket{\bs_{g}} ,
\end{equation}
where $b_{\bs} \in \mathbb{C}$. Since these macroscopic superpositions are unstable to $\mathbb{Z}_{2}$-odd measurements or decoherence, we will continue to regard $\mathcal{C}_{g}$ as a symmetry-protected quantum code, i.e., we restrict our attention to $\mathbb{Z}_{2}$-symmetric operations on all code states.

Given a code state $\ket{\psi_{g}} \in \mathcal{C}_g$, we apply the bit-flip channel
\begin{equation}
    \mathcal{E}_{v}(\rho) \equiv (1 - p) \rho + p X_{v} \rho X_{v}
\end{equation}
to each vertex $v$. This yields the decohered code state $\rho_{g} \equiv \mathcal{E}(\dyad{\psi_{g}})$, where $\mathcal{E} \equiv \prod_{v} \mathcal{E}_{v}$ is the total channel. As in the TFTC, we will attempt to recover $\ket{\psi_{g}}$ from $\rho_{g}$ by measuring the stabilizers $Z_{v} Z_{v'}$ of the unperturbed model. The measurement projector is now given by
\begin{equation}
    \hat{\Pi}_{\vs} \equiv \prod_{\expval{v v'}} \qty( \frac{1 + (-1)^{s_{v v'}} Z_{v} Z_{v'}}{2} ) ,
\end{equation}
where $\vs \equiv \qty{s_{v v'}}$ is the syndrome, with $s_{v v'} = 1$ ($s_{v v'} = 0$) denoting the presence (absence) of a domain wall between vertices $v$ and $v'$. The resulting postmeasurement state is $\rho_{g}(\vs) \equiv \hat{\Pi}_{\vs} \rho_{g} \hat{\Pi}_{\vs} / P_{\vs}$ with probability $P_{\vs} \equiv \tr[\hat{\Pi}_{\vs} \rho_{g}]$. 

The state $\rho_g(\vs)$ can be returned to the repetition code space $\mathcal{C}_{0}$ by a Pauli-$X$ unitary $X_{\vs}$ which annihilates the observed domain walls, i.e., such that $X_{\vs} Z_{v} Z_{v'} = (-1)^{s_{v v'}} Z_{v} Z_{v'} X_{\vs}$. There are only two choices for $X_{\vs}$ given the syndrome $\vs$, which differ by the $\mathbb{Z}_{2}$ symmetry $\hat{\eta}$. We fix $X_{\vs}$ by demanding that it has minimal Pauli weight---that is, we employ the ``majority vote'' decoder, where we flip the minimal number of spins such that $Z_{v} Z_{v'} = +1$ after the flip. Finally, we return to the original code space $\mathcal{C}_{g}$ via an adiabatic unitary $\mathcal{U}_{g} \equiv \sum_{\bs} \dyad{\bs_{g}}{\bs_{0}}$.

Our precise claim is that the average decoding fidelity $\overline{\mathcal{F}}$ using the majority vote decoder is exponentially close to unity, for all $g < g_{c}$ and all $p < \frac{1}{2}$, i.e.,
\begin{equation}
    \label{eq:tfim_decoding_fidelity}
    \begin{split}
        \overline{\mathcal{F}}(\ket{\psi_{g}}) &\equiv \sum_{\vs} P_{\vs} \bra{\psi_{g}} \mathcal{U}_{g} X_{\vs} \rho_{g}(\vs) X_{\vs} \mathcal{U}_{g}^{\dag} \ket{\psi_{g}} \\
        &= 1 - \mathcal{O}(e^{-(L/\xi)^{d}}) , 
    \end{split}
\end{equation}
for any $\ket{\psi_{g}} \in \mathcal{C}_{g}$, where $\xi$ is a length scale which diverges as $g \to g_{c}$. To demonstrate Eq.~\eqref{eq:tfim_decoding_fidelity}, first note that the bit-flip channel $\mathcal{E}$ exhibits a \emph{strong} $\mathbb{Z}_{2}$ symmetry, i.e.,
\begin{equation}
    \hat{\eta} \mathcal{E} (\rho) = \mathcal{E}(\hat{\eta} \rho), \quad \mathcal{E}(\rho) \hat{\eta} = \mathcal{E}(\rho \hat{\eta})
\end{equation}
for any $\rho$. The decoding unitary $\mathcal{U}_{g} X_{\vs}$ is similarly $\mathbb{Z}_{2}$-symmetric. Similar to previous sections (see also Appendix~\ref{app:strong_symm}), it is therefore sufficient to focus on decoding the specific spin-polarized code state $\ket{\vb{0}_g}$. A high decoding fidelity for this state implies a high decoding fidelity for an arbitrary code state.


%
%

\subsection{Magnetization in Decohered State}
%
As noted in Eq.~\eqref{eq:magnetization}, the spin-polarized state $\ket{\boldsymbol{0}_{g}}$ exhibits an extensive magnetization throughout the ferromagnetic phase $g < g_{c}$. As $g$ approaches $g_{c}$, $m$ decays to zero as $m(g) \sim (g_{c} - g)^{\beta}$, where $\beta$ is the magnetization critical exponent (not to be confused with an inverse temperature). Moreover, since $\ket{\vb{0}_g}$ is short-range entangled, the variance in $\hat{m}$ falls to zero as $1/N$:
\begin{align}
        \sigma^{2}_{m}(g) &\equiv \bra{\vb{0}_{g}} \hat{m}^{2} \ket{\vb{0}_{g}} - \bra{\vb{0}_{g}} \hat{m} \ket{\vb{0}_{g}}^{2} \notag \\
        & \hspace{-0.75cm} = \frac{1}{N^{2}} \sum_{v, v'} \qty[ \bra{\vb{0}_{g}} Z_{v} Z_{v'} \ket{\vb{0}_{g}} - \bra{\vb{0}_{g}} Z_{v} \ket{\vb{0}_{g}} \bra{\vb{0}_{g}} Z_{v'} \ket{\vb{0}_{g}} ] \notag \\
        &\hspace{-0.75cm} \sim \mathcal{O}(1/N) .
\end{align}
Therefore, in the thermodynamic limit, essentially all of the wavefunction amplitude in $\ket{\vb{0}_{g}}$ is contained in states of magnetization density approximately $m(g)$. A simple central limit theorem argument\footnote{Since $\ket{\vb{0}_g}$ is short-range correlated, it can be blocked into roughly independent blocks of linear size $\xi$, each of which exhibits average magnetization $m(g)$. We therefore expect $\hat{m}$ to be normally distributed with a variance decaying as $(L/\xi)^{-d}$. Alternatively, one can obtain a simple and rigorous bound using Chebyshev's inequality \cite{ross_first_2020}, $\mathbb{P}[\abs{\hat{m} - m(g)} \geq \varepsilon] \leq \frac{1}{\varepsilon^{2}} \sigma^{2}_{m}(g, p = 0) \sim N^{-1}$. Although this bound is sufficient for the remainder of the discussion and leads to a rigorous proof of decodability via syndrome measurements, we expect on physical grounds that this bound is actually very loose.} suggests that the probability of measuring $\hat{m}$ in the state $\ket{\vb{0}_g}$ and observing a value significantly different from $m(g)$ is exponentially suppressed in $N$:
\begin{equation}
\label{eq:clt}
    \mathbb{P} \Big[ \abs{\hat{m} - m(g)} \geq \varepsilon \Big] \sim \mathcal{O}( e^{- \varepsilon^{2} (L/\xi)^{d}} ) ,
\end{equation}
where $\xi$ is proportional to the correlation length.

The crucial observation is that, in the presence of bit-flip decoherence, the decohered state $\rho^{\vb{0}}_{g} \equiv \mathcal{E}(\dyad{\vb{0}_g})$ continues to exhibit a finite magnetization density:
%
\begin{equation}
    \begin{split}
        m(g, p) &\equiv \tr[\hat{m} \rho^{\vb{0}}_{g}] = \bra{\vb{0}_{g}} \mathcal{E}(\hat{m}) \ket{\vb{0}_{g}} \\
        &= (1-2p) \, m(g, p=0) ,
    \end{split}
\end{equation}
where we have used the fact that the Kraus operators of $\mathcal{E}$ are Hermitian, and that $\mathcal{E}(Z_{v}) =  (1 - 2p) Z_{v}$. We see that bit-flip decoherence reduces the magnetization density of $\ket{\vb{0}_{g}}$, but does not eliminate it for any $p < \frac{1}{2}$. Similarly, a straightforward calculation shows that the variance in the magnetization is modified to
\begin{equation}
    \sigma^{2}_{m}(g, p) = (1 - 2p)^{2} \sigma^{2}_{m}(g, p = 0) + \frac{1 - (1-2p)^{2}}{N} ,
\end{equation}
and therefore remains $\mathcal{O}(1/N)$. Altogether, we find that throughout the ferromagnetic phase $g < g_{c}$ and for all $p < \frac{1}{2}$, the two decohered states $\rho^{\bs}_{g} \equiv \mathcal{E}(\dyad{\bs_{g}})$ retain finite magnetizations of opposite signs, with vanishingly small variances in the thermodynamic limit. This is already a strong hint that perfect error correction will be possible for all $p < \frac{1}{2}$.

\subsection{Decoding via Stabilizer Measurements and Majority Vote}
\label{subsec:maj_vote}
Now, suppose that we measure the stabilizers $Z_{v} Z_{v'}$ in the decohered state $\rho_{g}^{\vb{0}} \equiv \mathcal{E}(\dyad{\vb{0}_{g}})$, resulting in the postmeasurement state $\rho_{g}^{\vb{0}}(\vs)$ with probability $P_{\vs}$. Since the measurement completely determines the state of the system up to a global spin-flip, the postmeasurement state is given by
\begin{equation}
    \rho^{\vb{0}}_{g}(\vs) = X_{\vs} \qty( \sum_{\bs, \bs' = \vb{0}, \vb{1}} M_{\bs \bs'}(\vs) \dyad{\bs_{0}}{\bs'_{0}} ) X_{\vs}, 
\end{equation}
where $X_{\vs}$ is chosen as the majority-vote decoder, described previously. By this choice, $X_{\vs} \ket{\vb{0}_0}$ exhibits positive magnetization, while $X_{\vs} \ket{\vb{1}_0}$ exhibits negative magnetization.

If we were to subsequently measure $Z_{v}$ anywhere, we would learn the entire spin configuration, and thereby obtain a measurement of the magnetization $\hat{m}$. From our choice of $X_{\vs}$, we can identify the two diagonal matrix elements $M_{\bs \bs}(\vs)$ as the \emph{conditional} probabilities of observing positive or negative magnetizations in the unmeasured state $\rho^{\vb{0}}_g$, given the observed syndrome:
\begin{equation}
    M_{\vb{0} \vb{0}}(\vs) = \mathbb{P}[ \hat{m} > 0 | \vs ], \quad M_{\vb{1} \vb{1}}(\vs) = \mathbb{P}[ \hat{m} < 0 | \vs ] .
\end{equation}
Using the majority vote decoder $X_{\vs}$, the decoding fidelity for the syndrome $\vs$ is given by
\begin{equation}
    \begin{split}
        \mathcal{F}_{\vs}(\ket{\vb{0}_g}) &= \bra{\vb{0}_0} X_{\vs} \rho^{\vb{0}}_{g}(\vs) X_{\vs} \ket{\vb{0}_0} = M_{\vb{0} \vb{0}}(\vs) ,
    \end{split}
\end{equation}
and the average decoding fidelity is given by
\begin{equation}
    \overline{\mathcal{F}}(\ket{\vb{0}_g}) = \sum_{\vs} P_{\vs} M_{\vb{0} \vb{0}}(\vs) = \mathbb{P}[\hat{m} > 0] ,
\end{equation}
i.e., $\overline{\mathcal{F}}$ is simply the \emph{unconditional} probability of observing a positive magnetization in the state $\rho^{\vb{0}}_{g}$. As we have already argued, this quantity is exponentially close to unity for all $g < g_{c}$ and $p < \frac{1}{2}$, leading to the result
\begin{equation}
\label{eq:ising_avg_fidelity}
    \overline{\mathcal{F}}(\ket{\vb{0}_g}) = 1 - \mathcal{O}(e^{- (L / \xi)^{d}}) .
\end{equation}
In turn, this implies that $\mathcal{F}_{\vs}$ is exponentially close to unity, with probability exponentially close to unity. We have therefore shown that majority-vote decoding recovers the code state $\ket{\vb{0}_{g}}$ (and thereby \emph{any} code state; see Appendix~\ref{app:strong_symm}) with high fidelity for all $g < g_{c}$ and all $p < \frac{1}{2}$. Consequently, the intrinsic error threshold for bit-flip decoherence is maximal ($p_{c} = \frac{1}{2}$) throughout the ferromagnetic phase of the TFIM, and majority vote serves as an (asymptotically) optimal decoder.

We note that, unlike our analysis of the TFTC, stoquasticity played essentially no role in the preceding analysis. The only requirements were (i) a pair of short-range entangled symmetry-broken ground states, related by the $\mathbb{Z}_{2}$ symmetry; and (ii) a nonzero expectation value of $Z_{v}$ in these states. Property (i) is guaranteed to hold for all $\mathbb{Z}_{2}$-symmetric gapped deformations of the repetition code Hamiltonian \cite{bachmann_automorphic_2012}, while property (ii) holds for \emph{generic}\footnote{An example of a non-generic deformation is obtained from the depth-one unitary circuit $\prod_{v} e^{i \theta X_{v}}$ with $\theta = \pi/4$, which rotates $Z_{v}$ into $Y_{v}$. For other values of $\theta$, both $Z_{v}$ and $Y_{v}$ exhibit nonzero expectation values and can be used in majority-vote decoding.} $\mathbb{Z}_{2}$-symmetric gapped deformations. Therefore, the majority-vote decoder can be used to successfully correct non-maximal bit-flip decoherence not just in the $g < g_c$ TFIM, but generically throughout the broader ferromagnetic phase obtained from $\mathbb{Z}_{2}$-symmetric gap-preserving local perturbations to the repetition code Hamiltonian $H_{g=0}$.

\begin{figure*}[t]
    \centering
    \includegraphics[width=\textwidth]{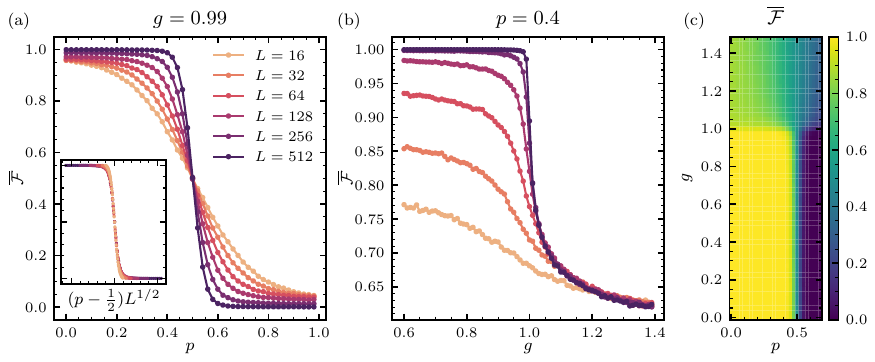}
    \caption{Average decoding fidelity $\overline{\mathcal{F}}$ in the (1+1)$d$ TFIM with transverse field strength $g$, subjected to bit-flip decoherence at strength $p$, and decoded using $Z_j Z_{j+1}$ measurements and majority vote. (a) Average decoding fidelity as a function of $p$, at fixed $g = 0.99$, for various system sizes $L$. For all $p < \frac{1}{2}$, $\overline{\mathcal{F}}$ approaches unity exponentially quickly in the system size $L$. Inset: finite-size scaling collapse using the scaling function~\eqref{eq:tfim_scaling_fixedg}, where for simplicity we set $\phi(L) = 2\overline{\mathcal{F}}(p = 0, L) - 1$. (b) Average decoding fidelity as a function of $g$, at fixed $p = 0.4$, for various system sizes $L$. Once more, $\overline{\mathcal{F}}$ approaches unity exponentially quickly in $L$ for all $g < g_c$. (c) Average decoding fidelity in the $(p, g)$ plane at the fixed system size $L = 256$. We see that $\overline{\mathcal{F}}$ is near unity throughout the region $(g, p) \in [0, 1) \times [0, \frac{1}{2})$, as predicted in Sec.~\ref{subsec:maj_vote}.}
    \label{fig:ising_bitflip}
\end{figure*}

\subsection{(1+1)\texorpdfstring{$d$}{d} Transverse-Field Ising Model: Numerical Results}
\label{subsec:tfim_numerics}
We now numerically demonstrate the preceding result in the (1+1)$d$ case, where efficient numerics are available via the Gaussian fermion covariance matrix formalism. Figure~\ref{fig:ising_bitflip} presents our main numerical results, which we discuss in the following. See Appendix~\ref{app:tfim_numerics} for further details on our numerical protocol, and additional numerical plots.


We begin by preparing a superposition of the two lowest-energy eigenstates of the (1+1)$d$ TFIM, which yields the symmetry-broken ground state $\ket{\vb{0}_g}$ in the ferromagnetic phase $g < g_c$. In the $g > g_c$ phase, our protocol prepares a superposition of the paramagnetic ground state and the first excited state; as we will see explicitly below, the code space defined by these two states does not robustly protect quantum information from bit-flip decoherence or stabilizer measurements. We employ \emph{open} boundary conditions for numerical convenience. 

For each qubit $j = 1, \ldots , L$, we randomly apply the bit-flip error $X_{j}$ with probability $p$ and leave the qubit unaffected with probability $1-p$; we denote the full error by $\ve \equiv \qty{e_{j}}$, where $e_{j} = 1$ ($e_{j} = 0$) whenever the $j$th qubit is flipped (unaffected). This stochastic unitary unraveling of the bit-flip decoherence channel will properly recover the decoding fidelity of the full channel after averaging over randomly sampled bit-flip errors, so long as our decoder does not utilize the information of which error occurred. Subsequently, we perform a projective measurement of $Z_{j} Z_{j+1}$ for $j = 1, \ldots , L-1$, resulting in a syndrome $\vs \equiv \qty{s_{j}}$. Finally, we perform the majority-vote correction $X_{\vs}$ by applying $X_j$ to each $j$th qubit belonging to the minority domain. Crucially, all of these operations preserve the Gaussianity of the state, and can therefore be implemented efficiently within the covariance matrix formalism.

Given the bit-flip realization $\ve$ and the observed syndrome $\vs$, all of the preceding operations map the initial state $\ket{\vb{0}_g}$ to the final state $\ket{\psi_g(\vs | \ve)} = b_{\vb{0}} \ket{\vb{0}_0} + b_{\vb{1}} \ket{\vb{1}_0}$. The recovery fidelity $\mathcal{F}_{\vs | \ve} = \abs{b_{\vb{0}}}^2$ can be easily extracted from the resulting state by computing the probability that $Z_1 = +1$. Therefore, the final recovery fidelity is given by
\begin{equation}
    \mathcal{F}_{\vs | \ve}(\ket{\vb{0}_g}) = \frac{1}{2} \qty[ 1 + \expval{Z_{1}}_{\vs, \ve} ] .
\end{equation}
Note that this single-site observable can be computed without implementing the majority vote on every site; we need only to flip site 1 whenever it belongs to the minority domain. Finally, the average fidelity $\overline{\mathcal{F}}$ is obtained by averaging over many sampled errors $\ve$ and subsequent syndromes $\vs$.

Figure~\ref{fig:ising_bitflip}(a) depicts the average decoding fidelity $\overline{\mathcal{F}}(\ket{\vb{0}_g})$ as a function of the bit-flip probability $p$, at a fixed value of the transverse field $g = 0.99$ just below the quantum critical point $g_c = 1$; see Fig.~\ref{fig:ising_bitflip_fixedg} for similar plots with other values of $g$, including $g = 0.999$. For all $p < p_c = \frac{1}{2}$, we find that $\overline{\mathcal{F}}$ relaxes towards unity exponentially quickly with increasing system size, as predicted in Eq.~\eqref{eq:ising_avg_fidelity}. For $p > \frac{1}{2}$, our numerical results obey the theoretically expected behavior $\overline{\mathcal{F}}(1 - p) = 1 - \overline{\mathcal{F}}(p)$ to within numerical precision. To obtain a scaling collapse, we assume that $\overline{\mathcal{F}}$ obeys the functional form
\begin{equation}
\label{eq:tfim_scaling_fixedg}
    \overline{\mathcal{F}}(p, L) = \frac{1}{2} + \phi(L) \Phi \qty((p - p_c) L^{d/2}) ,
\end{equation}
where $d = 1$ in this setting, $\Phi(x)$ is an antisymmetric function which asymptotes to $\mp \frac{1}{2}$ as $x \to \pm \infty$, and we expect $\phi(L) = 1 - \mathcal{O}(e^{-L/\xi})$. For simplicity, we set $\phi(L) = 2 \overline{\mathcal{F}}(0,L) - 1$ to achieve the collapse. The exponent $d/2$ follows from a similar central limit theorem argument to the one for Eq.~\eqref{eq:clt}: for a fixed $p < p_c = \frac{1}{2}$, the magnetization density $m(g,p)$ is expected to be normally distributed about a mean proportional to $(p_c - p)$ with a variance decaying as $1/L^d$, and so the probability of observing a negative magnetization is expected to be $\exp \qty{ -\mathcal{O} \qty( (p - p_c)^2 L^d) }$. 

Figure~\ref{fig:ising_bitflip}(b) depicts the same average decoding fidelity, now as a function of the transverse field strength $g$, at the fixed bit-flip probability $p = 0.4$; see Fig.~\ref{fig:ising_bitflip_fixedp} for other values of $p$. Once more, we see that $\overline{\mathcal{F}}$ tends to unity exponentially quickly in $L$ for all $g < 1$, as expected. In the paramagnetic phase $g > 1$, our numerical protocol does not initially prepare a ground state of the TFIM; instead, it prepares a superposition of the ground state and the first excited state. Interestingly, we find that our protocol still recovers the spin-polarized state $\ket{\vb{0}_0}$ with relatively high fidelity after measurements and majority vote are performed, but the magnitude of this fidelity appears to be independent of system size (in particular, it does not scale towards unity). A simple-minded intuition is that the initial state exhibits a magnetization density which scales as $\mathcal{O}(1/\sqrt{L})$ and a variance which scales as $\mathcal{O}(1/L)$; this is easily established in the extreme limit $g \to \infty$, and we expect the same scaling to hold throughout the $g > 1$ phase. We therefore expect an $\mathcal{O}(1)$ (i.e., asymptotically $L$-independent) probability greater than $\frac{1}{2}$ of measuring a positive magnetization in the initial state.

Finally, Fig.~\ref{fig:ising_bitflip}(c) depicts the same average decoding fidelity for the fixed system size $L = 256$, as a function of both $g$ and $p$. This figure can be qualitatively regarded as a phase diagram, with the bright yellow region depicting the phase of near-perfect QEC. As expected, we see that this phase coincides precisely with the region $g < 1$ and $p < \frac{1}{2}$.

%% file: sections/discussion.tex
\section{Discussion}
\label{sec:discussion}

In this work, we have investigated QEC protocols for near-critical topological quantum codes, defined by the ground-state manifold of the transverse-field toric code (TFTC). For small transverse fields $h$, it is natural to expect that the standard protocol of measuring and pairing anyons should remain an effective strategy for performing QEC against local Pauli decoherence; on the other hand, it is a priori unclear whether an explicit construction and implementation of the \emph{optimal} decoder for pairing the observed defects should be feasible for generic gapped deformations of the stabilizer toric code. Here we have shown that the TFTC admits conceptually simple and numerically efficient QEC algorithms, which are (asymptotically) optimal given that the stabilizers are first measured. Remarkably, these decoders exhibit thresholds against local decoherence which remain finite even in close proximity of the TFTC's quantum critical point. An analogous (and even simpler) result was obtained for the transverse-field Ising model (TFIM), viewed as a $\mathbb{Z}_{2}$ symmetry-protected quantum code: throughout the model's ferromagnetic phase, a simple majority-vote decoder was sufficient for correcting bit-flip decoherence with high fidelity for all non-maximal decoherence strengths.

We first considered stabilizer measurements of the pure (un-decohered) ground states of the TFTC, where we proved that the encoded information is preserved with high fidelity, with high probability in the measurement outcome, throughout the topological phase $h < h_{c}$. A simple physical intuition for this result arose by representing the TFTC code states using the quantum-classical mapping: logical errors arise only when anyon worldlines in imaginary time encircle the torus, and since such processes are exponentially rare in the unmeasured statistical ensemble, they remain exponentially rare in typical conditional ensembles arising from the measurements. We then explicitly demonstrated this result by constructing and numerically implementing a spacetime-isotropic formulation of the optimal postmeasurement decoder, which pairs the observed anyons to recover the stabilizer ground state. 

Building on the preceding result, we then investigated the optimal postmeasurement decoder for the TFTC subjected to independent bit-flip and phase-flip decoherence. Strikingly, the decoding threshold for bit-flip decoherence close to the quantum critical point is only slightly reduced from its value in the stabilizer limit (at least within our spacetime-isotropic numerical simulations), providing direct support for the field-theoretic prediction in Ref.~\cite{lip}. On the other hand, the threshold for phase-flip decoherence is \emph{exactly} equal to its value in the stabilizer toric code, in part because the transverse field does not generate fluctuations of the $e$ anyons. 

An interesting feature of our decoders is the scaling of its resource requirements as the critical point is approached: although we performed all numerical simulations on system sizes of fixed aspect ratio for simplicity, in principle the extent $L_{\tau}$ of the imaginary time direction needs only to be appreciably larger than the correlation length $\xi$. In particular, decoding at the stabilizer limit $h = 0$ requires simulating a two-dimensional statistical physics model, while nonzero $h$ requires simulating a quasi-2$d$ slab of thickness $L_{\tau} \sim \xi$. In the close vicinity of the quantum critical point ($|h_{c} - h| \sim L^{-1/\nu}$), a truly three-dimensional simulation is required. 


\subsection{Relation to Previous Works}
The present work is connected to several different themes in recent literature. Most directly, it relates to recent studies of mixed-state topological order~\cite{lu_detecting_2020,fan_diagnostics_2024,bao_mixedstate_2023,chenSeparabilityTransitionsTopological2024,chen_unconventional_2024,sangMixedStateQuantumPhases2024,wang_intrinsic_2025,sohal_noisy_2025,ellison_classification_2025,zhang_strong-to-weak_2025,zhou_finite-temperature_2025}, and mixed-state phases of matter more broadly~\cite{coserClassificationPhasesMixed2019,leeQuantumCriticalityDecoherence2023,lessa2024strongtoweakspontaneoussymmetrybreaking,sala2024spontaneousstrongsymmetrybreaking,sang_stability_2025,weinstein_efficient_2025,ma_topological_2025,gu_spontaneous_2025,ma_symmetry-protected_2025,lessa_mixed-state_2025,zhang_quantum_2025}. While much of this literature has focused on mixed states obtained by decohering fixed-point states (e.g. stabilizer states), our setting provides a tractable example in which substantial analytic and numerical control persists beyond the fixed-point limit. Such studies help to establish the broader phenomenology of mixed-state phases beyond particular solvable examples. The interplay between pure-state quantum critical phenomena and information-theoretic decoherence-induced transitions is of particular interest, and deserves further study.

There are also similarities between the present work and previous results on measurements~\cite{garrattMeasurementsConspireNonlocally2023a,weinsteinNonlocalityEntanglementMeasured2023a,yang_entanglement_2023,murciano_measurement_2023} and decoherence \cite{leeQuantumCriticalityDecoherence2023,zouChannelingQuantumCriticality2023} performed on critical quantum ground states. As in that setting, both measurements and decoherence appear within the present formalism as a disordered defect along a codimension-one slice of Euclidean spacetime. From this perspective, it is tempting to analyze the effect of stabilizer measurements near the quantum critical point using a similar approach: namely, by weakening the measurement projector $\hat{\Pi}_{\vs}$ to $\hat{\Pi}_{\vs}(\mu) \propto \exp \qty{ \mu \sum_v (-1)^{s_v} A_v + \mu \sum_p (-1)^{s_p} B_p }$, one can analyze the perturbative relevance or irrelevance of stabilizer measurements at the quantum critical point using a replica approach. Although measurements appear as an irrelevant perturbation in such an analysis~\cite{lip}, one cannot preclude the possibility of a measurement-induced transition at a finite value of $\mu$. In cases where conventional projective measurements of stabilizers destroy encoded information~\cite{bravyi_correcting_2018,venn_coherent_2023,behrends_surface_2024,behrends2024statisticalmechanicalmappingmaximumlikelihood,bao2024phasesdecodabilitysurfacecode,cheng_emergent_2025,yan_non-linear_2026}, it would be interesting to analyze whether arbitrarily weak stabilizer measurements similarly destroy the information, or if there might exist a transition driven by the measurement strength.

Along different lines, Ref.~\cite{liu_detecting_2025} previously proposed a relation between the pure-state decoding problem analyzed in Sec.~\ref{sec:meas_decoding} and the emergence of higher-form symmetries. In the simple case of the TFIM, the connection is particularly transparent. Specifically, the $d$-dimensional ferromagnetic TFIM exhibits an emergent $d$-form symmetry generated by the ``dressed'' order parameter $\widetilde{Z}_{v}$~\cite{hastings_quasiadiabatic_2005}, and any $\mathbb{Z}_{2}$-odd local observable (such as $Z_{v}$) generically exhibits a nonzero overlap with $\widetilde{Z}_{v}$. Any such local observable can therefore be used to construct an extensive ``bare'' magnetization (e.g. $\frac{1}{N} \sum_{v} Z_{v}$), which can subsequently be utilized to implement majority-vote decoding. From this perspective, the success of our majority vote decoder can be broadly attributed to the emergence of a $d$-form symmetry in the ferromagnetic phase; as discussed in Sec.~\ref{subsec:maj_vote}, this observation implies that majority-vote decoding will succeed for generic states adiabatically connected to the ferromagnetic fixed point, not just for the stoquastic TFIM deformation. 

On the other hand, the same logic does not obviously carry over to the case of the TFTC. In this setting, the code space exhibits an emergent ``logical-$Z$'' 1-form symmetry throughout the topological phase $h < h_{c}$ (in addition to the microscopic ``logical-$X$'' 1-form symmetry). Whereas the $d$-form symmetry generator of the TFIM is a local observable, the 1-form symmetry generators of the TFTC are one-dimensional objects. As a result, the ``bare'' logical-$Z$ operators exhibit exponentially small overlaps in $L$ with their dressed counterparts. From this perspective, the existence of the dressed logical-$Z$ operator does not obviously guarantee successful decoding via bare $B_{p}$ stabilizer measurements: indeed, previous studies have suggested that such measurements can destroy encoded information for a broad family of states within the topological phase, where such a dressed logical operator must emerge~\cite{hastings_quasiadiabatic_2005}. From this perspective, the precise connection between emergent higher-form symmetries and the success of QEC via stabilizer measurement remains partially unresolved\footnote{For clarity, we emphasize that Ref.~\cite{liu_detecting_2025} proposes to \emph{define} the emergence of a higher-form symmetry by successful QEC using bare stabilizer measurements. It is unclear to us whether this definition is equivalent to the existence of quasi-local topological symmetry operators, such as those obtained from quasi-adiabatic continuation~\cite{hastings_quasiadiabatic_2005}.}. It would be interesting to establish such a connection in further detail, and especially to extend this connection to the mixed-state setting: specifically, is there a sense in which the bit-flip decohered toric code (either at $h = 0$, or more generally for $h < h_{c}$) admits an emergent logical-$Z$ symmetry below its error threshold~\cite{rakovszky_dressed_2026}? 

There are also interesting potential connections to recent work on bottlenecks in the classical dynamics of $\mathbb{Z}_{2}$ gauge theories~\cite{stahl_slow_2026,thorngren2026perturbativelystableselfcorrectingclassical}. In that setting, the large free energy cost of long flux loops discussed in Sec.~\ref{subsec:ite} leads to long memory lifetimes of the gauge theory under local dissipative dynamics. Interestingly, Refs.~\cite{stahl_slow_2026,thorngren2026perturbativelystableselfcorrectingclassical} were able to extend this result beyond the pure LGT to include a coupling to matter, which breaks the microscopic one-form symmetry relating the code states; in some sense, this symmetry re-emerges \emph{entropically} within the topological phase, leading to well-defined code states separated by bottlenecks. It would be interesting to determine whether similar techniques can be applied to analyze QEC in the toric code perturbed by transverse and longitudinal fields (discussed in the following section), which is related to the LGT with matter by the quantum-classical mapping~\cite{tupitsyn_topological_2010}.

Finally, it is useful to compare our numerical method with previous work on QEC and measurement-induced phenomena. There are technical similarities with Ref.~\cite{baweja_postmeasurement_2025}, which showed how quantum Monte Carlo (QMC) can be used to study measurement-induced collective phenomena in ground states. In that work as well as here, post-measurement ground states are represented as imaginary-time path integrals with additional constraints or operator insertions at a particular imaginary time. Key features of the WA implemented here and the directed loop algorithm in Ref.~\cite{baweja_postmeasurement_2025} are that they (i) satisfy detailed balance with respect to the postmeasurement ensemble and (ii) remain ergodic in spite of the constraints associated with measurements. These features are shared with other recent applications of the WA in the context of QEC for stabilizer codes \cite{tobias_optimal_2026,vijay2026informationcriticalphasesdecoherence} and symmetry-enriched analogs \cite{temkin_charge-informed_2025}.

\subsection{Comments on QEC in the Fradkin-Shenker Model}
In developing our QEC protocols for the TFTC, many of the technical details were simplified greatly by the presence of an exact 1-form ``logical-$X$'' symmetry at all transverse field strengths. Given that the topological phase of the toric code is stable to \emph{all} local perturbations \cite{bravyi_topological_2010,bravyi_short_2011}, it is interesting to ask whether ``bare'' stabilizer measurements remain an effective decoding strategy when this symmetry is microscopically broken. A model of particular interest is the Fradkin-Shenker (FS) model~\cite{fradkinPhaseDiagramsLattice1979,tupitsyn_topological_2010,vidal_low-energy_2009,dusuel_robustness_2011,wu_phase_2012,verresenHiggsCondensatesAre2022}, i.e., the toric code perturbed by both transverse and longitudinal fields:
\begin{equation}
    H_{\vh} \equiv - \sum_{v} A_{v} - \sum_{p} B_{p} - h_{X} \sum_{\ell} X_{\ell} - h_{Z} \sum_{\ell} Z_{\ell} ,
\end{equation}
where $\vh \equiv (h_{X}, h_{Z})$. Whereas the transverse field $h_{X}$ provides quantum fluctuations to the $m$ anyons (i.e., $B_{p} = -1$ excitations), the longitudinal field $h_{Z}$ provides quantum fluctuations of the $e$ anyons. Crucially, $h_{Z}$ preserves the stoquasticity of the model in both the $Z$ basis and the $X$ basis, but \emph{not} in the stabilizer basis: any imaginary-time process in which an $e$ anyon encircles an $m$ anyon results in a phase of $-1$. The FS model therefore serves as a minimal model which incorporates both $e$ and $m$ anyon fluctuations, while preserving stoquasticity in the computational basis. 

On one hand, the presence of $h_{Z}$ makes a fully rigorous analysis of our decoding protocols significantly more complex than in the TFTC case with $h_{Z} = 0$. On the other hand, the phenomenology of $H_{\vh}$ and its ground states is well-understood~\cite{fradkinPhaseDiagramsLattice1979}, and so it is possible to make several well-founded conjectures on the decoding problem in this model. 


For example, it is straightforward to extend the construction of the sign-free ITE code states $\ket{\varphi^{\bs}_{\vh}} \propto e^{-\beta H_{\vh} / 2} \ket{\psi^{\bs}_{0}}$ to include the nonzero longitudinal field; just as in the $h_{Z} = 0$ case in Sec.~\ref{subsec:ite}, we expect imaginary-time evolution to furnish an approximately \emph{unitary} map from the stabilizer code space $\mathcal{C}_{0}$ to the code space $\mathcal{C}_{\vh}$ of the FS model. Specifically, we conjecture that
\begin{equation}
    \label{eq:fs_orthogonality}
    \bra{\varphi^{\bs}_{\vh}} \ket*{\varphi^{\bs'}_{\vh}} \equiv \frac{\bra{\psi^{\bs}_{0}} e^{-\beta H_{\vh}} \ket*{\psi^{\bs'}_{0}}}{\frac{1}{4} \tr[\hat{\Pi}_{0} e^{-\beta H_{\vh}}]} = \delta_{\bs \bs'} + \mathcal{O}(e^{-L / \xi}) .
\end{equation}
Indeed, similar to the discussion of Sec.~\ref{subsec:ite}, the processes leading to nonzero off-diagonal matrix elements can be regarded as arising from imaginary spacetime processes where $m$ anyons wrap the torus. On the other hand, non-uniformities of the diagonal matrix elements come from spacetime processes where $e$ anyons wrap the torus. Since both anyon species are gapped for $\vh$ within the topological phase, we expect all such processes to be exponentially suppressed in $L$. In fact, for a fixed $h_{X}$, we expect turning on a small $h_{Z}$ will only help to prevent $m$ anyons from wrapping the torus, due to the destructive interference of $e$ and $m$ worldlines. 

Given that both $e$ and $m$ worldlines are expected to remain contractible with high probability in the unmeasured ensemble, if one subsequently measures both $A_{v}$ and $B_{p}$ stabilizers, it is natural to further conjecture that the worldlines should remain contractible in the conditional ensemble with fixed syndrome along the $\tau = 0$ time slice. As a result, we expect that stabilizer measurements should again preserve quantum information in $\mathcal{C}_{\vh}$. In the $h_{Z} = 0$ case, we were able to \emph{prove} this to be the case using stoquasticity; see Sec.~\ref{subsec:stoquasticity}. With nonzero $h_{Z}$, we are unable to prove an analogous result by only assuming Eq.~\eqref{eq:fs_orthogonality} to be true. The key difference is that, whereas the postmeasurement basis states $\ket{\varphi^{\bs}_{h}(\vs)}$ for the TFTC remained strictly sign-free, the same is not true for nonzero $h_{Z}$, since any $e$ anyon measurements generally lead to unwanted minus signs. From a broader perspective, the obstruction is that the statistical ensemble containing both $e$ and $m$ worldlines is not sign-free---i.e., $H_{\vh}$ is not stoquastic in the stabilizer basis---and this makes it more challenging to infer properties of the conditional ensemble from that of the unconditional ensemble. On the other hand, while the destructive interference between $e$ and $m$ loops makes a rigorous analytical attack difficult, we also expect it to physically help prevent long loops of both species, and therefore improve (or at least not spoil) the capacity for stabilizer measurements to serve as part of a reliable decoder.


In the absence of analytical support for the protection of information under stabilizer measurements (either with or without decoherence), one strategy is to construct an optimal postmeasurement decoder and numerically evaluate its performance. Here we arrive at another stumbling block: such an optimal decoder involves conditioning on both the $e$ and $m$ anyon worldlines at once, and results in a statistical physics model with a sign problem. A natural sub-optimal decoder is to condition on only the $e$ syndromes or only the $m$ syndromes, in which case the resulting conditional ensemble admits a sign-free representation in terms of the 3$d$ classical Fradkin-Shenker model~\cite{tupitsyn_topological_2010}. Unfortunately, this representation does not (to our knowledge) admit a useful nonlocal Monte Carlo update like the WA which efficiently tunnels between the different worldline homology sectors. Local Monte Carlo updates like the Metropolis algorithm can be employed to simulate the model~\cite{tupitsyn_topological_2010,somoza_selfdual_2021,serna_worldsheet_2024}, but such algorithms are likely to face severe bottlenecks for estimating the statistics of the different homology sectors. 

In spite of these challenges, we expect that much more can be concretely said about the protection and recovery of quantum information in the Fradkin-Shenker model. For instance, perhaps it is possible to simultaneously utilize stoquasticity in both the Pauli-$X$ and Pauli-$Z$ bases to analyze the structure of the postmeasurement code states in greater theoretical detail. Alternatively, on the numerical side, we expect that the bottleneck in the sub-optimal decoder proposed above will be the least severe in close proximity to the quantum critical point, and so perhaps such a decoder can still be utilized to study the near-critical error threshold. We leave these interesting open directions for future work.

\subsection{Open Questions and Future Directions}
Much of the physical intuition, and efficient numerical methods, behind our QEC protocols arose from stoquasticity and the quantum-classical mapping. Nevertheless, one may reasonably ask if a more physical mechanism than stoquasticity can guarantee the protection of quantum information under stabilizer measurement. One specific example of a non-stoquastic deformation we expect to be decodable via stabilizer measurements is the toric code in a Pauli-$Y$ field, $H_{Y} = H_{0} - h_{Y} \sum_{\ell} Y_{\ell}$, which exhibits a large number of subsystem symmetries and is dual to the (stoquastic) plaquette Ising model~\cite{xu_strong-weak_2004,nussinov_discrete_2005,vidal_self-duality_2009}. More broadly, although we have recently argued~\cite{lip} that a finite \emph{intrinsic} error threshold should remain robust to generic gap-preserving deformations of the toric code, we expect that such deformed codes should generically \emph{not} be decodable via stabilizer measurements. It would be especially interesting to study the breakdown of our QEC protocol as one tunes away from a stoquastic limit.

A related open question concerns the optimality of stabilizer measurements for decoding bit-flip decoherence. Although we have (non-rigorously) argued that stabilizer measurements can be utilized as part of an optimal decoding strategy for the TFTC, it remains possible that the `intrinsic' threshold is slightly larger than the threshold of our optimal postmeasurement decoder. In principle, this question can be resolved by (for example) comparing the coherent information $I_{c}(R \rangle Q)$ (see footnote~\ref{fn:coherent_info}) to the \emph{measurement-averaged} coherent information, $\sum_{\vs} P_{{\vs}} I_{c}(R \rangle Q | \vs)$. While the latter is upper-bounded by the former due to the quantum data-processing inequality~\cite{schumacherQuantumDataProcessing1996,nielsenQuantumComputationQuantum2010}, we expect from the discussion of Sec.~\ref{subsec:optimality} that both quantities exhibit a transition from their maximal value to zero at the same critical decoherence strength. If, on the contrary, the two quantities can be shown to exhibit different transition points, then stabilizer measurements cannot serve as an optimal decoding protocol. 

Our work naturally invites further numerical studies in several directions. A straightforward extension of our numerical results on the isotropic LGT is to work directly in the limit of continuous imaginary time, where quantum Monte Carlo (QMC) via the WA \cite{huang_wormalgorithmtype_2020} can be employed to more quantitatively explore decoherence thresholds in the TFTC. Another possibility is to work with a generalized stochastic series expansion involving both Hamiltonian terms and local stabilizer measurement operators, as in Ref.~\cite{baweja_postmeasurement_2025}. In particular, it would be interesting to perform a more detailed finite-size scaling analysis of the decoherence-induced phase transitions at nonzero $h$, and thereby verify that the 2$d$ Nishimori RBIM universality class indeed persists for all $h < h_{c}$. The putative multicritical point at $h = h_{c}$ and $p = p^{\text{opt}}_{c}(h_{c})$ is of particular conceptual interest, where a disordered Nishimori 2$d$ transition and a clean 3$d$ transition meet at a single point. Given recent related works on higher Nishimori symmetry in classical Bayesian multicritical points~\cite{wiese_bayesian_2026,patil_higher_2026}, one might ask if the present multicritical point also exhibits a higher replica symmetry. 

The results of this work broadly open the door to further investigations of QEC in physically motivated non-stabilizer codes. While the TFTC and Fradkin-Shenker model are the most natural non-stabilizer topological codes from the perspective of classical statistical physics, recent work \cite{shackleton_twisted_2026} has demonstrated that a large class of twisted quantum double models, as well as their Hamiltonian deformations, can be efficiently simulated using QMC. It would be interesting to investigate the protection and recovery of quantum information in such models, and to understand the conditions under which the associated post-measurement states permit sign-free QMC calculations. Given the deep connections between QEC and topologically ordered phases of matter, such studies may provide an insightful new perspective on topological phases away from their well-understood solvable limits.

%% file: appendices/quantum_classical.tex
\section{Quantum-Classical Mapping}
\label{app:quantum_classical}
In this Appendix, we review the quantum-classical mapping for the transverse-field toric code $H_{h}$ [see Eq.~\eqref{eq:tftc}]. As discussed in Sec.~\ref{subsec:tftc}, the ground-state physics of $H_{h}$ is equivalent to that of a (2+1)$d$ $\mathbb{Z}_{2}$ LGT [Eq.~\eqref{eq:lgt}], and so the quantum-classical mapping of the latter model also describes the ground-state physics of the former model. We focus on the latter model for technical simplicity\footnote{At nonzero temperatures, the TFTC partition function $\tr e^{-\beta H_{h}}$ differs from that of the LGT in that it allows thermal excitations of the $e$ anyons. On the other hand, we are primarily interested in matrix elements of the form $\bra{\psi^{\bs}_{0}} e^{-\beta H_{h}} \ket*{\psi^{\bs'}_{0}}$, where such thermal excitations are strictly forbidden. Working directly with the LGT allows us to ignore these issues, which arise only from boundary conditions in the temporal direction.}.

The thermal partition function of the $\mathbb{Z}_{2}$ LGT is given by
\begin{equation}
    \mathcal{Z}_{\LGT} \equiv \tr[ \mathcal{P} e^{-\beta H_{\LGT}} ] ,
\end{equation}
where $H_{\LGT}$ is defined in Eq.~\eqref{eq:lgt}, and $\mathcal{P} \equiv \prod_{v} \qty( \frac{1 + A_{v}}{2} )$ projects onto the gauge-invariant sector of Hilbert space. In the limit $\beta \to \infty$, $e^{-\beta H_{\LGT}}$ acts as a projector onto the ground-state manifold of $H_{\LGT}$, and ground-state correlations (averaged over the degenerate ground-state manifold) can be computed using this thermal partition function.

As usual \cite{fradkin_order_1978,kogutIntroductionLatticeGauge1979,sachdevQuantumPhaseTransitions2011,fradkin_field_2013}, the quantum-classical mapping is derived by trotterizing $\beta \equiv \varepsilon M$ into $M$ imaginary time slices of size\footnote{As usual in the quantum-classical mapping \cite{sachdevQuantumPhaseTransitions2011}, one must formally take the limit $\varepsilon \to 0$ slower than $M^{-1} \to 0$ so that $\beta \to \infty$ diverges, i.e., so that our classical model describes the ground-state manifold of our quantum system.} $\varepsilon$, and inserting a resolution of the identity between each time slice. A minor novelty here is the inclusion of the projector $\mathcal{P}$. Since $\mathcal{P}$ commutes with $H_{\LGT}$, and $\mathcal{P}^{2} = \mathcal{P}$, we can insert as many factors of this projector along the thermal circle as we wish. It is convenient to insert this projector into each of the $M$ imaginary time slices, as follows:
\begin{equation}
    \mathcal{Z}_{\LGT} = \tr[(\mathcal{P} e^{-\varepsilon H_{\LGT}})^{M}] = \sum_{\qty{U}} \bra{\vb{U}_1} \mathcal{P} e^{-\varepsilon H_{\LGT}} \ket{\vb{U}_M} \ldots \bra{\vb{U}_2} \mathcal{P} e^{-\varepsilon H_{\LGT}} \ket{\vb{U}_1} ,
\end{equation}
where $\vb{U}_m \equiv \qty{U_{\ell, m}}$ denotes the total spin configuration in the $Z$ basis along the $m$th time slice, with $U_{\ell,m} = \pm 1$ on each link $\ell$ of the square lattice, and $\sum_{\qty{U}} \equiv \sum_{\vb{U}_{1}} \ldots \sum_{\vb{U}_{M}}$ is a sum over all spin configurations in every time slice. 

If $\varepsilon$ is taken very small, we can split up the $Z$ and $X$ terms in each matrix element:
\begin{equation}
    \begin{split}
        \bra{\vb{U}_m} \mathcal{P} e^{-\varepsilon H_{\LGT}} \ket{\vb{U}_{m-1}}  &\approx \bra{\vb{U}_m} e^{\varepsilon \sum_{p} B_{p}} \mathcal{P} e^{\varepsilon h \sum_{\ell} X_{\ell}} \ket{\vb{U}_{m-1}} \\
        &= e^{\varepsilon \sum_{p} \prod_{\ell \in p} U_{\ell, m}} \bra{\vb{U}_m} \mathcal{P} e^{\varepsilon h \sum_{\ell} X_{\ell}} \ket{\vb{U}_{m-1}} .
    \end{split}
\end{equation}
From the $Z$ terms in the Hamiltonian, we obtain a four-body interaction among spin variables $U_{\ell, m}$ on the four links surrounding a plaquette $p$, within each time slice $m$.

To evaluate the remaining matrix element, which contains entirely $X_{\ell}$ operators, it is useful to first note that $A_{v} \ket{\vb{U}_{m}}$ is a gauge transformation of the spin configuration $\vb{U}_m$ at vertex $v$: specifically, it flips $U_{\ell,m} \to -U_{\ell, m}$ for each link $\ell$ containing the vertex $v$. A more general gauge transformation of the form $U_{v v', m} \to s_{v} U_{v v',m} s_{v'}$, where $s_{v} = \pm 1$ on each vertex $v$ and $U_{v v', m} \equiv U_{\ell,m}$ for $\ell = \expval{v v'}$, is obtained by acting with $A_{v}$ on each vertex for which $s_{v} = -1$. The effect of the projector $\mathcal{P}$ acting to the left on $\bra{\vb{U}_m}$ is therefore to construct a sum over all gauge-equivalent states:
\begin{equation}
    \bra{\vb{U}_m} \mathcal{P} = \bra{\vb{U}_m} \prod_{v} \qty( \frac{1 + A_{v}}{2} ) = \frac{1}{2^{L^2}} \sum_{\vb{s}} \bra{\vb{s} \vb{U}_{m} \vb{s}} ,
\end{equation}
where $\vb{s} \equiv \qty{s_{v}}$, and $\vb{s} \vb{U}_{m} \vb{s} \equiv \qty{ s_{v} U_{v v',m} s_{v'} }$. The remaining matrix element then factorizes across links, and can be evaluated as follows:
\begin{equation}
    \begin{split}
        \bra{\vb{U}_m} \mathcal{P} e^{\varepsilon h \sum_{\ell} X_{\ell}} \ket{\vb{U}_{m-1}} &= \frac{1}{2^{L^2}} \sum_{\vb{s}} \prod_{\expval{v v'}} \bra{s_{v} U_{v v',m} s_{v'}} e^{\varepsilon h X_{v v'}} \ket{U_{v v',m-1}} \\
        &= \frac{1}{2^{L^2}} \sum_{\vb{s}} \prod_{\expval{v v'}} \qty[ \frac{\cosh(\varepsilon h)}{e^{J}} e^{J s_{v} U_{v v', m} s_{v'} U_{v v', m-1}} ] ,
    \end{split}
\end{equation}
with $J \equiv -\frac{1}{2} \log \tanh (\varepsilon h)$. In the first line, $X_{v v'} = X_{\ell}$ for $\ell = \expval{v v'}$. 

As a final step, let us regard the spins $s_{v}$ as ``timelike'' gauge field spins, $s_v \equiv U_{v, m - \frac{1}{2}}$, on links in the spacetime lattice connecting $(v,m)$ to $(v,m-1)$. We then find that the above matrix element again gives a four-body interaction among spins surrounding a plaquette, where the plaquette is now parallel to the imaginary time direction.

Altogether, we obtain the following partition function (up to trivial constants):
\begin{equation}
    \begin{split}
        \mathcal{Z}_{\LGT} &\propto \sum_{\qty{U}} \exp \qty{ \varepsilon \sum_{p, m} \prod_{\ell \in p} U_{\ell, m} + J \sum_{\expval{v v'}, m} U_{v, m - \frac{1}{2}} U_{v v', m} U_{v', m - \frac{1}{2}} U_{v v', m-1} } \\
        &= \sum_{\qty{U}} \exp \qty{ \varepsilon \sum_{[ijkl]_{\perp}} U_{ij} U_{jk} U_{kl} U_{li} + J \sum_{[ijkl]_{\parallel}} U_{ij} U_{jk} U_{kl} U_{li} } \\
        &\equiv \sum_{\qty{U}} e^{-S_{\LGT}[U]} ,
    \end{split}
\end{equation}
where in the second expression, $[ijkl]_{\perp}$ ($[ijkl]_{\parallel}$) denotes a spacetime plaquette with spacetime lattice sites $ijkl$, whose face lies perpendicular (parallel) to the imaginary time direction; see Figure~\ref{fig:background}(b). As expected, this is the extremely anisotropic limit of a classical 3$d$ LGT partition function \cite{wegnerDualityGeneralizedIsing1971,kogutIntroductionLatticeGauge1979}. Note in particular that no gauge condition has to be enforced on the classical model; here we can regard gauge redundancy as nothing more special than a local symmetry acting at a point $i = (v, m)$ in the spacetime lattice which flips the six link variables $U_{ij}$ surrounding it.

The preceding derivation was performed for the thermal partition function, i.e., with periodic boundary conditions in the imaginary time direction. It is simple to generalize the above mapping to evaluate matrix elements of the form $\bra{\psi^{\bs}_{0}} e^{-\beta H_{h}} \ket*{\psi^{\bs'}_{0}} \propto \bra{\psi^{\bs}_{0}} e^{-\beta H_{\LGT}} \ket*{\psi^{\bs'}_{0}}$, where $\ket{\psi^{\bs}_0}$ are the logical-$Z$ eigenstates of the stabilizer toric code [see Eq.~\eqref{eq:logical_Z}]. These are partition functions of LGTs with open boundary conditions at imaginary times $\tau = \pm \frac{\beta}{2}$, such that the timelike boundaries are ``flux free'' -- that is, there are no $m$ anyons present in the system at $\tau = \pm \frac{\beta}{2}$. Additionally, to map the state $\ket*{\psi^{\bs'}_{0}}$ to $\ket{\psi^{\bs}_{0}}$, we require the anyon worldlines generated by imaginary-time evolution to implement the appropriate logical-$X$ operator. This means that we require the anyon worldlines to be globally \emph{contractible} when $\bs = \bs'$, and \emph{non-contractible} when $\bs \neq \bs'$.

The required worldline homologies for each matrix element $\bra{\psi^{\bs}_{0}} e^{-\beta H_{h}} \ket*{\psi^{\bs'}_{0}}$ can be understood directly from the boundary conditions on $S_{\LGT}$ obtained via the quantum-classical mapping, but are particularly transparent if we instead express these matrix elements using time-dependent perturbation theory\footnote{Recall the derivation of this expression: we write $e^{-\tau H_{h}} = e^{- \tau H_{0}} e^{\tau H_{0}} e^{- \tau H_{h}}$, and notice that $e^{\tau H_{0}} e^{-\tau H_{h}}$ satisfies the differential equation $\partial_{\tau} (e^{\tau H_{0}} e^{- \tau H_{h}}) = e^{\tau H_{0}} (h \sum_{\ell} X_{\ell}) e^{-\tau H_{0}} (e^{\tau H_{0}} e^{-\tau H_{h}})$. Integrating both sides from $0$ to $\beta$, and repeatedly substituting the left-hand side into the right-hand side, gives the desired result.}:
\begin{equation}
    \bra{\psi^{\bs}_{0}} e^{-\beta H_{h}} \ket*{\psi^{\bs'}_{0}} = \sum_{n = 0}^{\infty} h^{n} \sum_{\ell_{1} \ldots \ell_{n}} \int_{0 \leq \tau_{1} < \ldots < \tau_{n} \leq \beta} \hspace{-1.75cm} \dd{\tau_{1}} \ldots \dd{\tau_{n}} \bra{\psi^{\bs}_{0}} e^{-(\beta - \tau_{n}) H_{0}} X_{\ell_{n}} e^{- (\tau_{n} - \tau_{n-1}) H_{0}} \ldots e^{- (\tau_{2} - \tau_{1}) H_{0}} X_{\ell_{1}} e^{-\tau_{1} H_{0}} \ket*{\psi^{\bs'}_{0}} . 
\end{equation}
In the above perturbative expansion, each fixed order $n$ in perturbation theory, and spacetime location $\qty{(\tau_{\alpha}, \ell_{\alpha})}$ of $m$ anyon hopping events, corresponds precisely to an anyon worldline trajectory from the quantum-classical mapping. We see from this expression that each worldline trajectory applies a Pauli-$X$ string $X_{\ell_{n}} \ldots X_{\ell_{1}}$ to the original state $\ket*{\psi^{\bs'}_{0}}$, while the intermediate time evolution $e^{-(\tau_{\alpha} - \tau_{\alpha - 1}) H_{0}}$ simply contributes to the classical action of the trajectory. Each worldline trajectory can therefore be regarded as carrying the factor $\bra{\psi^{\bs}_{0}} X_{\ell_{n}} \ldots X_{\ell_{1}} \ket*{\psi^{\bs'}_{0}}$, which is nonzero only when $X_{\ell_{n}} \ldots X_{\ell_{1}}$ gives the correct logical-$X$ operator to map $\ket*{\psi^{\bs'}_{0}}$ to $\ket{\psi^{\bs}_{0}}$.

%% file: appendices/ite_overlaps.tex
\section{ITE states are approximately true code states}
\label{app:ite_truecodestates}
In this Appendix, we prove that stabilizer code states evolved by ITE lie approximately within the TFTC code space $\mathcal{C}_{h}$, up to exponentially small corrections. As described in more detail below, the proof relies on two physically reasonable assumptions: first, we assume that the gap $\Delta$ from the ground-state manifold $\mathcal{C}_{h}$ to the excited states remains finite as $L \to \infty$; and second, we assume that the projection of a stabilizer code state $\ket{\psi_{0}} \in \mathcal{C}_{0}$ onto $\mathcal{C}_{h}$ scales with $L$ no worse than $\mathcal{O}(e^{-\text{poly}(L)})$. The first assumption is well-known to hold for all $h < h_{c}$, while the second assumption can be reasonably justified throughout the $h < h_{c}$ phase using standard physical arguments~\cite{gu_fidelity_2010}.

We first fix some notation. Let $\ket{\chi^{\bh}_{h}}$ denote the \emph{exact} four lowest-energy eigenstates of $H_{h}$, for $h < h_{c}$. These states are eigenstates of the logical-$X$ operators $\overline{X}_{\mu}$, with $\bh \equiv (\eta_{x}, \eta_{y})$ and $\eta_{\mu} = \pm $ denoting the corresponding eigenvalues: $\overline{X}_{\mu} \ket{\chi^{\bh}_{h}} = \eta_{\mu} \ket{\chi^{\bh}_{h}}$. We denote their eigen-energies by $E^{\bh}_{\gs}$, such that $H_{h} \ket{\chi^{\bh}_{h}} = E^{\bh}_{\gs} \ket{\chi^{\bh}_{h}}$. From the Perron-Frobenius theorem\footnote{Explicitly, since $H_{h}$ is stoquastic in the computational basis, the Perron-Frobenius theorem implies that the exact ground state must be sign-free in the computational basis. Since the logical-$X$ operators $\overline{X}_{\mu}$ act as permutations on the computational basis states, a sign-free eigenstate of $\overline{X}_{\mu}$ must have eigenvalue $+1$.}~\cite{meyer_matrix_2023}, the absolute ground state is $\ket*{\chi^{(+,+)}_{h}}$ with energy $E^{(+,+)}_{\gs}$, which we will (without loss of generality) set to zero by redefining $H_{h} \to H_{h} - E^{(+,+)}_{\gs}$. Then, assuming that the gap $\Delta$ remains open throughout the topological phase, Ref.~\cite{hastings_quasiadiabatic_2005} proved that (for sufficiently large $L$) the energetic splitting $\delta$ between the code states is exponentially small in $L$:
\begin{equation}
    \delta \equiv \max_{\bh} E^{\bh}_{\gs} \leq \mathcal{O}(e^{-L / \xi}) ,
\end{equation}
where $\xi$ is a finite constant which diverges as $\Delta \to 0$. 

For each stabilizer code state $\ket{\psi_{0}} \in \mathcal{C}_{0}$, we define a corresponding imaginary time-evolved state $\ket{\varphi_{h}}$ via Eq.~\eqref{eq:imag_time_evo}, repeated here for convenience:
\begin{equation}
    \label{eq:imag_time_evo_app}
    \ket{\varphi_{h}} \equiv \frac{e^{-\beta H_{h} / 2} \ket{\psi_{0}}}{\sqrt{\bra{\psi_{0}} e^{-\beta H_{h}} \ket{\psi_{0}}}} .
\end{equation}
Our goal in this section is to show that $\ket{\varphi_{h}}$ approximately lies within $\mathcal{C}_{h}$, i.e.,
\begin{equation}
    \label{eq:proj_into_Ch}
    \sum_{\bh} \abs{\bra{\chi^{\bh}_{h}} \ket{\varphi_{h}}}^{2} \geq 1 - \mathcal{O}(e^{-L / \xi}) .
\end{equation}
To this end, let us denote the remaining excited energy eigenstates of $H_{h}$ for a fixed value of $h$ by $\qty{\ket{n}}_{n = 1}^{\infty}$. The corresponding eigen-energies are denoted $E_{n}$, such that $\Delta = E_{1} \leq E_{2} \leq \ldots$. We can then write the left side of Eq.~\eqref{eq:proj_into_Ch} as
\begin{equation}
    \begin{split}
        \sum_{\bh} \abs{\bra{\chi^{\bh}_{h}} \ket{\varphi_{h}}}^{2} &= \frac{\sum_{\bh} \abs{\bra{\chi^{\bh}_{h}} e^{-\beta H_{h} / 2} \ket{\psi_{0}}}^{2}}{\bra{\psi_{0}} e^{-\beta H_{h}} \ket{\psi_{0}}} \\
        &= \frac{\sum_{\bh} e^{-\beta E^{\bh}_{\gs}} \abs{\bra{\chi^{\bh}_{h}} \ket{\psi_{0}}}^{2}}{\sum_{\bh} e^{-\beta E^{\bh}_{\gs}} \abs{\bra{\chi^{\bh}_{h}} \ket{\psi_{0}}}^{2} + \sum_{n \geq 1} e^{-\beta E_{n}} \abs{\bra{n} \ket{\psi_{0}}}^{2}} \\
        &\geq \frac{e^{-\beta \delta} \Omega}{\Omega + e^{-\beta \Delta} (1 - \Omega)} \\
        &= \frac{e^{-\beta \delta}}{1 + e^{-\beta \Delta}(\Omega^{-1} - 1)},
    \end{split}
\end{equation}
where we have defined $\Omega$ as the norm of the projection of $\ket{\psi_{0}}$ into $\mathcal{C}_{h}$:
\begin{equation}
    \Omega \equiv \sum_{\bh} \abs{\bra{\chi^{\bh}_{h}} \ket{\psi_{0}}}^{2} .
\end{equation}
So long as $\Omega$ is not too small, we can choose $\beta$ such that $e^{-\beta \Delta} \Omega^{-1}$ is vanishingly small while $e^{-\beta \delta}$ remains close to unity. Specifically, since $\delta \leq \mathcal{O}(e^{- L / \xi})$, we have $e^{-\beta \delta} \geq 1 - \mathcal{O}(e^{-L/\xi})$ so long as $\beta \sim \mathcal{O}(\text{poly}(L))$. On physical grounds, we expect that $\Omega$ vanishes with $L$ no faster than
\begin{equation}
    \Omega \geq \mathcal{O}(e^{- \text{poly}(L)}) ,
\end{equation}
and in particular that $\Omega \sim e^{- \alpha_{h} L^{2}}$ throughout the topological phase $h < h_{c}$ \cite{gu_fidelity_2010}. With this assumption, Eq.~\eqref{eq:proj_into_Ch} is obtained by choosing $\beta$ to satisfy
\begin{equation}
    \Delta^{-1} \log \Omega^{-1} \ll \beta \ll \delta^{-1} .
\end{equation}

%% file: appendices/decoding_fidelities.tex
\section{Decoding Fidelities of Nearby States}
\label{app:decoding_fidelities}

In the main text, we have focused on decoding algorithms for the ITE states $\ket{\varphi_{h}}$ defined in Eq.~\eqref{eq:imag_time_evo}. Since these states form a four-dimensional vector space and lie within the true code space $\mathcal{C}_{h}$ up to exponentially small corrections (see Appendix~\ref{app:ite_truecodestates}), they must approximately span $\mathcal{C}_{h}$; that is, for any $\ket{\psi_{h}} \in \mathcal{C}_{h}$, there exists an ITE state $\ket{\varphi_{h}}$ such that $\abs{\bra{\varphi_{h}} \ket{\psi_{h}}} \geq 1 - \mathcal{O}(e^{-L / \xi})$. In particular, if $\ket{\psi_{h}} = \mathcal{U}_{h} \ket{\psi_{0}}$, then $\ket{\varphi_{h}} \propto e^{-\beta H_{h} / 2} \ket{\psi_{0}}$ is an excellent approximation to $\ket{\psi_{h}}$ (see Appendix~\ref{app:imag_time_adiabatic}.)

In this Appendix, we will show that any QEC protocol succeeds with high fidelity on the true code space if and only if it succeeds on the ITE states. This result is quite natural, since the ITE states and true code states are essentially indistinguishable from each other. If the QEC protocol were to succeed on one class of states and not the other, it could be used as a channel which distinguishes between the two states.



We will handle the cases of Sec.~\ref{sec:meas_decoding} (without Pauli decoherence) and Sec.~\ref{sec:pauli_meas_decoding} (with Pauli decoherence) simultaneously; indeed, the former case is a special case of the latter, with decoherence strength $p = 0$ [see Eq.~\eqref{eq:decoherence}]. First recall that the average decoding fidelity $\overline{\mathcal{F}}(\ket{\psi_{h}})$ for a given code state $\ket{\psi_{h}} \in \mathcal{C}_{h}$ is given by [see Eq.~\eqref{eq:avg_decoding_fidelity}]
\begin{equation}
    \begin{split}
        \overline{\mathcal{F}} (\ket{\psi_{h}}) &\equiv \sum_{\vs} P_{\vs} \bra{\psi_{h}} V_{\vs} \rho_{h}(\vs) V_{\vs}^{\dag} \ket{\psi_{h}} \\
        &= \sum_{\vs} \bra{\psi_{h}} V_{\vs} \hat{\Pi}_{\vs} \rho_{h} \hat{\Pi}_{\vs} V_{\vs}^{\dag} \ket{\psi_{h}} \\
        &= \bra{\psi_{h}} \mathcal{M}(\rho_{h}) \ket{\psi_{h}} \\
        &= \mathcal{F} \Big( \dyad{\psi_{h}}, \mathcal{M}(\rho_{h}) \Big) ,
    \end{split}
\end{equation}
where $\rho_{h} \equiv \mathcal{E}(\dyad{\psi_{h}})$ and $\rho_h(\vs) \equiv \hat{\Pi}_{\vs} \rho_h \hat{\Pi}_{\vs} / P_{\vs}$, in the third line we have introduced the \emph{decoding channel},
\begin{equation}
    \label{eq:decoding_channel}
    \mathcal{M}(\rho) \equiv \sum_{\vs} V_{\vs} \hat{\Pi}_{\vs} \rho \hat{\Pi}_{\vs} V_{\vs}^{\dag} ,
\end{equation}
and in the final line we have defined the Uhlmann fidelity $\mathcal{F}(\rho, \omega)$ between two density matrices $\rho$ and $\omega$ \cite{nielsenQuantumComputationQuantum2010},
\begin{equation}
    \label{eq:uhlmann}
    \mathcal{F}(\rho, \omega) = \qty( \tr \sqrt{\sqrt{\rho} \omega \sqrt{\rho}} )^{2}. 
\end{equation}
The Uhlmann fidelity satisfies two important properties for our purposes. First, it satisfies a \emph{quantum data processing inequality} \cite{schumacherQuantumDataProcessing1996,nielsenQuantumComputationQuantum2010}: for any quantum channel $\mathcal{N}$, we have the inequality
\begin{equation}
\label{eq:quantum_data_processing}
    \mathcal{F}(\mathcal{N}(\rho), \mathcal{N}(\omega)) \geq \mathcal{F}(\rho, \omega) ,
\end{equation}
i.e., two states $\rho$ and $\omega$ can only become less distinguishable under the action of a channel. Second, the quantity $A(\rho, \omega) \equiv \cos^{-1}[ \sqrt{\mathcal{F}(\rho, \omega)} ]$ satisfies the triangle inequality $A(\rho, \omega) \leq A(\rho, \eta) + A(\eta, \omega)$ for any three states $\rho,\omega,\eta$~\cite{nielsenQuantumComputationQuantum2010}. Note that since $\cos^{-1}(x)$ is monotonically decreasing for $x \in [0, 1]$, we also have the quantum data processing inequality $A(\mathcal{N}(\rho), \mathcal{N}(\omega)) \leq A(\rho, \omega)$.

With these properties in mind, we can derive a lower bound on the average decoding fidelity for true code states $\ket{\psi_{h}} \in \mathcal{C}_{h}$ using the average decoding fidelity of an ITE state $\ket{\varphi_{h}}$. Letting $\mathcal{N} \equiv \mathcal{M} \circ \mathcal{E}$ denote the composition of the decoherence channel $\mathcal{E}$ with the decoding channel $\mathcal{M}$, two applications of the triangle inequality give
\begin{align}
    A \Big( \dyad{\psi_{h}} , \mathcal{N}(\dyad{\psi_{h}}) \Big) &\leq A \Big( \dyad{\psi_{h}}, \dyad{\varphi_{h}} \Big) + A \Big( \dyad{\varphi_{h}}, \mathcal{N}(\dyad{\varphi_{h}}) \Big) + A \Big( \mathcal{N}(\dyad{\varphi_{h}}), \mathcal{N}(\dyad{\psi_{h}}) \Big) \notag \\
    &\leq 2 A \Big( \dyad{\psi_{h}}, \dyad{\varphi_{h}} \Big) + A \Big( \dyad{\varphi_{h}}, \mathcal{N}(\dyad{\varphi_{h}}) \Big) \notag \\
    &= 2 \cos^{-1} (\abs{\bra{\varphi_{h}} \ket{\psi_{h}}}) + A \Big( \dyad{\varphi_{h}}, \mathcal{N}(\dyad{\varphi_{h}}) \Big) ,
\end{align}
where we have used the quantum data processing inequality to obtain the second expression. By computing the cosine of both sides, straightforward algebra\footnote{Explicitly, denote $x \equiv \overline{\mathcal{F}}(\ket{\psi_{h}})$, $y \equiv \overline{\mathcal{F}}(\ket{\varphi_{h}}) $, $\theta \equiv \cos^{-1} \sqrt{y}$ and $\varepsilon \equiv 2 \cos^{-1}(\abs{\bra{\varphi_{h}} \ket{\psi_{h}}})$. Initially, we have $\cos^{-1} \sqrt{x} \leq \varepsilon + \cos^{-1} \sqrt{y}$, i.e., $\sqrt{x} \geq \cos( \varepsilon + \theta)$. Subtracting $\sqrt{y} = \cos \theta$ from both sides, we obtain $\sqrt{x} - \sqrt{y} \geq \cos(\theta + \varepsilon) - \cos \theta = -2 \sin(\theta + \frac{\varepsilon}{2}) \sin \frac{\varepsilon}{2} \geq - 2\sin \frac{\varepsilon}{2}$. Recalling that $\sin (\cos^{-1} z) = \sqrt{1 - z^{2}}$, we obtain $\sqrt{x} \geq \sqrt{y} - 2 \sqrt{ 1 - \abs{\bra{\varphi_{h}} \ket{\psi_{h}}}^{2} }$. Finally squaring both sides and using $y \leq 1$ leads to the desired result.} derives the inequality
\begin{equation}
    \overline{\mathcal{F}}(\ket{\psi_{h}}) \geq \overline{\mathcal{F}}(\ket{\varphi_{h}}) - 4 \sqrt{ 1 - \abs{\bra{\varphi_{h}} \ket{\psi_{h}}}^{2} } .
\end{equation}
We can similarly perform the same calculation with $\ket{\psi_{h}}$ and $\ket{\varphi_{h}}$ exchanged, leading to a similar inequality. Combining these, we find the two-sided bound
\begin{equation}
    \abs{ \overline{\mathcal{F}}(\ket{\psi_{h}}) - \overline{\mathcal{F}}(\ket{\varphi_{h}}) } \leq 4 \sqrt{1 - \abs{ \bra{\varphi_{h}} \ket{\psi_{h}} }^{2}} .
\end{equation}
As discussed in Sec.~\ref{subsec:ite}, we can approximate any true code state $\ket{\psi_{h}} \in \mathcal{C}_{h}$ by an ITE state $\ket{\varphi_{h}}$ with fidelity exponentially close to unity. Therefore, we can always choose $\ket{\varphi_{h}}$ such that the right-hand side is $\mathcal{O}(e^{-L / \xi})$. The true code states $\ket{\psi_{h}}$ therefore exhibit high decoding fidelities if and only if the corresponding ITE states $\ket{\varphi_{h}}$ exhibit high decoding fidelities, as expected.


%% file: appendices/imag_adiabatic.tex
\section{Equivalence of Imaginary Time Evolution and Adiabatic Unitary Evolution}
\label{app:imag_time_adiabatic}
Given a stabilizer state $\ket{\psi_{0}} \in \mathcal{C}_{0}$, there are two natural methods by which we can attempt to construct a TFTC state $\ket{\psi_{h}} \in \mathcal{C}_{h}$. The most natural approach is via adiabatic unitary evolution: starting from $\ket{\psi_{0}}$, we evolve by $H_{h(t)}$, where $h(t)$ continuously interpolates between $h(0) = 0$ and $h(T) = h$, for some very long time $T$:
\begin{equation}
    \label{eq:adiabatic_app}
    \ket{\psi_{h}} \equiv \mathcal{U}_{h} \ket{\psi_{0}} = \mathcal{T} \exp \qty{ -i \int_{0}^{T} \dd{t} H_{h(t)} } \ket{\psi_{0}} ,
\end{equation}
where $\mathcal{T}$ is the time-ordering symbol. The adiabatic theorem~\cite{albash_adiabatic_2018} then guarantees that $\ket{\psi_{h}}$ lies approximately within $\mathcal{C}_{h}$, so long as $T$ is taken to be very large compared to the squared inverse gap $\Delta^{-2}$. On the other hand, taking $T$ too large will incur undesirable dynamical phases $e^{-i \int_{0}^{T} \dd{t} E^{\bh}_{\gs}(t)}$ between the near-degenerate ground states. While precise error estimates depend on the choice of adiabatic ramp~\cite{albash_adiabatic_2018}, we will consider the limit $\Delta^{-2} \text{poly}(L) \ll T \ll \exp(L)$ where both types of errors are well-controlled.

While conceptually appealing, the preceding approach has the drawback that the states $\ket{\psi_{h}}$ are difficult to describe analytically. The second approach, employed frequently throughout the main text, is to \emph{project} into $\mathcal{C}_{h}$ via imaginary time evolution:
\begin{equation}
    \ket{\varphi_{h}} \equiv \frac{e^{-\beta H_{h} / 2} \ket{\psi_{0}}}{\sqrt{\bra{\psi_{0}} e^{-\beta H_{h}} \ket{\psi_{0}}}} ,
\end{equation}
where $\beta \sim \text{poly}(L)$ is taken sufficiently large that any amplitude outside of $\mathcal{C}_{h}$ is exponentially suppressed, but not so large that $e^{-\beta H_{h}}$ can resolve the exponentially small energetic splitting between the four ground states. As discussed in Sec.~\ref{subsec:ite}, standard statistical physics arguments strongly suggest that the above nonunitary evolution effectively maps $\mathcal{C}_{0}$ to $\mathcal{C}_{h}$ unitarily, up to small corrections, so long as $h < h_{c}$ is within the topological phase.

In this Appendix, we will argue that these two approaches are essentially equivalent throughout the topological phase; that is,
\begin{equation}
    \ket{\psi_{h}} = e^{-i \phi(h, T)} \ket{\varphi_{h}} + \mathcal{O}(e^{-L / \xi}) ,
\end{equation}
where $\phi(h, T)$ is a trivial dynamical phase. Our only assumptions are Eq.~\eqref{eq:Z_defn_cost} (i.e., that non-contractible $m$ loops are exponentially suppressed for $h < h_{c}$) and the mild assumptions of Appendix~\ref{app:ite_truecodestates}. We will not treat errors in the adiabatic evolution fully rigorously here; instead, we will assume $\ket{\psi_{h}}$ stays within $\mathcal{C}_{h}$ exactly, and that we can ignore the exponentially small differences in dynamical phases within $\mathcal{C}_{h}$ altogether. The only effect of adiabatic evolution is then to rotate within the instantaneous four-dimensional code space $\mathcal{C}_{h(t)}$. In particular, if $\ket{\psi^{a}_{h}}$ denotes an orthonormal basis of $\mathcal{C}_{h}$ for each $h$, then we will define $\ket{\psi_{h}}$ using the non-abelian Berry connection~\cite{wilczek_appearance_1984}:
\begin{equation}
    \label{eq:nonabelian_berry}
    \ket{\psi_{h}} = e^{-i \phi(h, T)} \sum_{a} b_{a}(h) \ket{\psi^{a}_{h}}, \quad b_{a}(h) = \sum_{a'} \qty[ \mathcal{P} \exp \qty{ i \int_{0}^{h} \dd{h}' A(h') } ]_{aa'} b_{a'}(0) ,
\end{equation}
where $\mathcal{P}$ stands for path-ordering, and $A(h)$ is the matrix-valued non-abelian Berry connection
\begin{equation}
    [A(h)]_{a a'} \equiv \bra{\psi^{a}_{h}} i \pdv{h} \ket*{\psi^{a'}_{h}} .
\end{equation}

Our goal is therefore to compute $A(h)$ with a basis constructed from imaginary-time evolution. To this end, recall that our ITE basis states $\ket{\varphi^{\bs}_{h}}$ in Eq.~\eqref{eq:logicalZ_ITE} do not perfectly lie within $\mathcal{C}_{h}$, and are not perfectly orthonormal. It will be briefly convenient to imagine promoting these states to an \emph{exact} orthonormal basis of $\mathcal{C}_{h}$, denoted $\ket{\widetilde{\varphi}^{\bs}_{h}}$, by first projecting into $\mathcal{C}_{h}$ and then orthonormalizing via the Gram-Schmidt procedure~\cite{hoffmanLinearAlgebra1971}. Both of these procedures modify the states by exponentially small corrections, i.e., $\bra{\varphi^{\bs}_{h}} \ket{\widetilde{\varphi}^{\bs}_{h}} \geq 1 - \mathcal{O}(e^{-L / \xi})$. We imagine performing this procedure for all $h$ along the path $h(t)$, so that we obtain a smoothly varying orthonormal basis of each instantaneous $\mathcal{C}_{h(t)}$. The Berry connection is then given in this basis by
\begin{equation}
    [A(h)]_{\bs \bs'} = \bra{\widetilde{\varphi}^{\bs}_{h}} i \pdv{h} \ket*{\widetilde{\varphi}^{\bs'}_{h}} = \bra{\varphi^{\bs}_{h}} i \pdv{h} \ket*{\varphi^{\bs'}_{h}} + \mathcal{O}(e^{-L / \xi}) .
\end{equation}
To compute the first term, we note the following identity\footnote{The general identity, for a set of matrices $M(x)$ which depends continuously on a parameter $x$, is
\begin{equation}
    \pdv{x} e^{M(x)} = \int_{0}^{1} \dd{s} e^{s M(x)} \pdv{M}{x} e^{(1 - s) M(x)} ,
\end{equation}
which can be derived by writing $e^{M(x)} = \lim_{n \to \infty} [1 + \frac{1}{n} M(x)]^n$ and differentiating using the product rule.
}:
\begin{equation}
    \pdv{h} e^{-\beta H_{h} / 2} = \int_{0}^{\beta / 2} \dd{\tau} e^{-\tau H_{h}} \qty( \sum_{\ell} X_{\ell} ) e^{\tau H_{h}} e^{-\beta H_{h}/ 2} =  \int_{0}^{\beta / 2} \dd{\tau} \sum_{\ell} X_{\ell}(-\tau) e^{-\beta H_{h} / 2} ,
\end{equation}
where $X_{\ell}(-\tau) \equiv e^{-\tau H_{h}} X_{\ell} e^{\tau H_{h}}$. Using this identity, and the definition~\eqref{eq:logicalZ_ITE} of the ITE states $\ket{\varphi^{\bs}_{h}}$, we obtain
\begin{equation}
    \begin{split}
        \bra{\varphi^{\bs}_{h}} \pdv{h} \ket*{\varphi^{\bs'}_{h}} &= \int_{0}^{\beta / 2} \dd{\tau} \sum_{\ell} \bra{\varphi^{\bs}_{h}} X_{\ell}(-\tau) \ket*{\varphi^{\bs'}_{h}} - \frac{1}{2} \int_{-\beta / 2}^{\beta / 2} \dd{\tau} \sum_{\ell} \bra*{\varphi^{\bs'}_{h}} X_{\ell}(-\tau) \ket*{\varphi^{\bs'}_{h}} \bra{\varphi^{\bs}_{h}} \ket*{\varphi^{\bs'}_{h}} \\
        &= \int_{0}^{\beta / 2} \dd{\tau} \sum_{\ell} \qty{ \bra{\varphi^{\bs}_{h}} X_{\ell}(-\tau) \ket*{\varphi^{\bs'}_{h}} -  \bra*{\varphi^{\bs'}_{h}} X_{\ell}(-\tau) \ket*{\varphi^{\bs'}_{h}} \bra{\varphi^{\bs}_{h}} \ket*{\varphi^{\bs'}_{h}} } ,
    \end{split}
\end{equation}
where the second term in both lines comes from differentiating the denominator $\mathcal{Z}_h$. For $\bs = \bs'$, the two terms are equal and the entire expression vanishes. For $\bs \neq \bs'$, the second term is $\mathcal{O}(e^{-L / \xi})$ by Eq.~\eqref{eq:ortho_deformed_basis}, while the first term in brackets can be expressed as
\begin{equation}
    \bra{\varphi^{\bs}_{h}} X_{\ell}(-\tau) \ket*{\varphi^{\bs'}_{h}} = \frac{1}{\mathcal{Z}_{h}} \bra{\psi^{\bs}_{0}} e^{- (\tau + \beta / 2) H_{h}} X_{\ell} e^{(\tau - \beta / 2) H_{h}} \ket*{\psi^{\bs'}_{0}} .
\end{equation}
We can interpret the numerator as an LGT partition function in which, at imaginary time $-\tau$, two $m$ anyons are nucleated on the plaquettes adjacent to $\ell$. Due to the initial and final boundary conditions, this matrix element requires $m$ anyon worldlines to nontrivially wrap the torus, similar to the overlap $\bra{\varphi^{\bs}_{h}} \ket*{\varphi^{\bs'}_{h}} \propto \bra{\psi^{\bs}_{0}} e^{-\beta H_{h}} \ket*{\psi^{\bs'}_{0}}$ without the insertion of $X_{\ell}$ at time $-\tau$. We once again expect this matrix element to be exponentially suppressed in $L$ for all $h < h_{c}$. We therefore find that all matrix elements of $A(h)$ are exponentially small in $L$, and so $b_{\bs}(h) = b_{\bs}(0) + \mathcal{O}(e^{-L / \xi})$. Altogether, letting $\ket{\psi_{0}} = \sum_{\bs} b_{\bs}(0) \ket{\psi^{\bs}_{0}}$, we find
\begin{equation}
    \ket{\psi_{h}} = e^{-i \phi(h, T)} \sum_{\bs} b_{\bs}(0) \ket{\varphi^{\bs}_{h}} + \mathcal{O}(e^{- L / \xi}) = e^{-i \phi(h, T)} \frac{e^{-\beta H_{h} / 2} \ket{\psi_{0}}}{\sqrt{\bra{\psi_{0}} e^{-\beta H_{h}} \ket{\psi_{0}}}} + \mathcal{O}(e^{-L / \xi}) ,
\end{equation}
as desired.

%% file: appendices/pruning.tex
\section{Pruning}
\label{app:pruning}
In this Appendix, we discuss an alternative method to that of Sec.~\ref{subsec:ite} for promoting ITE into an exactly unitary transformation. For reasons to become clear, we refer to this method as ``pruning''. 

First recall that the ITE basis states $\ket{\varphi^{\bs}_{h}}$ are defined via ITE from the stabilizer logical-$Z$ eigenstates:
\begin{equation}
    \ket{\varphi^{\bs}_{h}} \equiv \frac{1}{\sqrt{\mathcal{Z}_{h}}} e^{-\beta H_{h} / 2} \ket{\psi^{\bs}_{0}}, \quad \mathcal{Z}_{h} \equiv \bra{\psi^{\bs}_{0}} e^{-\beta H_{h}} \ket{\psi^{\bs}_{0}} .
\end{equation}
As described in Sec.~\ref{subsec:ite}, so long as $h < h_{c}$, these states are orthonormal up to exponentially small corrections in $L$ [see Eq.~\eqref{eq:ortho_deformed_basis}]. Furthermore, because $H_{h}$ is stoquastic, each of these states is strictly non-negative in the computational basis (see Sec.~\ref{subsec:stoquasticity}). It follows from these two observations together that the four states $\ket{\varphi^{\bs}_{h}}$ must exhibit essentially non-overlapping support in the computational basis. By discarding the exponentially small portions of these states where they do share overlapping support, we will arrive at a basis of exactly orthogonal and strictly sign-free code states. We will refer to these code states as ``pruned'' code states.

Explicitly, divide the full set of computational basis states $\vb{z} \equiv \qty{z_{\ell}}$ into four disjoint subsets $\mathcal{S}^{\bs}_{h}$, based on their relative contributions to the four states $\ket{\varphi^{\bs}_{h}}$:
\begin{equation}
    \label{eq:pruned_subsets}
    \mathcal{S}^{\bs}_{h} \equiv \qty{ \vb{z} : \operatorname*{argmax}_{\bs'} \bra{\vb{z}} \ket*{\varphi^{\bs'}_{h}} = \bs  } ,
\end{equation}
i.e., $\vb{z} \in \mathcal{S}^{\bs}_{h}$ if and only if $\bra{\vb{z}} \ket{\varphi^{\bs}_{h}} > \bra{\vb{z}} \ket*{\varphi^{\bs'}_{h}}$ for all $\bs' \neq \bs$. In the rare cases where $\bra{\vb{z}} \ket{\varphi^{\bs}_{h}} = \bra{\vb{z}} \ket*{\varphi^{\bs'}_{h}}$ for $\bs \neq \bs'$, $\vb{z}$ does not belong to any of the four partitions and is ``pruned away''. We next define the unnormalized states $\ket*{\widetilde{\mathcal{P}}^{\bs}_{h}}$ via the wavefunction elements
\begin{equation}
    \bra{\vb{z}} \ket*{\widetilde{\mathcal{P}}^{\bs}_{h}} = \begin{cases}
        \bra{\vb{z}} \ket{\varphi^{\bs}_{h}}, & \vb{z} \in \mathcal{S}^{\bs}_{h} , \\
        0, & \vb{z} \not\in \mathcal{S}^{\bs}_{h} ,
    \end{cases}
\end{equation}
i.e., $\ket*{\widetilde{\mathcal{P}}^{\bs}_{h}}$ is the projection of $\ket{\varphi^{\bs}_{h}}$ onto the vector subspace spanned by elements of $\mathcal{S}^{\bs}_{h}$. Clearly, the four states $\ket*{\widetilde{\mathcal{P}}^{\bs}_{h}}$ are mutually orthogonal and non-negative in the computational basis. Finally, we simply normalize these states to obtain our orthonormal basis of ``pruned'' code states $\ket*{\mathcal{P}^{\bs}_{h}}$:
\begin{equation}
   \ket*{\mathcal{P}^{\bs}_{h}} \equiv \frac{\ket*{\widetilde{\mathcal{P}}^{\bs}_{h}}}{\norm{\ket*{\widetilde{\mathcal{P}}^{\bs}_{h}}}} .
\end{equation}

We now show that, throughout the topological phase $h < h_{c}$, the pruned states $\ket*{\mathcal{P}^{\bs}_{h}}$ have high fidelity with the original ITE basis states $\ket{\varphi^{\bs}_{h}}$. First note the overlap between $\ket*{\mathcal{P}^{\bs}_{h}}$ and $\ket{\varphi^{\bs}_{h}}$ is given by $\norm*{\ket*{\widetilde{\mathcal{P}}^{\bs}_{h}}}$, and so our goal is to show that the states $\ket*{\widetilde{\mathcal{P}}^{\bs}_{h}}$ have norm close to unity. To this end, we can lower-bound the norm of $\ket*{\widetilde{\mathcal{P}}^{\bs}_{h}}$ using Eq.~\eqref{eq:ortho_deformed_basis} as follows:
\begin{equation}
    \begin{split}
        1 - \bra*{\widetilde{\mathcal{P}}^{\bs}_{h}}\ket*{\widetilde{\mathcal{P}}^{\bs}_{h}} &= \sum_{\vb{z} \not\in \mathcal{S}^{\bs}_{h}} \abs{\bra{\vb{z}} \ket{\varphi^{\bs}_{h}}}^{2} \\
        &\leq \sum_{\vb{z} \not\in \mathcal{S}^{\bs}_{h}} \sum_{\bs' : \bs' \neq \bs} \bra{\varphi^{\bs}_{h}} \ket{\vb{z}} \bra{\vb{z}} \ket*{\varphi^{\bs'}_{h}} \\
        &\leq \sum_{\bs' : \bs' \neq \bs} \bra{\varphi^{\bs}_{h}} \ket*{\varphi^{\bs'}_{h}} \\
        &\leq \mathcal{O}(e^{-L/\xi}) .
    \end{split}
\end{equation}
In the first line, we have used the fact that $\ket{\varphi^{\bs}_{h}}$ is normalized; in the second line, we have noted that for all $\vb{z} \not\in \mathcal{S}^{\bs}_{h}$, there exists a $\bs'$ such that $\bra{\vb{z}} \ket{\varphi^{\bs}_{h}} \leq \bra{\vb{z}} \ket*{\varphi^{\bs'}_{h}}$; in the third line, we have extended the first sum to run over all $\vb{z}$; and in the final line, we have used Eq.~\eqref{eq:ortho_deformed_basis}. This proves that the pruned code states have high overlap with the original ITE code states,
\begin{equation}
    \bra{\varphi^{\bs}_{h}} \ket{\mathcal{P}^{\bs}_{h}} \geq 1 - \mathcal{O}(e^{-L/\xi}), 
\end{equation}
which in turn implies that the pruned states $\ket{\mathcal{P}^{\bs}_{h}}$ lie within $\mathcal{C}_{h}$ up to exponentially small corrections (i.e., their projection into $\mathcal{C}_{h}$ has norm exponentially close to unity). Note the central role played by stoquasticity in every component of this construction, which crucially relies on the fact that $\bra{\vb{z}} \ket{\varphi^{\bs}_{h}}$ is non-negative.

An interesting fact about the pruned code states is that they remain \emph{exactly} orthonormal upon measuring the stabilizers $B_{p}$. Specifically, since the four states $\ket{\mathcal{P}^{\bs}_{h}}$ exhibit completely disjoint support in the computational basis, they satisfy
\begin{equation}
    \bra{\mathcal{P}^{\bs}_{h}} \hat{\Pi}_{\vs} \ket*{\mathcal{P}^{\bs'}_{h}} = P_{\vs} \delta_{\bs \bs'}
\end{equation}
exactly, where $P_{\vs} = \bra{\mathcal{P}^{\bs}_{h}} \hat{\Pi}_{\vs} \ket*{\mathcal{P}^{\bs}_{h}}$. As a consequence, whenever $P_{\vs}$ is nonzero, the measurement operator $\hat{\Pi}_{\vs}$ acts exactly unitarily on states of the form $\ket{\mathcal{P}_h} \equiv\sum_{\bs} b_{\bs} \ket{\mathcal{P}^{\bs}_{h}}$, and so information encoded within the span of the pruned code states is perfectly preserved by stabilizer measurement. In particular, the probability $P_{\vs}$ of measuring the syndrome $\vs$ is entirely independent of the measured pruned state $\ket{\mathcal{P}_h}$.

Similarly, once decoherence is added, the probability $P_{\vs}$ of measuring the syndrome $\vs$ is once again independent of decohered pruned state $\mathcal{E}(\dyad{\mathcal{P}_h})$. This follows from the exact identity
\begin{equation}
    \bra{\mathcal{P}^{\bs}_h} \mathcal{E}(\hat{\Pi}_{\vs}) \ket*{\mathcal{P}^{\bs'}_h} = P_{\vs} \delta_{\bs \bs'} ,
\end{equation}
similar to the pure-state case. An immediate consequence is that stabilizer measurements of a bit-flip-decohered pruned state $\ket{\mathcal{P}_h}$ result in an exactly incoherent error channel on the corresponding stabilizer state $\ket{\psi_0} = \sum_{\bs} b_{\bs} \ket{\psi^{\bs}_0}$, 
\begin{equation}
    \frac{\hat{\Pi}_{\vs} \mathcal{E}(\dyad{\mathcal{P}_h}) \hat{\Pi}_{\vs}}{P_{\vs}} = \sum_{\ve} P_{\ve | \vs} X_{\vs | \ve} \dyad{\psi_0} X_{\vs | \ve} ,
\end{equation}
in contrast to the approximate action of the incoherent channel $\mathcal{M}_{\vs}$ in Eq.~\eqref{eq:M_channel}. Here $P_{\ve | \vs} = p_{\ve} \bra{\mathcal{P}_h} X_{\ve} \hat{\Pi}_{\vs} X_{\ve} \ket{\mathcal{P}_h} / P_{\vs}$ is the probability that the bit-flip error $\ve$ occurred, given that the syndrome $\vs$ was observed, while $X_{\vs | \ve}$ is a Pauli-$X$ operator defined such that $\hat{\Pi}_{\vs} X_{\ve} \ket{\mathcal{P}_h} \propto X_{\vs | \ve} \ket{\psi_0}$.

%% file: appendices/strong_symm.tex
\section{Decoding Fidelity and Strong Symmetry}
\label{app:strong_symm}
Throughout this work, our quantitative measure of successful QEC has been the measurement-averaged decoding fidelity,
\begin{equation}
    \overline{\mathcal{F}}(\ket{\psi_{h}}) = \bra{\psi_{h}} \mathcal{N}(\dyad{\psi_{h}}) \ket{\psi_{h}} .
\end{equation}
Here $\mathcal{N} \equiv \mathcal{M} \circ \mathcal{E}$ is the composition of a Pauli decoherence channel $\mathcal{E}$ and a decoding channel $\mathcal{M}$ [see Eq.~\eqref{eq:decoding_channel}], the latter of which involves measuring stabilizers $A_{v}, B_{p}$, annihilating the observed anyons to recover a stabilizer toric code state in $\mathcal{C}_{0}$, and finally adiabatically restoring the transverse field to obtain a TFTC state in $\mathcal{C}_{h}$. In the special case where $\mathcal{E}$ describes a bit-flip (i.e., Pauli-$X$) channel, the total channel $\mathcal{N}$ exhibits a \emph{strong} symmetry with respect to the logical-$X$ operators $\overline{X}_{\mu}$, i.e.,
\begin{equation}
    \overline{X}_{\mu} \mathcal{N}(\rho) = \mathcal{N}(\overline{X}_{\mu} \rho), \quad \mathcal{N}(\rho) \overline{X}_{\mu} = \mathcal{N}(\rho \overline{X}_{\mu}) 
\end{equation}
for any $\rho$. This observation leads to a useful simplification: rather than considering the decoding fidelity of an arbitrary TFTC state, it is sufficient to focus on decoding the specific ITE state $\ket*{\varphi^{(0,0)}_{h}} \propto e^{-\beta H_{h} / 2} \ket*{\psi^{(0,0)}_{0}}$ obtained from the logical-$Z$ stabilizer state $\ket*{\psi^{(0,0)}_{0}}$. We will show in this Appendix that our QEC protocol succeeds with high fidelity on an arbitrary code state if and only if it succeeds with high fidelity on this particular state. 

To be precise, let\footnote{As usual, we use adiabatic evolution $\mathcal{U}_{h}$ only for conceptual simplicity. We could alternatively use the ITE isometry $\mathcal{V}_{h}$ constructed in Sec.~\ref{subsec:ite}, or a suitable quasiadiabatic generator.} $\ket{\psi^{\bs}_{h}} \equiv \mathcal{U}_{h} \ket{\psi^{\bs}_{0}}$ be a basis of $\mathcal{C}_{h}$ which transforms covariantly under the logical-$X$ operators: $\overline{X}^{\bs} \ket*{\psi^{\bs'}_{h}} = \ket*{\psi^{\bs \oplus \bs'}_{h}}$. We will prove that, if $\overline{\mathcal{F}}(\ket*{\psi^{(0,0)}_{h}}) \geq 1 - \varepsilon$, then
\begin{equation}
        \label{eq:strong_symm_inequality}
    \min_{\ket{\psi_{h}} \in \mathcal{C}_{h}} \overline{\mathcal{F}}(\ket{\psi_{h}}) \geq 1 - 8 \sqrt{\varepsilon} .
\end{equation}
Consequently, an arbitrary TFTC state $\ket{\psi_{h}} = \sum_{\bs} b_{\bs} \ket{\psi^{\bs}_{h}}$ is decodable with high fidelity if and only if $\ket*{\psi^{(0,0)}_{h}}$ is decodable with high fidelity. Combined with the results of Appendices~\ref{app:decoding_fidelities} and~\ref{app:imag_time_adiabatic}, we can instead focus our attention on developing decoders for the ITE state $\ket*{\varphi^{(0,0)}_{h}}$, which is virtually indistinguishable from $\ket*{\psi^{(0,0)}_{h}}$. Note that an analogous result holds for the TFIM under bit-flip decoherence in Sec.~\ref{sec:tfim}.

To prove Eq.~\eqref{eq:strong_symm_inequality}, consider the \emph{entanglement fidelity} $\mathcal{F}_{e}$, defined by
\begin{equation}
    \mathcal{F}_{e} = \bra{\Psi_{QR}} \mathcal{N} (\dyad{\Psi_{QR}}) \ket{\Psi_{QR}} ,
\end{equation}
where $\ket{\Psi_{QR}} \equiv \frac{1}{\sqrt{4}} \sum_{\bs} \ket{\psi^{\bs}_{h}}_{Q} \otimes \ket{\bs}_{R}$ is a maximally entangled state between the principal TFTC system $Q$ and a two-qubit reference $R$, and the channel $\mathcal{N}$ is understood to only act on subsystem $Q$. Using the strong symmetry of $\mathcal{N}$, the entanglement fidelity can be reduced to
\begin{equation}
    \mathcal{F}_{e} = \frac{1}{4^2} \sum_{\bs, \bs'} \bra{\psi^{\bs}_{h}} \mathcal{N}(\dyad*{\psi^{\bs}_{h}}{\psi^{\bs'}_{h}}) \ket*{\psi^{\bs'}_{h}} = \bra*{\psi^{(0,0)}_{h}} \mathcal{N}(\dyad*{\psi^{(0,0)}_{h}}) \ket*{\psi^{(0,0)}_{h}} = \overline{\mathcal{F}}(\ket*{\psi^{(0,0)}_{h}}) ,
\end{equation}
i.e., the entanglement fidelity is equal to the decoding fidelity of the single state $\ket*{\psi^{(0,0)}_{h}}$. Supposing that the latter is lower-bounded by $1 - \varepsilon$, the Fuchs-van de Graaf inequality \cite{nielsenQuantumComputationQuantum2010} gives
\begin{equation}
    \norm{ \mathcal{N}(\dyad{\Psi_{QR}}) - \dyad{\Psi_{QR}} }_{1} \leq 2 \sqrt{1 - \bra*{\psi^{(0,0)}_{h}} \mathcal{N}(\dyad*{\psi^{(0,0)}_{h}}) \ket*{\psi^{(0,0)}_{h}} } \leq 2 \sqrt{\varepsilon} .
\end{equation}
This further implies that the channel $\mathcal{N}$ must be close to the identity channel in the diamond distance\footnote{Recall that the diamond distance $\norm{\mathcal{N} - \mathcal{N}'}_{\diamond} \equiv \sup_{\rho} \norm{(\mathcal{N} \otimes \text{id}_{k})(\rho) - (\mathcal{N}' \otimes \text{id}_{k})(\rho)}_{1}$ is a measure of distinguishability between the two channels $\mathcal{N}$ and $\mathcal{N}'$; here the maximization is taken over all states $\rho$ on the original Hilbert space tensored with an auxiliary $k$-dimensional Hilbert space. The bound in Eq.~\eqref{eq:diamond_2} follows immediately from this definition. For the bound in Eq.~\eqref{eq:diamond_1}, see Exercise 3.6 of Ref.~\cite{watrous_theory_2018}.}:
\begin{equation}
    \label{eq:diamond_1}
    \norm{\mathcal{N} - \text{id}}_{\diamond} \leq (\dim \mathcal{C}_{h}) \norm{ \mathcal{N}(\dyad{\Psi_{QR}}) - \dyad{\Psi_{QR}} }_{1} \leq 8 \sqrt{\varepsilon} .
\end{equation}
Correspondingly, $\mathcal{N}(\dyad{\psi_{h}})$ must be close to $\dyad{\psi_{h}}$ for any code state:
\begin{equation}
    \label{eq:diamond_2}
    \norm{\mathcal{N}(\dyad{\psi_{h}}) - \dyad{\psi_{h}}}_{1} \leq \norm{\mathcal{N} - \text{id}}_{\diamond} \leq 8 \sqrt{\varepsilon} .
\end{equation}
Finally, one more application of the Fuchs-van de Graaf inequality yields
\begin{equation}
    \mathcal{F}(\ket{\psi_{h}}) = \bra{\psi_{h}} \mathcal{N}(\dyad{\psi_{h}}) \ket{\psi_{h}} \geq \qty( 1 - \frac{1}{2} \norm{\mathcal{N}(\dyad{\psi_{h}}) - \dyad{\psi_{h}}}_{1} )^{2} \geq 1 - 8 \sqrt{\varepsilon}.
\end{equation}

%% file: appendices/coherent_decoding.tex
\section{Decoding Coherent Errors}
\label{app:coherent_decoding}
In this Appendix, we prove a necessary and sufficient condition for coherent errors to be correctable. These conditions are valid for general coherent errors on a general quantum error-correcting code, so the present discussion is abstracted away from the specific case of the toric code.

Let $\qty{\ket{\psi_{0}^{a}}}$ be a basis for the code space $\mathcal{C}$, and let $\ket{\psi^{a}_{\mathcal{U}}} \equiv \mathcal{U} \ket{\psi^{a}_{0}}$ be the state $\ket{\psi^{a}_{0}}$ corrupted by a coherent error (i.e., a unitary evolution) $\mathcal{U}$. Additionally, let $P^{a}_{s} \equiv \bra{\psi^{a}_{\mathcal{U}}} \hat{\Pi}_{s} \ket{\psi^{a}_{\mathcal{U}}}$ be the probability of measuring syndrome $s$ in the corrupted code state $\ket{\psi^{a}_{\mathcal{U}}}$, and let $\ket{\psi^{a}_{\mathcal{U}}(s)} = \frac{1}{\sqrt{P^{a}_{s}}} \hat{\Pi}_{s} \ket{\psi^{a}_{\mathcal{U}}}$ be the corresponding postmeasurement state. Here we assume that $\hat{\Pi}_{s}$ is a complete set of orthogonal projectors, so that $\sum_{s} \hat{\Pi}_{s} = 1$. Then, the following two conditions are necessary and sufficient for quantum error correction to be possible upon observing the syndrome $s$:
\begin{enumerate}
    \item All of the measurement probabilities $P^{a}_{s}$ are equal, i.e., $P^a_{s} = P_s$ is independent of $a$;
    \item All of the states $\ket{\psi^{a}_{\mathcal{U}}(s)}$ remain orthonormal after the syndrome measurement, i.e., $\bra{\psi^{a}_{\mathcal{U}}(s)} \ket{\psi^{b}_{\mathcal{U}}(s)} = \delta_{a b}$.
\end{enumerate}
If desired, these conditions can be written together in a basis-independent manner as follows: for any two code states $\ket{\psi_{0}}, \ket{\psi'_{0}} \in \mathcal{C}$, we have
\begin{equation}
\label{eq:coherent_decodability_app}
    \bra{\psi_{0}} \mathcal{U}^{\dag} \hat{\Pi}_{s} \mathcal{U} \ket{\psi_{0}'} = P_{s} \bra{\psi_{0}} \ket{\psi_{0}'} .
\end{equation}
Note that this is simply the Knill-Laflamme condition \cite{knill_theory_1997,knill_theory_2000,nielsenQuantumComputationQuantum2010} for an error channel with Kraus operators $\hat{\Pi}_{s} \mathcal{U}$.

The proof of this statement is rather trivial. First, let us denote the principal system as $Q$, and introduce an auxiliary reference system $R$ which is initially maximally entangled with $Q$:
\begin{equation}
    \ket{\Psi_{QR}} \equiv \frac{1}{\sqrt{\mathcal{D}}} \sum_{a = 1}^{\mathcal{D}} \ket{\psi^{a}_{0}}_{Q} \otimes \ket{a}_R ,
\end{equation}
where $\mathcal{D} = \dim \mathcal{C}$ is the dimension of the code space. Following the coherent error and the measurement of syndrome $s$, we obtain the state
\begin{equation}
    \ket{\Psi_{QR}(s)} = \frac{1}{\sqrt{\mathcal{D} P_{s}}} \sum_{a = 1}^{\mathcal{D}} \sqrt{P^{a}_{s}} \ket{\psi^{a}_{\mathcal{U}}(s)}_{Q} \otimes \ket{a}_{R} ,
\end{equation}
where $P_{s} = \frac{1}{\mathcal{D}} \sum_{a = 1}^{\mathcal{D}} P^{a}_{s}$ is the probability of observing the syndrome $s$ in the full state $\ket{\Psi_{QR}}$.

Any two purifications of a mixed state with the same dimension differ at most by a unitary on the purifying system~\cite{nielsenQuantumComputationQuantum2010}. Viewing $Q$ as a purification of $R$, there exists a unitary $V_{s}$ which maps $\ket*{\Psi_{QR}(s)}$ back to $\ket{\Psi_{QR}}$ if and only if $Q$ and $R$ remain maximally entangled after the measurement. To wit, the reduced density matrix on the reference $R$ is
\begin{equation}
    \rho_{R}(s) \equiv \tr_{Q} \dyad{\Psi_{QR}(s)} = \frac{1}{\mathcal{D} P_{s}} \sum_{a, b = 1}^{\mathcal{D}} \qty( \sqrt{P^{a}_{s} P^{b}_{s}} \bra{\psi^{b}_{\mathcal{U}}(s)} \ket{\psi^{a}_{\mathcal{U}}(s)}) \dyad{a}{b}_{R} .
\end{equation}
This state is maximally mixed on $R$, and thereby $R$ is maximally entangled with its complement, if and only if $\bra{\psi^{a}_{\mathcal{U}}(s)} \ket{\psi^{b}_{\mathcal{U}}(s)} = \delta_{ab}$, and $P^{a}_{s} = P_{s}$ is independent of $a$. This proves that a unitary correction $V_{s}$ exists if and only if the two conditions are satisfied. 

Indeed, it is now simple to explicitly construct the desired recovery unitary:
\begin{equation}
    V_{s} = \sum_{a = 1}^{\mathcal{D}} \dyad{\psi^{a}_{0}}{\psi^{a}_{\mathcal{U}}(s)} ,
\end{equation}
which acts unitarily on the space of postmeasurement states by condition 2. If we consider an arbitrary initial state $\ket{\psi_{0}} = \sum_{a = 1}^{\mathcal{D}} c_{a} \ket{\psi^{a}_{0}}$ in the code space, then by condition 1, the postmeasurement state is simply
\begin{equation}
    \ket{\psi_{\mathcal{U}}(s)} \equiv \frac{1}{\sqrt{P_{s}}} \hat{\Pi}_{s} \mathcal{U} \ket{\psi_{0}} = \sum_{a = 1}^{\mathcal{D}} c_{a} \ket{\psi_{\mathcal{U}}^{a}(s)} = V^{\dag}_{s} \ket{\psi_{0}},
\end{equation}
i.e., the measurement has effectively implemented the unitary $V^{\dag}_{s} \mathcal{U}^{\dag}$. Finally, we recover the initial code state $\ket{\psi_{0}}$ by simply applying $V_{s}$.

More generally, suppose that Eq.~\eqref{eq:coherent_decodability_app} is only \emph{approximately} satisfied, in the sense that
\begin{equation}
\label{eq:coherent_decodability_app_approx}
    \abs{ \frac{\bra{\psi_0} \mathcal{U}^{\dag} \hat{\Pi}_{s} \mathcal{U} \ket{\psi_0'} - Q_s \bra{\psi_0} \ket{\psi_0'}}{Q_s} } \leq \varepsilon
\end{equation}
for some $Q_{s} > 0$. By setting $\ket{\psi'_{0}} = \ket{\psi_{0}}$, we see immediately that the probability $P_{s}$ of observing syndrome $s$ is bounded between $(1 - \varepsilon) Q_{s}$ and $(1 + \varepsilon) Q_{s}$; i.e., $P_{s}$ is approximately equal to $Q_{s}$ up to a small \emph{relative} error, for any code state $\ket{\psi_{0}} \in \mathcal{C}_{0}$. 

To construct a decoding unitary in this setting, we perform a polar decomposition of the operator $\hat{\Pi}_s \mathcal{U} \hat{\Pi}_{0}$, where $\hat{\Pi}_0$ is the projector onto $\mathcal{C}$, to obtain
\begin{equation}
    \frac{1}{\sqrt{Q_s}}\hat{\Pi}_s \mathcal{U} \hat{\Pi}_{0} = \mathcal{V}_s \mathcal{I}_s^{1/2} , \quad \mathcal{I}_s \equiv \frac{\hat{\Pi}_0 \mathcal{U}^{\dag} \hat{\Pi}_s \mathcal{U} \hat{\Pi}_0}{Q_s} ,
\end{equation}
where $\mathcal{V}_s$ is an exact isometry from $\mathcal{C}$ to the subspace corresponding to syndrome $s$. In the exact setting, $\mathcal{I}_s = \hat{\Pi}_0$ and $\mathcal{V}_s = V_s^{\dag}$. In the approximate setting, the bound~\eqref{eq:coherent_decodability_app_approx} implies that the Hermitian operator $\mathcal{I}_s$ has spectrum bounded between $1 - \varepsilon$ and $1 + \varepsilon$. Taking $\mathcal{V}_s^{\dag}$ as our decoding unitary, we obtain a recovery fidelity
\begin{equation}
    \mathcal{F}_s(\ket{\psi_0}) \equiv \abs{\bra{\psi_0} \mathcal{V}_s^{\dag} \ket{\psi_{\mathcal{U}}(s)}}^2 = \frac{\abs{\bra{\psi_0} \mathcal{V}_s^{\dag} \hat{\Pi}_s \mathcal{U} \ket{\psi_0}}^2}{\bra{\psi_0} \mathcal{U}^{\dag} \hat{\Pi}_s \mathcal{U} \ket{\psi_0}} = \frac{\abs{\bra{\psi_0} \mathcal{I}_s^{1/2} \ket{\psi_0}}^2}{\bra{\psi_0} \mathcal{I}_s \ket{\psi_0}} \geq \frac{1 - \varepsilon}{1 + \varepsilon} \geq 1 - 2\varepsilon.
\end{equation}
Therefore, the approximate satisfaction of Eq.~\eqref{eq:coherent_decodability_app} (relative to the size of $P_s$) implies the existence of a high-fidelity approximate recovery unitary.

In the context of stabilizer codes, an important caveat to this theorem is that the recovery $V_{s}$ may not necessarily be a Pauli unitary. When \textit{incoherent} Pauli decoherence is applied to a stabilizer code at a rate below its error threshold, the original encoded state can always be recovered with high fidelity using syndrome measurement followed by a Pauli unitary. However, when a \textit{coherent} error is applied, it is possible in principle that a nontrivial coherent unitary rotation is induced on the code space, and error correction by Pauli operations alone is insufficient. In the toric code context, this is to say that incoherent errors performed below the threshold can always be corrected purely by pairing anyons; for coherent errors, it is possible in principle that the encoded information remains intact following syndrome measurement, but the original encoded state \textit{cannot} be recovered simply by pairing anyons together. This situation is known to arise in the case of the coherent error $\mathcal{U} = \prod_{\ell} e^{i \theta X_{\ell}}$, for $\theta > \theta_{c}$ performed on the surface code, for certain system sizes and boundary conditions \cite{bravyi_correcting_2018}. In the context of the present work, we have shown in Sec.~\ref{subsec:bitflip_meas_decoding} that this situation does not arise for the TFTC due to stoquasticity.

%% file: appendices/worm.tex
\section{The Worm Algorithm}
\label{app:worm_algo}
In this Appendix, we discuss the worm algorithm (WA) \cite{prokofev_worm_1998,prokofev_worm_2001,deng_dynamic_2007,collevecchio_worm_2016} as an efficient Monte Carlo implementation of the optimal postmeasurement decoder in both of Secs.~\ref{sec:meas_decoding} and \ref{sec:pauli_meas_decoding}. A useful ancillary result is that the worm algorithm serves as an efficient optimal decoder for the stabilizer toric code under independent bit-flip and phase-flip decoherence, as recently discussed in Ref.~\cite{tobias_optimal_2026}. 

\subsection{The Worm Algorithm and \texorpdfstring{$\mathbb{Z}_{2}$}{Z2} Lattice Gauge Theory}
\label{app:worm_algo_lgt}
To start, we describe how the WA is used to efficiently simulate the low-temperature expansion of the classical $\mathbb{Z}_{2}$ LGT [see Eqs.~\eqref{eq:lgt_classical_isotropic}~and~\eqref{eq:lgt_classical_isotropic_partition}]. As discussed in Sec.~\ref{sec:background}, we can regard the 3$d$ LGT as a model of $\mathbb{Z}_{2}$-valued closed flux loops, which physically represent the worldlines of $m$ anyons in imaginary spacetime. We can therefore express the isotropic partition function in Eq.~\eqref{eq:lgt_classical_isotropic_partition} as follows:
\begin{equation}
    \widetilde{\mathcal{Z}}_{\lambda} \equiv \sum_{\qty{U}} e^{-\widetilde{S}_{\LGT}[U]} \propto \sum_{\text{Closed Loops } \mathcal{L}} e^{-2 \abs{\mathcal{L}} / \lambda} ,
\end{equation}
where $\mathcal{L}$ is a collection of $\mathbb{Z}_{2}$-valued closed loops in the dual lattice of the 3$d$ spacetime cubic lattice. The global homology of the loops $\mathcal{L}$ (i.e., whether non-contractible loop configurations are allowed) depends on the boundary conditions of $\widetilde{S}_{\LGT}$. To mimic the quantum matrix elements $\mathcal{Z}_{h} \equiv \bra{\psi^{\bs}_{0}} e^{-\beta H_{h}} \ket{\psi^{\bs}_{0}}$, we will require that the loops $\mathcal{L}$ are globally contractible, and cannot pass through the ``hard-wall'' boundary conditions at imaginary times $\tau = \pm \beta / 2$.


Given an initial loop configuration $\mathcal{L}$ (for example, the zero-temperature configuration with no loops), the WA constructs a new loop configuration $\mathcal{L}'$ as follows \cite{prokofev_worm_2001}:
\begin{enumerate}
    \item Pick a random site $\tilde{i}$ of the dual lattice, and instantiate a ``head'' variable $\mathsf{h}$ and a ``tail'' variable $\mathsf{t}$, each initially with the value $\tilde{i}$. 
    \item Allow the head variable to perform a random walk through the dual lattice by iterating the following steps:
    \begin{enumerate}
        \item Pick a random direction $\mu \in \qty{\pm x, \pm y, \pm \tau}$. Denote the unit vector in the direction $\mu$ as $e_{\mu}$.
        \item If the dual lattice link connecting $\mathsf{h}$ to $\mathsf{h} + e_{\mu}$ is not already occupied by a segment of flux, then with probability $e^{-2 / \lambda}$, add a segment of flux connecting $\mathsf{h}$ to $\mathsf{h} + e_{\mu}$ and update $\mathsf{h} \leftarrow \mathsf{h} + e_{\mu}$.
        \item Otherwise, if the dual lattice link connecting $\mathsf{h}$ to $\mathsf{h} + e_{\mu}$ is already occupied by a segment of flux, delete that segment with unit probability and update $\mathsf{h} \leftarrow \mathsf{h} + e_{\mu}$.
        \item If $\mathsf{h} = \mathsf{t}$, stop the iteration.
    \end{enumerate}
\end{enumerate}
In this way, $\mathcal{L}'$ is obtained from $\mathcal{L}$ by temporarily leaving the state space of closed loops and allowing a single open-string ``worm''. The head of the worm executes a random walk, drawing or eliminating flux lines along its trajectory using simple Metropolis acceptance rules, until it eventually arrives back at the tail and its trajectory forms a closed loop\footnote{In order to ensure that each sampled $\mathcal{L}$ is contractible,whenever a non-contractible configuration is encountered, we simply keep iterating the WA until we arrive back at a contractible configuration.}. It is straightforward to verify that this multi-step process satisfies detailed balance,
\begin{equation}
    \frac{\mathbb{P}[\mathcal{L} \to \mathcal{L}']}{\mathbb{P}[\mathcal{L}' \to \mathcal{L}]} = \frac{e^{-2|\mathcal{L}'|/ \lambda}}{e^{-2 |\mathcal{L}| / \lambda}} ,
\end{equation}
and therefore the Markov process associated with this algorithm converges to the Gibbs distribution.

The WA is extraordinarily efficient within the topological phase $\lambda < \lambda_{c}$, where typical loops are short. Empirically, the WA exhibits almost no critical slowing down as the critical point is approached \cite{prokofev_worm_2001,deng_dynamic_2007}. In the trivial phase $\lambda > \lambda_{c}$, where flux loops exhibit vanishing coarse-grained line tension, it typically takes $\mathcal{O}(L^{3})$ time for each worm to close on itself; nevertheless, since flux loops are large and system-spanning, it takes far fewer individual worms to thermalize the system. Indeed, it has been rigorously proven that the WA for Ising models on bipartite graphs is rapidly mixing \cite{collevecchio_worm_2016}, and the same result extends to the 3$d$ LGT via Wegner duality. 

In Sec.~\ref{subsec:meas_decoding_numerics}, we utilized the WA to numerically implement the optimal decoder for the TFTC under stabilizer measurements. We first use the WA to sample a loop configuration $\mathcal{L}$ from thermal equilibrium, which provides the syndrome $\vs \equiv \qty{s_{p}}$ from the locations of fluxes passing through the $\tau = 0$ surface. We then continue to sample additional loop configurations from the \emph{conditional} probability distribution by forbidding worms from passing through the $\tau = 0$ surface, thereby leaving the syndrome intact. Crucially, the WA efficiently samples over all four homology classes of loops within the $\tau \neq 0$ space, allowing for us to numerically estimate the probabilities of each homology class by their relative sampling frequencies. The optimal decoding protocol is to select the homology class with the largest probability, i.e., the one sampled with the largest frequency.

In an actual decoding experiment, the observer does not have access to an initially sampled loop configuration giving rise to the syndrome $\vs$. However, as long as the chosen Monte Carlo algorithm is ergodic, \emph{any} initial loop configuration consistent with $\vs$ is a valid starting configuration. We can therefore arbitrarily choose the initial loop configuration (for example, with minimal-weight perfect matching) and subsequently run the WA to sample from different homology classes. We have checked that the observed relative frequencies of each homology class are indeed independent of the initial starting configuration.

\subsection{The Worm Algorithm for the Stabilizer Toric Code under Bit-Flip Decoherence}
\label{app:worm_algo_stab}
The WA can also be used as an efficient implementation of the optimal decoder for the stabilizer toric code under bit-flip decoherence \cite{tobias_optimal_2026,temkin_charge-informed_2025}. Since this will serve as an intermediate step towards the optimal postmeasurement decoding of the TFTC under bit-flip decoherence, we start by explaining the WA decoder in the stabilizer case.

Let us first briefly review optimal decoding of i.i.d. bit-flip decoherence in the stabilizer toric code \cite{dennis_topological_2002}. Starting from a code state $\ket{\psi_{0}} \in \mathcal{C}_{0}$, we apply the bit-flip decoherence channel~\eqref{eq:bitflip_expansion}, resulting in the state
\begin{equation}
    \rho \equiv \mathcal{E}(\dyad{\psi_{0}}) = \sum_{\ve} p_{\ve} X_{\ve} \dyad{\psi_{0}} X_{\ve} .
\end{equation}
We can visualize each bit-flip error $\ve$ as a collection of open strings in the dual lattice, where $m$ anyons are created at the endpoints $\partial \ve \equiv \qty{\partial \ve_{p}}$ of these strings. Therefore, when the stabilizers $B_{p}$ are measured, the above sum is collapsed to contain only those strings whose endpoints agree with the observed syndrome $\vs \equiv \qty{s_{p}}$:
\begin{equation}
    \rho(\vs) \equiv \frac{\hat{\Pi}_{\vs} \rho \hat{\Pi}_{\vs}}{P_{\vs}} = \frac{1}{P_{\vs}} \sum_{\ve : \partial \ve = \vs} p_{\ve} X_{\ve} \dyad{\psi_{0}} X_{\ve}, 
\end{equation}
where $P_{\vs}$ is the total probability of observing syndrome $\vs$,
\begin{equation}
    P_{\vs} \equiv \sum_{\ve : \partial \ve = \vs} p_{\ve} = (1 - p)^{N} \sum_{\ve : \partial \ve = \vs} \qty(\frac{p}{1-p})^{\abs{\ve}} .
\end{equation}

The set of all errors $\ve$ satisfying $\partial \ve = \vs$ can be further partitioned into just four \emph{homology classes} $\bs | \vs$. Fixing a particular error $X_{\vs}$ consistent with the observed syndrome, we say that $\ve \in \bs | \vs$ if $\partial \ve = \vs$ and $X_{\ve} X_{\vs} \ket{\psi_{0}} = \overline{X}^{\bs} \ket{\psi_{0}}$, i.e., if $X_{\ve} X_{\vs}$ is equal to the logical-$X$ operator $\overline{X}^{\bs}$ up to stabilizers $A_{v}$; consequently, two errors $\ve$ and $\ve'$ within the same homology class differ by stabilizers, which can be graphically visualized as \emph{contractible} closed loops in the dual lattice. On the other hand, two non-homologous errors $\ve, \ve'$ with the same syndrome differ by a logical-$X$ operator, which can be visualized as a non-contractible loop in the dual lattice.

Since homologous errors act identically on code states $\ket{\psi_{0}}$, $\rho(\vs)$ can be reorganized as follows:
\begin{equation}
    \rho(\vs) = \frac{1}{P_{\vs}} X_{\vs} \qty( \sum_{\bs | \vs} \mathcal{Q}_{\bs}(\vs) \overline{X}^{\bs} \dyad{\psi_{0}} \overline{X}^{\bs} ) X_{\vs} ,
\end{equation}
where $\mathcal{Q}_{\bs}(\vs)$ is the total probability of the homology class $\bs | \vs$:
\begin{equation}
    \mathcal{Q}_{\bs}(\vs) \equiv \sum_{\ve \in \bs | \vs} p_{\ve} = (1 - p)^{N} \sum_{\ve \in \bs | \vs} \qty(\frac{p}{1-p})^{\abs{\ve}} .
\end{equation}
Note that $\sum_{\bs} \mathcal{Q}_{\bs}(\vs) = P_{\vs}$. By choosing $X_{\vs}$ such that $\text{argmax}_{\bs} \, \mathcal{Q}_{\bs}(\vs) = (0,0)$, the optimal postmeasurement decoder is simply given by $X_{\vs}$, and the decoding fidelity is given by
\begin{equation}
    \mathcal{F}_{\vs} = \bra{\psi_{0}} X_{\vs} \rho(\vs) X_{\vs} \ket{\psi_{0}} = \frac{\mathcal{Q}_{(0,0)}(\vs)}{P_{\vs}} = \frac{1}{P_{\vs}} \max_{\bs} \mathcal{Q}_{\bs}(\vs) .
\end{equation}
Therefore, optimal decoding succeeds with high fidelity if, for the sampled syndrome $\vs$, almost all typical errors $\ve$ consistent with $\vs$ lie within the same homology class.

In order to implement the optimal decoder, it is necessary to compute the four probabilities $\mathcal{Q}_{\bs}(\vs)$ for the observed syndrome $\vs$. Each of these quantities can be cast as the partition function of a random-bond Ising model (RBIM) \cite{dennis_topological_2002}, which can be computed exactly using free fermion methods or approximately via tensor network contraction \cite{bravyi_efficient_2014}. Here we will describe how the ratios $\mathcal{Q}_{\bs}(\vs) / P_{\vs}$ can be efficiently estimated using classical Monte Carlo simulation via the WA. First, given the syndrome $\vs$, we pick an initial error $\ve$ consistent with $\vs$, i.e., satisfying $\partial \ve = \vs$. Next, to obtain a subsequent error $\ve'$ consistent with $\vs$, we add a closed loop to $\ve$ via the WA as follows:
\begin{enumerate}
    \item Pick a random plaquette $q$ (i.e., a site of the dual lattice; we use $q$ here instead of $p$ to avoid confusion with the bit-flip probability $p$). Instantiate a head variable $\mathsf{h}$ and a tail variable $\mathsf{t}$, each initially with the value $q$. 
    \item Allow the head variable to perform a random walk by iterating the following steps:
    \begin{enumerate}
        \item Pick a random direction $\mu \in \qty{\pm x, \pm y}$. 
        \item If the dual lattice link connecting $\mathsf{h}$ to $\mathsf{h} + e_{\mu}$ is not already occupied by an error, then with probability $p / (1-p)$, add an error to this link, then update $\mathsf{h} \leftarrow \mathsf{h} + e_{\mu}$.
        \item Otherwise, if the dual lattice link connecting $\mathsf{h}$ to $\mathsf{h} + e_{\mu}$ is already occupied by an error, delete it with unit probability and update $\mathsf{h} \leftarrow \mathsf{h} + e_{\mu}$.
        \item If $\mathsf{h} = \mathsf{t}$, stop the iteration.
    \end{enumerate} 
\end{enumerate}
It is clear that this algorithm preserves the syndrome $\vs$, and samples errors $\ve$ with $\partial \ve = \vs$ with relative probabilities $[p/(1-p)]^{\abs{\ve}}$. Crucially, loops added by  the WA can be either contractible or non-contractible, so that the above algorithm efficiently samples between all four homology classes at once. We can therefore estimate each $\mathcal{Q}_{\bs}(\vs)$ by simply tracking each sampled error's homology class, and computing each class's observed relative frequency.

\begin{figure}[t]
    \centering
    \includegraphics[width = 0.5\textwidth]{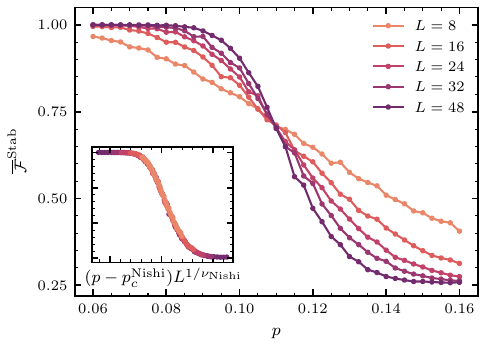}
    \caption{Optimal decoding fidelity for the stabilizer toric code subjected to i.i.d. bit-flip decoherence, implemented via the WA. We obtain an excellent crossing at $p = p_{c}^{\text{Nishi}} \approx 0.109$ and a scaling collapse with the correlation length exponent $\nu_{\text{Nishi}} \approx 1.534$, in agreement with previous numerical studies on the Nishimori RBIM~\cite{honecker_universality_2001,wan_revisiting_2025}.}
    \label{fig:bitflip_decoding}
\end{figure}
Figure~\ref{fig:bitflip_decoding} depicts our numerical verification that the WA can be used as an efficient implementation of the optimal decoder for the stabilizer toric code under i.i.d. bit-flip decoherence. As is well-known~\cite{dennis_topological_2002}, the critical error threshold and its nearby critical phenomena are described by the 2$d$ random-bond Ising model on its Nishimori line. We observe an excellent crossing at the latter model's known critical point $p_{c}^{\text{Nishi}}$. Using the correlation length $\nu_{\text{Nishi}} \approx 1.534$ computed numerically in Ref.~\cite{wan_revisiting_2025}, we observe an excellent finite-size scaling collapse.

\subsection{The Worm Algorithm for the Transverse-Field Toric Code under Bit-Flip Decoherence}
Finally, we describe our algorithm for the optimal postmeasurement decoding of the TFTC under bit-flip decoherence. As described in Sec.~\ref{subsec:bitflip_meas_numerics}, our goal is effectively to sample anyon worldlines $\mathcal{L}$ and bit-flip errors $\ve$ from the joint conditional probability distribution described by the partition function
\begin{equation}
    \widetilde{Z}_{\lambda,p}(\vs) \equiv \sum_{\mathcal{L}, \ve | \vs} t^{\abs{\mathcal{L}}} w^{\abs{\ve}} ,
\end{equation}
where $t \equiv e^{-2/\lambda}$ and $w \equiv p / (1-p)$. Our Monte Carlo algorithm consists of three independent steps:
\begin{enumerate}
    \item First, we update $\mathcal{L} \to \mathcal{L}'$ using the WA described in Sec.~\ref{app:worm_algo_lgt}. This update preserves the syndrome $\vs_{\mathcal{L}}$ of $\mathcal{L}$ alone, and therefore preserves $\vs = \vs_{\mathcal{L}} \oplus \vs_{\ve}$.
    \item Second, we update $\ve \to \ve'$ using the WA described in Sec.~\ref{app:worm_algo_stab}. This update preserves the syndrome $\vs_{\ve}$ of $\ve$ alone, and therefore also preserves $\vs$.
    \item The above two updates are not ergodic alone, since they do not mix between all the different pairs of syndromes $(\vs_{\mathcal{L}}, \vs_{\ve})$ which are together consistent with the global syndrome $\vs$. Our third update remedies this by simultaneously toggling the error $e_{\ell}$ at a given link $\ell$, and adding or removing a closed flux loop encircling the link $\ell$ in spacetime. Denoting the resulting new flux configuration as $\mathcal{L}''$ and the resulting new error configuration as $\ve''$, this update is accepted with probability
    \begin{equation}
        \min \qty{1, \frac{t^{\abs{\mathcal{L}''}} w^{\abs{\ve}''}}{t^{\abs{\mathcal{L}'}} w^{\abs{\ve'}}}} ,
    \end{equation}
    which can be computed locally from the single bit-flip $e_{\ell}$ and the four flux values encircling $\ell$. This update is performed on every link $\ell$ independently.
    
\end{enumerate}

%% file: appendices/more_numerics.tex
\section{Additional Numerical Results}
\label{app:more_numerics}
In this Appendix, we collect a number of additional numerical results.

\subsection{Winding Number Statistics}
\label{app:winding_number}
In Section~\ref{subsec:stoquasticity}, we asserted that non-contractible $m$ anyon worldlines are exponentially suppressed in the matrix elements $\bra{\psi^{\bs}_{0}} e^{-\beta H_{h}} \ket*{\psi^{\bs'}_{0}}$ for all $h < h_{c}$. Here we numerically verify this statement in the isotropic limit. In other words, we numerically study the partition function
\begin{equation}
    \widetilde{\mathcal{Z}}^{\text{Winding}}_{\lambda} \equiv \sum_{\text{Closed Loops } \mathcal{L}} e^{-2 \abs{\mathcal{L}} / \lambda} ,
\end{equation}
where each $\mathcal{L}$ is a collection of closed flux loops in the dual lattice of an $L \times L \times L$ cubic lattice, with periodic boundary conditions in all three directions for convenience. Our boundary conditions are such that the flux loops $\mathcal{L}$ are allowed to be either contractible or non-contractible in each direction. Formally, $\widetilde{\mathcal{Z}}^{\text{Winding}}_{\lambda}$ is equivalent to the partition function of a 3$d$ Ising model at inverse temperature $\beta^{\text{Ising}} = \tanh^{-1}(e^{-2/\lambda})$.

\begin{figure}
    \centering
    \includegraphics[width = \columnwidth]{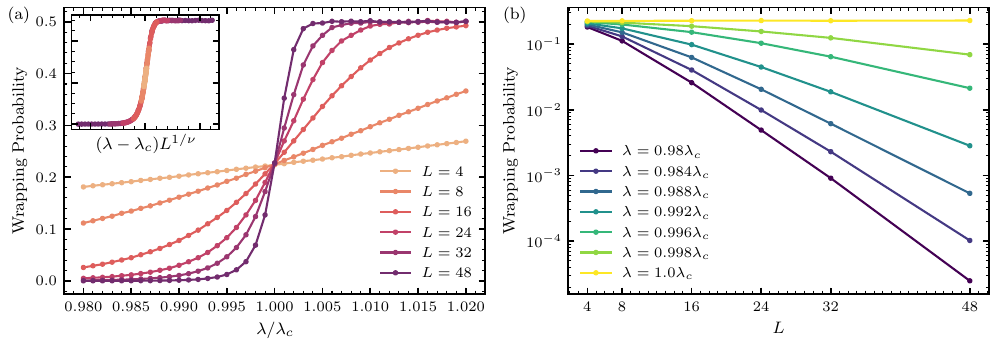}
    \caption{Wrapping probabilities in the partition function $\widetilde{\mathcal{Z}}^{\text{Winding}}_{\lambda}$. (a) Wrapping probability as a function of the temperature $\lambda$, at various linear system sizes $L$. Throughout the topological phase $\lambda < \lambda_{c}$, the wrapping probability is dramatically suppressed with increasing system size. In the trivial phase $\lambda > \lambda_{c}$, the wrapping probability asymptotes to $1/2$ with increasing system size. (b) Wrapping probabilities as a function of the linear system size $L$, for several temperatures within the topological phase. We observe a suppression consistent with $\mathcal{O}(e^{-L/\xi})$ scaling, with $\xi$ diverging as $\lambda \to \lambda_{c}$. An identical result is expected to hold in the extremely anisotropic limit, which implies that the deformed code states $\ket{\varphi^{\bs}_{h}}$ [see Eq.~\eqref{eq:imag_time_evo}] are orthonormal up to exponentially small corrections for all $h < h_{c}$.}
    \label{fig:wrapping_probabilities}
\end{figure}

As usual, we study $\widetilde{\mathcal{Z}}^{\text{Winding}}_{\lambda}$ using the WA. The wrapping probability $W_{\mu}$ is defined as the probability that $\mathcal{L}$ exhibits a non-contractible loop about the cycle $\mu \in \qty{x, y, z}$. Figure \ref{fig:wrapping_probabilities}(a) depicts the average of the three wrapping probabilities, $\frac{1}{3} \sum_{\mu = x, y, z} W_{\mu}$, computed from the frequency of sampling non-contractible closed-loop configurations, as a function of the temperature $\lambda$. Figure \ref{fig:wrapping_probabilities}(b) depicts the same quantity as a function of $L$, for several different values of $\lambda$ below the critical temperature $\lambda_{c}$.

As expected, the wrapping probability tends to zero exponentially quickly in the linear system size throughout the topological phase $\lambda < \lambda_{c}$, and rapidly tends to 1/2 throughout the trivial phase $\lambda > \lambda_{c}$. This is consistent with our physical picture of $m$ anyons condensing as one tunes $\lambda$ through the critical point. We observe a crossing at the critical point $\lambda_{c} \approx 1.313$ of the LGT, and obtain an excellent scaling collapse with $\nu \approx 0.630$ obtained from the 3$d$ Ising model \cite{ferrenberg_pushing_2018}.

An identical result is expected to hold for the extremely anisotropic limit of this model, which implies that the ITE logical-$Z$ states $\ket{\varphi^{\bs}_{h}}$ [see Eq.~\eqref{eq:logicalZ_ITE}] are orthonormal up to exponentially small corrections for all $h < h_{c}$.

\subsection{Additional Numerical Plots for the Transverse-Field Toric Code under Bit-Flip Decoherence}
\label{app:lgt_bitflip_moreplots}
Here we collect a few additional plots for the optimal postmeasurement decoding fidelity of the TFTC subjected to bit-flip decoherence. 
\begin{figure}
    \centering
    \includegraphics[width=\textwidth]{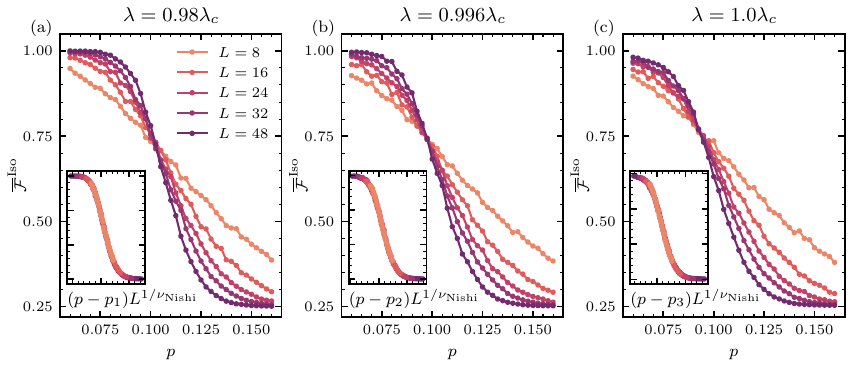}
    \caption{Average optimal postmeasurement decoding fidelity $\overline{\mathcal{F}}^{\text{Iso}}$ for the spacetime-isotropic TFTC under bit-flip decoherence, as a function of the bit-flip strength $p$, for three fixed values of the LGT temperature $\lambda$ and several linear system sizes $L$. Insets: finite-size scaling collapses using the Nishimori RBIM correlation length exponent $\nu_{\text{Nishi}} \approx 1.534$~\cite{wan_revisiting_2025}, and the error thresholds $p_1 \approx 0.102$, $p_2 \approx 0.0975$, and $p_3 \approx 0.0931$.}
    \label{fig:lgt_bitflip_app_2}
\end{figure}

Figure~\ref{fig:lgt_bitflip_app_2} depicts the average optimal postmeasurement decoding fidelity $\overline{\mathcal{F}}^{\text{Iso}}$ at three representative values $\lambda = (0.98 \lambda_c, 0.996 \lambda_c, 1.0\lambda_c)$ of the LGT temperature $\lambda$ [see Eq.~\eqref{eq:lgt_classical_isotropic}], computed as described in Sec.~\ref{subsec:bitflip_meas_numerics}. In each case, we see a clear crossing with an excellent finite-size scaling collapse, indicating error thresholds at $p_1 \approx 0.102$, $p_2 \approx 0.0975$, and $p_3 \approx 0.0931$. Each scaling collapse is performed with the correlation length exponent $\nu_{\text{Nishi}}$ of the Nishimori RBIM. Our numerics are therefore consistent with the scenario that the entire phase boundary $p_c^{\text{opt}}(\lambda_c)$ exhibits a renormalization group flow back to the Nishimori fixed point at $p_c^{\text{opt}}(\lambda = 0)$. However, we caution that we have not systematically optimized a best-fit value of $\nu$, and our numerics are visually consistent with a fairly broad range of values for $\nu$.

Remarkably, Fig.~\ref{fig:lgt_bitflip_app_2}(c) shows that our decoding protocol exhibits a nonzero error threshold even at $\lambda = \lambda_c$. Although the theoretical formalism we have developed in this work only strictly applies to the $\lambda < \lambda_c$ regime, this numerical observation is suggestive that the critical point could also robustly protect encoded information from bit-flip decoherence, and that stabilizer measurements can serve as part of an effective QEC protocol in this setting.

\begin{figure}
    \centering
    \includegraphics[width=\textwidth]{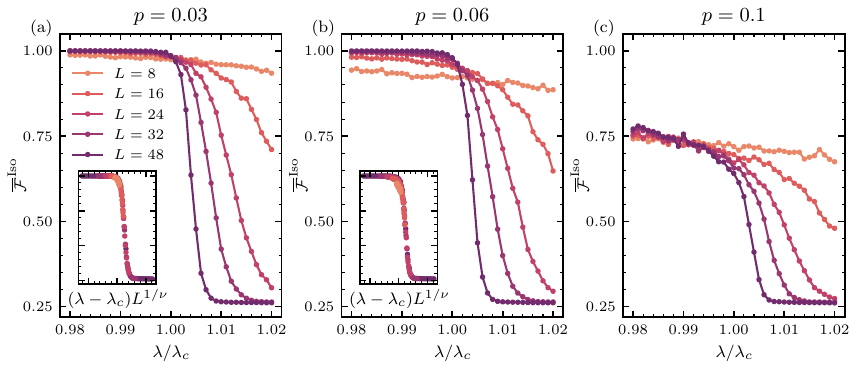}
    \caption{Average optimal postmeasurement decoding fidelity $\overline{\mathcal{F}}^{\text{Iso}}$ for the spacetime-isotropic TFTC under bit-flip decoherence, as a function of the LGT temperature $\lambda$, for three fixed values of the bit-flip strength $p$ and several linear system sizes $L$. (a) at $p = 0.03$, $\overline{\mathcal{F}}^{\text{Iso}}$ is barely modified compared to the $p = 0$ case in Fig.~\ref{fig:meas_decoding_numerics}. Inset: finite-size scaling collapse using the 3$d$ Ising exponent $\nu \approx 0.630$~\cite{ferrenberg_pushing_2018}. (b) At $p = 0.06$, noticeable finite-size effects are present, but there nevertheless remains a clear transition at $\lambda_c$ Inset: finite-size scaling collapse using the same 3$d$ Ising exponent. (c) AT $p = 0.1$, above the near-critical optimal decoding threshold [see Fig.~\ref{fig:lgt_bitflip_decoding}], the decoding transition arises at some $\lambda < \lambda_c$.}
    \label{fig:lgt_bitflip_app_1}
\end{figure}

Figure~\ref{fig:lgt_bitflip_app_1} depicts the optimal postmeasurement decoding fidelity at several representative values of the bit-flip strength $p$, as a function of the LGT temperature $\lambda$. At small values of $p$, such as $p = 0.03$ in Fig.~\ref{fig:lgt_bitflip_app_1}(a), the decoding fidelity is essentially indistinguishable from the $p = 0$ case from Sec.~\ref{subsec:meas_decoding_numerics}. In particular, there is a clear crossing at $\lambda = \lambda_c$ with an excellent finite-size scaling collapse using the 3$d$ Ising exponent $\nu \approx 0.630$~\cite{ferrenberg_pushing_2018}.

At intermediate values of $p$, such as $p = 0.06$ in Fig.~\ref{fig:lgt_bitflip_app_1}(b), we see that there is a strong finite-size drift in the location of the crossing, and in particular the crossings between adjacent system sizes arise at $\lambda > \lambda$; nevertheless, these pairwise crossings appear to drift back towards $\lambda_c$ with increasing system size, and our finite-size scaling collapse is ultimately once more consistent with a transition at $\lambda = \lambda_c$ in the 3$d$ Ising* universality class.

Finally, Figure~\ref{fig:lgt_bitflip_app_1}(c) depicts the optimal postmeasurement decoding fidelity at a value of $p$ beyond the decoding threshold $p_c^{\text{opt}}(\lambda_c) = p_3 \approx 0.0931$, but below the threshold $p_c^{\text{Nishi}} \approx 0.109$ from the stabilizer limit. We see that the crossing has shifted to a nonzero value of $\lambda$ below $\lambda_c$, as expected. We have not attempted a finite-size scaling collapse as a function of $\lambda$ for this transition.

\subsection{Minimal-Weight Perfect Matching}
\label{app:mwpm}
Minimal-weight perfect matching (MWPM) is a widely utilized and highly efficient suboptimal decoding algorithm \cite{dennis_topological_2002,wang_confinement-higgs_2003,fowler_minimum_2014,higgott_sparse_2023}. In the stabilizer toric code subjected to bit-flip decoherence, it exhibits near-optimal performance \cite{wang_confinement-higgs_2003}. In this Appendix, we will investigate MWPM as a suboptimal postmeasurement decoder for the TFTC, both without and with bit-flip decoherence. 

As in the optimal decoder, we first sample a bulk worldline configuration $\mathcal{L}$ from WA Monte Carlo simulations of the isotropic 3$d$ LGT, as well as a bit-flip error $\ve$ i.i.d. on each edge of the $2d$ lattice with probability $p$. These yield a bulk syndrome $\vs_{\mathcal{L}}$ and a surface syndrome $\vs_{\ve}$ respectively, which in turn provide a joint syndrome $\vs = \vs_{\mathcal{L}} \oplus \vs_{\ve}$. We then utilize the PyMatching python package~\cite{higgott_sparse_2023} to construct a MWPM recovery for the joint syndrome $\vs$. Each MWPM recovery is deemed successful if it lies within the same joint homology class as the pair $(\mathcal{L}, \ve)$ (see the discussion of Sec.~\ref{subsec:bitflip_meas_decoding}), and is unsuccessful otherwise. The MWPM decoding fidelity $\overline{\mathcal{F}}^{\text{MWPM}}$ is defined as the fraction of successful recoveries using this protocol. 

\begin{figure}
    \centering
    \includegraphics[width = \textwidth]{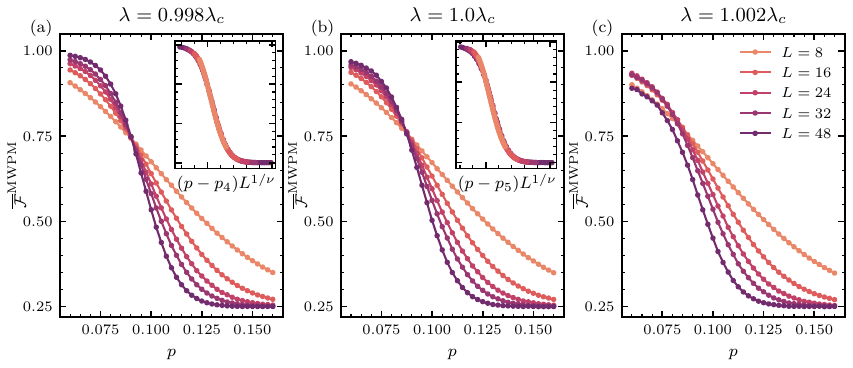}
    \caption{MWPM decoding fidelity $\overline{\mathcal{F}}^{\text{MWPM}}$ for the spacetime-isotropic TFTC at three different temperatures $\lambda$, as a function of the bit-flip strength $p$, at various linear system sizes $L$. (a) At $\lambda = 0.998 \lambda_{c}$, we see clear evidence of a decoding threshold at $p_{4} \approx 0.0904$. We observe a good scaling collapse using the correlation length exponent $\nu = \nu_{\text{MWPM}} = 1.46$~\cite{wang_confinement-higgs_2003}. (b) At the bulk critical point $\lambda = \lambda_{c}$, we again observe clear evidence of a decoding threshold, now at $p_{5} \approx 0.0874$. (c) In the confined phase $\lambda > \lambda_{c}$, we find no evidence of a decoding threshold.}
    \label{fig:mwpm_2}
\end{figure}

Figure~\ref{fig:mwpm_2}(a) depicts the MWPM decoding fidelity $\overline{\mathcal{F}}^{\text{MWPM}}$ as a function of the bit-flip decoherence strength $p$, at the fixed bulk LGT temperature $\lambda = 0.998 \lambda_c$ [see Eq.~\eqref{eq:lgt_classical_isotropic}], just barely within the topological phase. We observe a clear crossing at the threshold $p_4 \equiv p_c^{\text{MWPM}}(0.998\lambda_c) \approx 0.0904$, with an excellent scaling collapse using the correlation length exponent $\nu_{\text{MWPM}} \approx 1.46$ obtained from previous numerics on MWPM in the stabilizer limit $\lambda = 0$~\cite{wang_confinement-higgs_2003}. The threshold $p_4$ lies about 12\% below the corresponding threshold $p_c^{\text{MWPM}}(\lambda = 0) \approx 0.103$~\cite{wang_confinement-higgs_2003}, a noticeable but insubstantial reduction.

Figure~\ref{fig:mwpm_2}(b) depicts the MWPM decoding fidelity directly at the critical point $\lambda = \lambda_c$. Remarkably, we once again see a clear crossing, now at the threshold $p_5 \equiv p_c^{\text{MWPM}}(\lambda_c) \approx 0.0874$. We observe a decent scaling collapse using the same exponent $\nu_{\text{MWPM}}$ from the stabilizer limit; we have not attempted to carefully extract a best-fit exponent $\nu$.

Figure~\ref{fig:mwpm_2}(c) depicts the MWPM decoding fidelity at $\lambda = 1.002\lambda_c$, just within the confined phase. We see no evidence of a decoding threshold in this setting; in fact, for sufficiently large linear system sizes $L$, $\overline{\mathcal{F}}^{\text{MWPM}}$ appears to systematically decrease with increasing $L$ even at the smallest decoherence strengths. This observation is consistent with our expectations that the confined phase cannot safely encode quantum information even in the absence of decoherence.

\begin{figure}
    \centering
    \includegraphics[width = \textwidth]{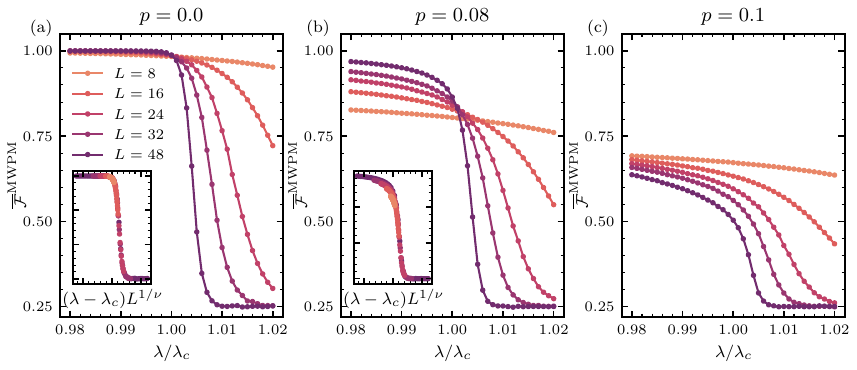}
    \caption{MWPM decoding fidelity $\overline{\mathcal{F}}^{\text{MWPM}}$ for the spacetime-isotropic TFTC at three fixed bit-flip strengths $p$, as a function of the temperature $\lambda$, at various linear system sizes $L$. (a) In the $p = 0$ limit, MWPM achieves high-fidelity decoding for all $\lambda < \lambda_{c}$, identically to the optimal postmeasurement decoder. (b) At $p = 0.08$, $\overline{\mathcal{F}}^{\text{MWPM}}$ systematically increases towards unity for all $\lambda < \lambda_{c}$, although there is a strong finite-size drift in the location of the crossing between system sizes. (c) For $p = 0.1$, the critical threshold $\lambda_{c}(p = 0.1)$ lies below $0.98 \lambda_{c}$.}
    \label{fig:mwpm_1}
\end{figure}

Figure~\ref{fig:mwpm_1}(a) depicts the MWPM decoding fidelity in the absence of bit-flip decoherence, $p = 0$, as a function of the bulk LGT temperature $\lambda$. We note that the observed data is essentially identical to the optimal decoding fidelity in the decoherence-free protocol of Sec.~\ref{sec:meas_decoding}, depicted in Fig.~\ref{fig:meas_decoding_numerics}. In particular, MWPM appears to yield a decoding fidelity exponentially close to unity throughout the topological phase $\lambda < \lambda_{c}$, to the best of our numerical resolution. We are unaware of any theoretical reason that MWPM should perform as well as the optimal postmeasurement decoder in the $p = 0$ limit.

Figure~\ref{fig:mwpm_1}(b) depicts the MWPM decoding fidelity at the fixed decoherence strength $p = 0.08$, which lies below all $\lambda$-dependent decoding thresholds $p_c^{\text{MWPM}}(\lambda)$ depicted in Fig.~\ref{fig:lgt_bitflip_decoding}(c). We see that $\overline{\mathcal{F}}^{\text{MWPM}}$ systematically increases towards unity for all $\lambda < \lambda_c$. Unlike the $p = 0$ case, there is a strong finite-size drift in the location of the crossing between different system sizes; in particular, the crossing begins well beyond $\lambda_c$ in the smallest system sizes, and appears to systematically drift towards $\lambda = \lambda_c$. We expect that in increasingly larger system sizes, the position of the crossing will continue to decrease and settle at $\lambda_c$ (for this value of $p$).

Finally, Fig.~\ref{fig:mwpm_1}(c) depicts the MWPM decoding fidelity at decoherence strength $p = 0.1$. $\overline{\mathcal{F}}^{\text{MWPM}}$ systematically decreases with increasing $L$ over the range of $\lambda$ shown. In fact, there \emph{must} exist some crossing at smaller values of $\lambda$, since $p = 0.1$ is below the known stabilizer MWPM threshold $p_{c}^{\text{MWPM}}(\lambda = 0) \approx 0.103$~\cite{wang_confinement-higgs_2003}. We have checked that our numerical method reproduces this value in the $\lambda = 0$ limit. As a further sanity check, we found a MWPM threshold at $\lambda = 0.98 \lambda_{c}$ of $p_{c}^{\text{MWPM}}(\lambda = 0.98 \lambda_{c}) \approx 0.0968$, so that $p = 0.1$ lies above the threshold at the leftmost point of Fig.~\ref{fig:mwpm_1}(c).

%% file: appendices/tfim_numerics.tex
\section{Free-Fermion Numerics for Decoding the Transverse-Field Ising Model}
\label{app:tfim_numerics}
In this Appendix, we provide details for our numerical investigation of the 1$d$ TFIM under bit-flip decoherence. 

\subsection{Setup}
We consider the 1$d$ TFIM, with \emph{open} boundary conditions for numerical convenience. The Hamiltonian is
\begin{equation}
    \begin{split}
        H &\equiv - g \sum_{j = 1}^{L} X_{j} - \sum_{j = 1}^{L-1} Z_{j} Z_{j+1} \\
        &= -i g \sum_{j = 1}^{L} \gamma_{2j-1} \gamma_{2j} - i \sum_{j = 1}^{L-1} \gamma_{2j} \gamma_{2j+1} ,
    \end{split}
\end{equation}
where in the second line, we have performed a Jordan-Wigner transformation using the Majorana operators
\begin{equation}
    \gamma_{2j-1} \equiv \qty[ \prod_{i = 1}^{j-1} X_{i} ] Z_{j}, \quad \gamma_{2j} \equiv \qty[ \prod_{i = 1}^{j-1} X_{i} ] Y_{j} ,
\end{equation}
which satisfy the anticommutation relations $\acomm{\gamma_{j}}{\gamma_{j'}} = 2 \delta_{j j'}$, as well as the useful relations $i \gamma_{2j-1} \gamma_{2j} = X_{j}$ and $i \gamma_{2j} \gamma_{2j+1} = Z_{j} Z_{j+1}$. We see that $H$ is a quadratic Majorana fermion Hamiltonian, of the general form
\begin{equation}
    H = \frac{i}{4} \sum_{i, j = 1}^{2L} \gamma_{i} A_{ij} \gamma_{j} ,
\end{equation}
where $A$ is a real antisymmetric matrix. Its ground-state covariance matrix $G_{ij} \equiv i \expval{\gamma_{i} \gamma_{j}} - i \delta_{ij}$ can then be efficiently computed by block-diagonalizing $A$, from which all higher correlations can be computed via Wick's theorem \cite{molinari2023noteswickstheoremmanybody}; see Appendix~A of Ref.~\cite{weinsteinNonlocalityEntanglementMeasured2023a} for a review. After an orthogonal transformation, $H$ is written in the form
\begin{equation}
    H = - \frac{i}{2} \sum_{\alpha = 1}^{L} \varepsilon_{\alpha} \eta_{2\alpha-1} \eta_{2\alpha}, 
\end{equation}
where $\varepsilon_{\alpha} > 0$ are the single-particle energies, ordered such that $\varepsilon_{1} \leq \varepsilon_{2} \leq \ldots$, and $\eta_{\alpha} = \sum_{j} R_{\alpha j} \gamma_{j}$ are a set of Majorana operators related to the original operators $\gamma_{j}$ by an orthogonal transformation $R$. 

In the ferromagnetic phase $g < 1$, the smallest single-particle energy $\varepsilon_{1}$ is exponentially small in $L$, while all other energies $\varepsilon_{\alpha}$ are of order unity. Furthermore, after a suitable choice of basis, the Majorana modes $\eta_{\ell} \equiv \eta_{1}$ and $\eta_{r} \equiv \eta_{2}$ are exponentially localized on the left and right ends of the chain. In the notation of Sec.~\ref{sec:tfim}, the absolute ground state $i \eta_{\ell} \eta_{r} = +1$ corresponds to the parity-even ground state $\ket{\boldsymbol{+}_{g}} \equiv\frac{1}{\sqrt{2}} (\ket{\vb{0}_g} + \ket{\vb{1}_g})$, while the first excited state $i \eta_{\ell} \eta_{r} = -1$ corresponds to the parity-odd ground state $\ket{\boldsymbol{-}_{g}} \equiv \frac{1}{\sqrt{2}} (\ket{\vb{0}_g} - \ket{\vb{1}_g})$. The two states $i \eta_{\ell} \eta_{r} = \pm 1$ therefore define a basis of two code states which are highly robust to local parity-symmetric decoherence, whose nontrivial matrix elements in the ground-state manifold are exponentially small\footnote{Explicitly, any local operator $O$ acts within the ground-state manifold as $\hat{\Pi}(g) O \hat{\Pi}(g) = a + b \eta_{\ell} + c \eta_{r} + d \, i \eta_{\ell} \eta_{r}$. $\mathbb{Z}_{2}$ symmetry of the operator $O$ requires $b = c = 0$, while locality of the operator demands that $d$ is exponentially small in $L$.}. More generally, we will regard the two-dimensional subspace spanned by $i \eta_{1} \eta_{2} = \pm 1$ as defining a code space even in the $g \geq g_{c}$ regime, allowing us to explore how the protection of information breaks down as $g$ is tuned through $g_c$.

\subsection{Numerical Protocol}
Our goal is to compute the average decoding fidelity $\overline{\mathcal{F}}(\ket{\vb{0}_g})$, given by
\begin{equation}
    \overline{\mathcal{F}}(\ket{\vb{0}_g}) \equiv \bra{\vb{0}_g} \mathcal{M} \circ \mathcal{E}(\dyad{\vb{0}_g}) \ket{\vb{0}_g},
\end{equation}
where $\mathcal{E}(\cdot) \equiv \sum_{\ve} p_{\ve} X_{\ve} (\cdot) X_{\ve}$ is the bit-flip decoherence channel, with $X_{\ve} \equiv \prod_{j} X_{j}^{e_{j}}$ and $p_{\ve} \equiv (1 - p)^{L - \abs{\ve}} p^{\abs{\ve}}$, and $\mathcal{M}(\cdot) \equiv \sum_{\vs} V_{\vs} \hat{\Pi}_{\vs} (\cdot) \hat{\Pi}_{\vs} V_{\vs}^{\dag}$ is the decoding channel, which uses the measurement operators $\hat{\Pi}_{\vs} \propto \prod_{j} [1 + (-1)^{s_{j}} Z_{j} Z_{j+1}]$ and recovery unitaries $V_{\vs} = \mathcal{U}_{g} X_{\vs}$. The latter consists of a majority-vote bit-flip operator $X_{\vs}$ which pairwise annihilates all domain walls, and a subsequent adiabatic unitary $\mathcal{U}_{g}$. As discussed in Appendix~\ref{app:strong_symm}, the decoding fidelity for the symmetry-broken state $\ket{\vb{0}_g}$ can be used to lower-bound the decoding fidelity for an arbitrary code state $\ket{\psi_{g}} = \sum_{\bs} b_{\bs} \ket{\bs_{g}}$.

To compute this quantity, we will randomly sample a bit-flip error $\ve$ with probability $p_{\ve}$, and subsequently sample a syndrome outcome $\vs$ with probability $P_{\vs | \ve} = \bra{\vb{0}_g} X_{\ve} \hat{\Pi}_{\vs} X_{\ve} \ket{\vb{0}_{g}}$. Conditioned on $\ve$ and $\vs$, we compute the fidelity
\begin{equation}
    \mathcal{F}_{\vs | \ve}(\ket{\vb{0}_g}) = \bra{\vb{0}_0} X_{\vs} \qty[ \frac{1}{P_{\vs | \ve}} \hat{\Pi}_{\vs} \qty( X_{\ve} \dyad{\vb{0}_g} X_{\ve} ) \hat{\Pi}_{\vs}  ] X_{\vs} \ket{\vb{0}_0} ,
\end{equation}
where the quantity in square brackets denotes the postmeasurement state after measuring the stabilizers $Z_{j} Z_{j+1}$. By averaging over many sampled errors $\ve$ and subsequent syndromes $\vs$, we recover the average decoding fidelity $\overline{\mathcal{F}}$. We now explain how to compute $\mathcal{F}_{\vs | \ve}(\ket{\vb{0}_g})$ using the covariance matrix formalism.

First, we require a representation of the state $\ket{\vb{0}_g}$. This is naively challenging, since $\ket{\vb{0}_g} = \frac{1}{\sqrt{2}} (\ket{\boldsymbol{+}_{g}} + \ket{\boldsymbol{-}_{g}})$ is a superposition of two Gaussian fermionic states and is not a Gaussian state itself (in particular, it does not have definite fermion parity). This issue is circumvented by introducing a zeroth auxiliary qubit, represented by the pair of Majoranas $\chi_{\ell}, \chi_{r}$ such that $X_{0} = i \chi_{r} \chi_{\ell}$, and initializing the state\footnote{An alternative understanding of this protocol is that we are maximally entangling our code space with a single-qubit reference and computing the entanglement fidelity; as discussed in Appendix~\ref{app:strong_symm}, the resulting two quantities are equal.}
\begin{equation}
    \ket*{\widetilde{\vb{0}}_g} \equiv \frac{1}{\sqrt{2}} \qty( \ket{+}_{0} \ket{\boldsymbol{+}_{g}} + \ket{-}_{0} \ket{\boldsymbol{-}_{g}} ) = \exp \qty{ -i \frac{\pi}{4} i \chi_{r} \eta_{\ell} } \ket{+}_{0} \ket{\boldsymbol{+}_{g}}
\end{equation}
by imposing the Majorana pairing $i \chi_{\ell} \eta_{\ell} = i \chi_{r} \eta_{r} = +1$. Since $\ket*{\widetilde{\vb{0}}_g}$ is obtained from a manifestly Gaussian state $\ket{+}_0 \ket{\boldsymbol{+}_{g}}$ via the Gaussian fermionic unitary $\exp \qty{ \frac{\pi}{4} \chi_{r} \eta_{\ell} }$, $\ket*{\widetilde{\vb{0}}_g}$ is itself Gaussian; moreover, it is readily verified that $\bra*{\widetilde{\vb{0}}_g} O \ket*{\widetilde{\vb{0}}_g} = \bra{\vb{0}_g} O \ket{\vb{0}_g}$ for any parity-even observable $O$, while $\bra*{\widetilde{\vb{0}}_g} Z_{0} O \ket*{\widetilde{\vb{0}}_g} = \bra{\vb{0}_g} O \ket{\vb{0}_g}$ for any parity-odd observable $O$. Therefore, we can compute any parity-even or parity-odd\footnote{Note that parity-odd local observables in the TFIM exhibit Jordan-Wigner strings in the fermionic representation, and are increasingly difficult to compute as one moves to the right in the chain. Luckily, our protocol only requires access to $Z_1$, and therefore is highly efficient.} expectation value of $\ket{\vb{0}_g}$ by applying the Gaussian-state formalism to represent $\ket*{\widetilde{\vb{0}}_{g}}$ using $2(L+1)$ fermion modes in total. For example, we can immediately compute\footnote{Note that in terms of the original Majoranas $\eta_{\alpha}$, there is a priori a U(1) ambiguity in the definition of each pair $\eta_{2\alpha - 1}, \eta_{2\alpha}$, corresponding to the undetermined phase of the complex fermion mode. We fix this ambiguity up to a sign by demanding that $\eta_{\ell}$ and $\eta_{r}$ are localized on the left and right edges of the system respectively, and the overall sign is determined by demanding that $\acomm{\eta_{\ell}}{\gamma_1}$ is positive. With this choice, $\eta_{\ell}$ and $\eta_{r}$ smoothly approach $\gamma_1$ and $\gamma_{2L}$ respectively as $g \to 0$.}
\begin{equation}
        \bra{\vb{0}_g} Z_{1} \ket{\vb{0}_g} = \bra*{\widetilde{\vb{0}}_{g}} Z_{0} Z_{1} \ket*{\widetilde{\vb{0}}_{g}} = \bra*{\widetilde{\vb{0}}_g} i \chi_{\ell} \gamma_{1} \ket*{\widetilde{\vb{0}}_g} = \frac{1}{2} \sum_{\alpha} \acomm{\eta_{\alpha}}{\gamma_{1}} \bra*{\widetilde{\vb{0}}_{g}} i \chi_{\ell} \eta_{\alpha} \ket*{\widetilde{\vb{0}}_{g}} = \frac{1}{2} \acomm{\eta_{\ell}}{\gamma_{1}} ,
\end{equation}
where we've noted that $\gamma_{j} = \sum_{\alpha} R_{\alpha j} \eta_{\alpha}$, and $R_{\alpha j} = \frac{1}{2} \acomm{\eta_{\alpha}}{\gamma_{j}}$.

Next, we sample a bit-flip error $X_{\ve}$ by independently sampling $e_{j} = 0, 1$ on each qubit. Each bit-flip operation preserves the Gaussianity of the state; a bit-flip operation $X_{k}$ on site $k$ modifies the covariance matrix as follows:
\begin{equation}
    G_{ij} \mapsto (-1)^{\delta_{i,2k-1} + \delta_{i, 2k} + \delta_{j, 2k-1} + \delta_{j, 2k}} G_{ij} .
\end{equation}
Subsequently, we measure each stabilizer $Z_{k} Z_{k+1} = i \gamma_{2k} \gamma_{2k+1}$. Such a measurement also preserves Gaussianity; each syndrome outcome $s_{k} = 0, 1$ occurs with probability $\frac{1}{2}[ 1 + (-1)^{s_{k}} G_{2k, 2k+1} ]$, and modifies the covariance matrix as
\begin{equation}
    G_{ij} \mapsto G_{ij} + \frac{G_{i,2k+1} G_{j,2k} - G_{i,2k} G_{j,2k+1}}{(-1)^{s_{k}} + G_{2k, 2k+1}} ,
\end{equation}
where we have used Wick's theorem \cite{molinari2023noteswickstheoremmanybody} to obtain the latter expression. We measure each stabilizer subsequently from the left end of the chain to the right end, updating the covariance matrix at each step.

Finally, we implement the majority vote $X_{\vs}$ and measure the fidelity $\mathcal{F}_{\vs | \ve}$. Assuming for a moment that the majority vote has been performed, the final state $\ket{\psi_{g}(\vs | \ve)} \equiv P_{\vs | \ve}^{-1/2} X_{\vs} \hat{\Pi}_{\vs} X_{\ve} \ket{\vb{0}_g}$ lies within the repetition code space, and so can be expanded as $b_{\vb{0}} \ket{\vb{0}_0} + b_{\vb{1}} \ket{\vb{1}_0}$. The desired overlap is then given by
\begin{equation}
    \mathcal{F}_{\vs | \ve}(\ket{\vb{0}_g}) = \abs{b_{\vb{0}}}^{2} = \frac{1}{2} \qty[ 1 + \bra{\psi_{g}(\vs | \ve)} Z_{1} \ket{\psi_{g}(\vs | \ve)} ] .
\end{equation}
In other words, $\mathcal{F}_{\vs | \ve}$ can be obtained by computing the expectation value of $Z_{1} = i \chi_{\ell} \gamma_{1}$ in the final state. Notably, the value of $Z_{1}$ in the final state depends only on whether the first qubit was flipped by $X_{\vs}$ or not, i.e., whether it belonged to the majority domain. We can therefore implement the majority vote simply by computing the total magnetization \emph{relative} to $Z_{1}$ using the syndrome outcomes $s_{j}$:
\begin{equation}
    m_{1} = \sum_{j = 2}^{L} Z_{1} Z_{j} = \sum_{j = 1}^{L-1} (-1)^{\sum_{i \leq j} s_{i}} .
\end{equation}
Qubit 1 belongs to the majority (minority) domain if $m_{1} > 0$ ($m_{1} < 0$), and so we simply flip the value of $Z_{1}$ whenever $m_{1} < 0$. In our numerics we demand that $L$ is even, so that $m_{1}$ is never zero.

\subsection{Additional Numerical Plots}
For completeness, here we provide a few supplementary numerical plots for the decoding protocol described above and in Sec.~\ref{subsec:tfim_numerics}.

\begin{figure}
    \includegraphics[width=\textwidth]{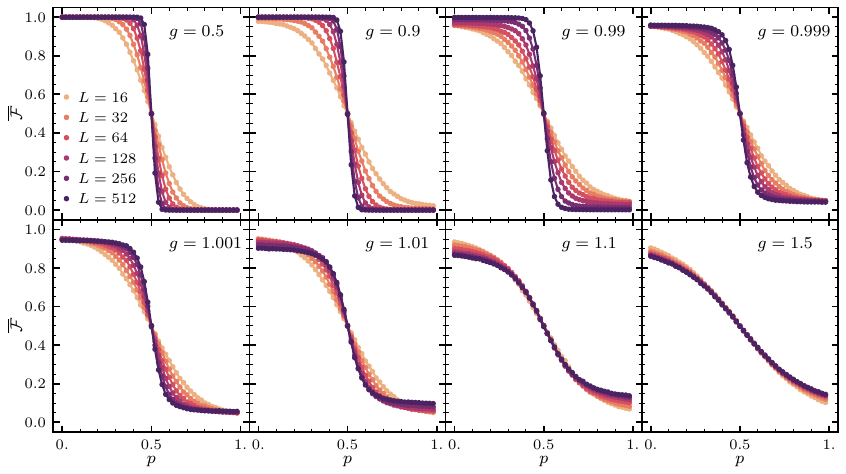}
    \caption{Average decoding fidelity $\overline{\mathcal{F}}$ for the (1+1)$d$ TFIM subjected to bit-flip decoherence and decoded with $Z_j Z_{j+1}$ stabilizer measurements and majority vote, as a function of the decoherence strength $p$, for several values of the transverse field $g$.}
    \label{fig:ising_bitflip_fixedg}
\end{figure}

Figure~\ref{fig:ising_bitflip_fixedg} depicts the decoding fidelity as a function of the bit-flip probability $p$, for several different values of $g$ below and above the critical point $g_c = 1$, and for several different system sizes $L$. For $g < 1$, we see qualitatively similar behavior as in Fig.~\ref{fig:ising_bitflip}(a); in particular, for all $p < \frac{1}{2}$, $\overline{\mathcal{F}}$ systematically increases towards unity with increasing system size. For $g > 1$, we instead see at small $p$ that $\overline{\mathcal{F}}$ systematically decreases with increasing system size, and appears to approach an $L$-independent curve with non-maximal $\overline{\mathcal{F}}$ everywhere. Note that the crossing at $p = \frac{1}{2}$ in these $g > 1$ plots is entirely due to the antisymmetry property $\overline{\mathcal{F}}(1-p) = 1 - \overline{\mathcal{F}}(p)$, and is not necessarily indicative of a transition. At very large values of $g$, we see that $\overline{\mathcal{F}}$ is almost completely independent of $L$.

\begin{figure}
    \includegraphics[width=\textwidth]{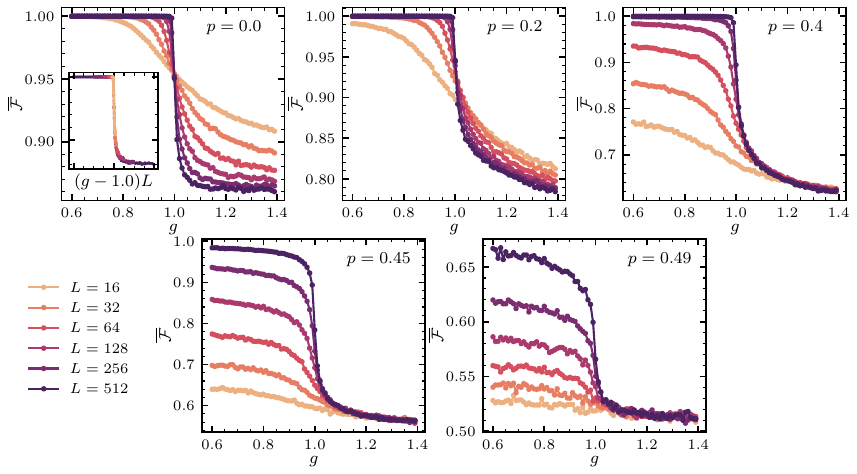}
    \caption{Average decoding fidelity $\overline{\mathcal{F}}$ for the (1+1)$d$ TFIM subjected to bit-flip decoherence and decoded with $Z_j Z_{j+1}$ stabilizer measurements and majority vote, as a function of the transverse field strength $g$, for several values of the decoherence strength $p$ and at various system sizes $L$.}
    \label{fig:ising_bitflip_fixedp}
\end{figure}

Figure~\ref{fig:ising_bitflip_fixedp} depicts the decoding fidelity as a function of the transverse field $g$, for several different values of the bit-flip probability $p$. Our $p = 0$ results align well with previous numerical results obtained in Ref.~\cite{lavasani_stability_2025}; in particular, an excellent finite-size scaling collapse is achieved using the correlation length exponent $\nu = 1$ of the (1+1)$d$ TFIM. Interestingly, the behavior at nonzero values of $p$ appears qualitatively different than at $p = 0$: specifically, $\overline{\mathcal{F}}$ does not appear to exhibit a crossing for nonzero values of $p$, and at large values of $p$ instead immediately collapses to an $L$-independent curve for $g > 1$. Given the somewhat ad-hoc definition of our decoding protocol in the $g > 1$ regime, we have not attempted to explain this feature of our numerical results.

\begin{figure}
    \centering
    \includegraphics[width=\textwidth]{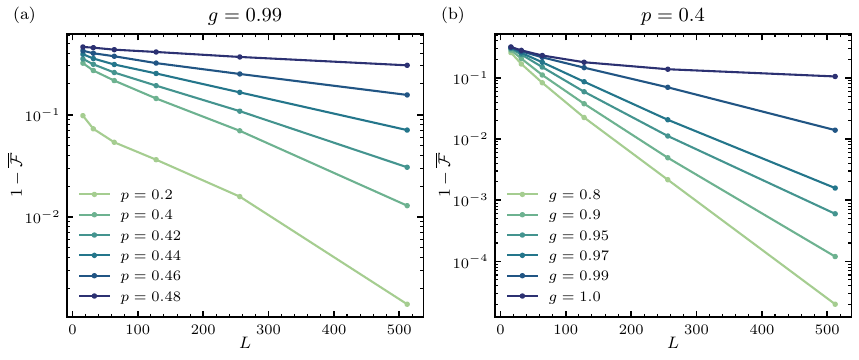}
    \caption{Deviation of the average decoding fidelity $\overline{\mathcal{F}}$ in the (1+1)$d$ TFIM from its maximal value of unity, as a function of system size $L$, for several different values of the transverse field $g$ and decoherence strength $p$. In all cases observed, $1 - \overline{\mathcal{F}}$ decreases with $L$ exponentially quickly.}
    \label{fig:ising_bitflip_Lscaling}
\end{figure}

Finally, Figure~\ref{fig:ising_bitflip_Lscaling} depicts the decoding fidelity as a function of system size; Fig.~\ref{fig:ising_bitflip_Lscaling}(a) depicts several values of $p$ at fixed $g = 0.99$, while Fig.~\ref{fig:ising_bitflip_Lscaling}(b) depicts several values of $g$ at fixed $p = 0.4$. We see that the fidelity relaxes to its maximal value of unity exponentially quickly for all $g < 1$ and $p < \frac{1}{2}$, as expected from the discussion of Sec.~\ref{sec:tfim}. 

